\documentclass[preprint,number,twocolumn,5p]{elsarticle}
\usepackage{subcaption}
\usepackage{amssymb}
\usepackage{booktabs}
\usepackage{upgreek} 
\usepackage{mhchem}
\usepackage[dvipsnames]{xcolor}
\usepackage{hyperref}
\usepackage[capitalise,noabbrev]{cleveref}
\usepackage[separate-uncertainty=true]{siunitx}
\DeclareSIUnit{\electron}{e^-}
\begin{document}
\journal{Nuclear Instrumentation and Methods in Physics Research A}

\begin{frontmatter}



\title{Characterization of the H2M Monolithic CMOS Sensor}

\author[cern]{Rafael~Ballabriga}
\author[bnl]{Eric~Buschmann}
\author[cern]{Michael~Campbell}
\author[ifae]{Raimon~Casanova~Mohr}
\author[cern]{Dominik~Dannheim}
\author[desy]{Jona~Dilg}
\author[cern]{Ana~Dorda}
\author[desy]{Ono~Feyens}
\author[desy]{Finn~King}
\author[cern]{Philipp~Gadow}
\author[desy]{Ingrid-Maria~Gregor}
\author[desy]{Karsten~Hansen}
\author[desy]{Yajun~He}
\author[desy]{Lennart~Huth}
\author[cern]{Iraklis~Kremastiotis}
\author[desy]{Stephan~Lachnit}
\author[cern,iphc]{Corentin~Lemoine}
\author[desy]{Stefano~Maffessanti}
\author[desy]{Larissa~Mendes}
\author[cern]{Younes~Otarid}
\author[desy]{Christian~Reckleben}
\author[cern]{Sébastien~Rettie}
\author[desy]{Manuel~Alejandro~del~Rio~Viera}
\author[desy]{Sara~Ruiz~Daza\corref{mycorrespondingauthor}}
\cortext[mycorrespondingauthor]{Corresponding author}
\ead{sara.ruiz.daza@desy.de}
\author[desy]{Judith~Schlaadt}
\author[desy]{Adriana~Simancas}
\author[cern]{Walter~Snoeys}
\author[desy]{Simon~Spannagel}
\author[desy]{Tomas~Vanat}
\author[desy]{Anastasiia~Velyka}
\author[desy]{Gianpiero~Vignola}
\author[desy]{H{\aa}kan~Wennl{\"o}f}

\affiliation[cern]{
    organization={CERN},
    addressline={Esplanade des Particules 1},
    city={CH-1211 Geneva 23},
    country={Switzerland}
}
\affiliation[bnl]{
    organization={Brookhaven National Laboratory (BNL)},
    addressline={New York 11973-5000},
    city={Upton},
    country={USA}
}
\affiliation[ifae]{
    organization={Institut de Física d’Altes Energies (IFAE)},
    addressline={Edifici CN, UAB campus},
    city={08193 Bellaterra (Barcelona)},
    country={Spain}
}
\affiliation[desy]{
    organization={Deutsches Elektronen-Synchrotron DESY},
    addressline={Notkestr. 85},
    city={22607 Hamburg},
    country={Germany}
}
\affiliation[iphc]{
    organization={IPHC, Université de Strasbourg},
    addressline={23 rue du Loess},
    city={Strasbourg},
    country={France}
}

\begin{abstract}

The H2M (Hybrid-to-Monolithic) is a monolithic pixel sensor manufactured in a modified \SI{65}{\nano\meter}~CMOS imaging process with a small collection electrode.
Its design addresses the challenges of porting an existing hybrid pixel detector architecture into a monolithic chip, using a digital-on-top design methodology, and developing a compact digital cell library. 
Each square pixel integrates an analog front-end and digital pulse processing with an 8-bit counter within a \SI{35}{\micro\meter}~pitch. 

This contribution presents the performance of H2M based on laboratory and test beam measurements, including a comparison with analog front-end simulations in terms of gain and noise. 
A particular emphasis is placed on backside thinning in order to reduce material budget, down to a total chip thickness of \SI{21}{\micro\meter} for which no degradation in MIP detection performance is observed.
For all investigated samples, a MIP detection efficiency above \SI{99}{\%} is achieved below a threshold of approximately \SI{205}{} electrons. At this threshold, the fake-hit rate corresponds to a matrix occupancy of fewer than one pixel per the \SI{500}{\nano\second} frame.

Measurements reveal a non-uniform in-pixel response, attributed to the formation of local potential wells in regions with low electric field. 
%--- an effect intensified by the relatively large pixel pitch. These wells slow down charge collection, and due to the fast response of the CSA, the signal amplitude is reduced for slower charges, degrading the hit detection efficiency. 
A simulation flow combining technology computer-aided design, Monte Carlo, and circuit simulations is used to investigate and describe this behavior, and is applied to develop mitigation strategies for future chip submissions with similar features.
\end{abstract}


\end{frontmatter}

\tableofcontents


\section{Introduction}

Monolithic active pixel sensors (MAPS) with a small collection electrode are a promising sensor type for future high-energy physics (HEP) experiments. This includes tracking or vertexing applications at a lepton collider like CLIC~\cite{clic-report} or FCC-ee~\cite{fccee}. MAPS come with the benefit of integrating the readout electronics and sensitive volume in the same die, which inherently reduces the material budget but also removes the necessity of the costly bump bonding process. A small collection electrode layout results in a small input capacitance to the first amplifier stage (order of \SI{}{\femto\farad}), which reduces power consumption for a given signal charge~\cite{snoeys2012}. The TPSCo \SI{65}{\nano\meter} ISC (Image Sensor CMOS) process~\cite{walter} is a promising technology in that regard. It allows for a higher density of circuit elements than larger feature size processes, and technology demonstrators such as APTS~\cite{apts} or DPTS~\cite{dpts} have qualified it for applications in HEP.

The hybrid-to-monolithic (H2M) test chip is a MAPS produced in this \SI{65}{\nano\meter} ISC process. It explores complex per-pixel pulse processing functionality for time and amplitude measurements, at the cost of a pitch of \SI{35}{\micro\meter}, which is \SI{40}{\percent} larger than other prototypes developed for High Energy Physics in this process. 
%The architecture of the chip is ported from a Timepix-like hybrid pixel detector. 
The architecture of the chip is based on readout concepts from earlier hybrid readout ASICs~\cite{LLOPART2007485,clicpix, clicpix2}, with a clock distributed over the matrix and pulse-processing counters in every pixel.
This porting process is interesting to explore, as the large set of hybrid chips for HEP applications may speed up the development of future monolithic sensors. The design applied a digital-on-top workflow, which is a contemporary strategy for the design of complex integrated circuitry. The digital logic is synthesized around the analog building blocks such that certain constraints (e.g. on timing) are fulfilled. The design effort included the composition of a compact digital cell library comprising a set of standard circuit elements with a footprint reduced by about \SI{25}{\percent} compared to the standard cells.
%With these efforts, the pitch is 37\% smaller than that of Timepix (\SI{55}{\micro\meter}), from which the equivalent functionality originates.

Preliminary results from the characterization and simulation of H2M are presented in~\cite{h2m_measurements,h2m_simulations}, respectively. \cref{chip} of this article describes the H2M chip design, its features, and the data acquisition system utilised for the presented measurements. Laboratory characterization and calibration are presented in \cref{lab}, while the methodology for test-beam measurements and the resulting performance in the detection of minimum-ionizing particles (MIPs) are detailed in \cref{sec:tb_method} and~\cref{sec:tb_results}. \cref{sec:simulation} summarizes the employed simulation procedure, which has proven to be a valuable tool to understand the results of the measurements.
Conclusions and an outlook to further plans are provided in~\cref{sec:summary}.

\section{The H2M Chip}\label{chip}
% general descriptions (floor plan, dimension, etc.)
The H2M chip matrix consists of 64$\times$16 square pixels with a pitch of \SI{35}{\micro\meter}, resulting in a total sensitive area of 2.24$\times$\SI{0.56}{\milli\meter\squared}. 
The total chip size is 3$\times$\SI{1.5}{\milli\meter\squared}, including the 
periphery, differential-to-CMOS receivers/drivers, and I/O connectors at the bottom of the pixel matrix.
Each pixel features a collection electrode, an analog front-end, and digital logic capable of 8-bit pulse processing. 
The following section describes the sensor design, including the analog front-end and digital logic, as well as the chip periphery and readout architecture. 

\subsection{Sensor Design}
\begin{figure}[tbp]
    \centering
    \includegraphics[width=1\linewidth]{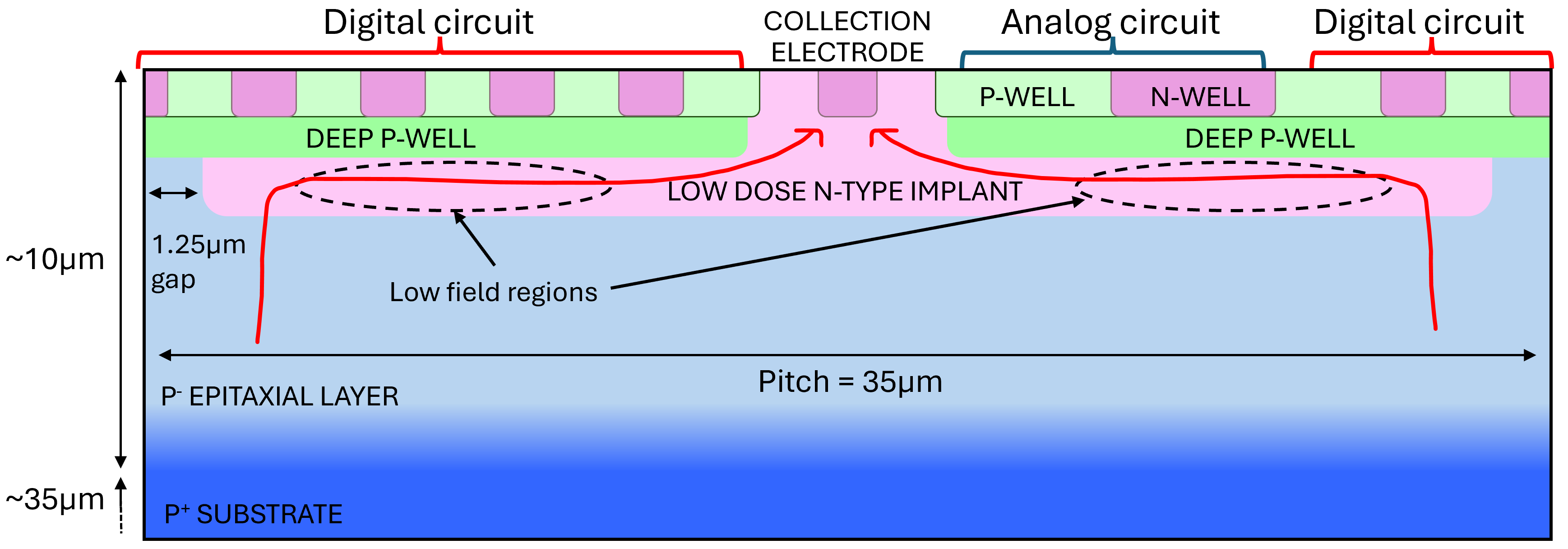}
    \caption{H2M cross section sketch of one pixel, with drift paths of electrons indicated by red arrows. The approximately \SI{5}{\micro\meter} metal interconnect layers are not shown in the schematic.} 
    \label{fig:h2m_cross_section}
\end{figure}
The sensor is manufactured in a modified TPSCo \SI{65}{\nano\meter} ISC process~\cite{walter}, and has been thinned on wafer level to a total thickness of approximately \SI{50}{\micro\meter}.
This includes around \SI{5}{\micro\meter} of metal interconnect layers, a \SI{10}{\micro\meter} p-type epitaxial layer, and a \SI{35}{\micro\meter} p-type substrate. 
Due to the comparatively high doping concentration of the substrate, the lifetime of charge carriers is significantly reduced in this region, leading to rapid recombination before they can contribute to signal formation. Therefore, mainly the epitaxial layer serves as an active volume for the signal formation.

In order to enhance the depletion within the epitaxial layer, a low-dose n-type implant is introduced, as illustrated in \cref{fig:h2m_cross_section}. This implant has a \SI{2.5}{\micro\meter} gap at the pixel boundaries, forming a vertical pn-junction that introduces a lateral component of the electric field. 
As a result, the charge collection efficiency and timing performance of the sensor are improved. 
This design is known as \emph{modified with gap} layout~\cite{walter}.

\begin{figure}[tbp]
    \centering
    \includegraphics[width=0.6\linewidth]{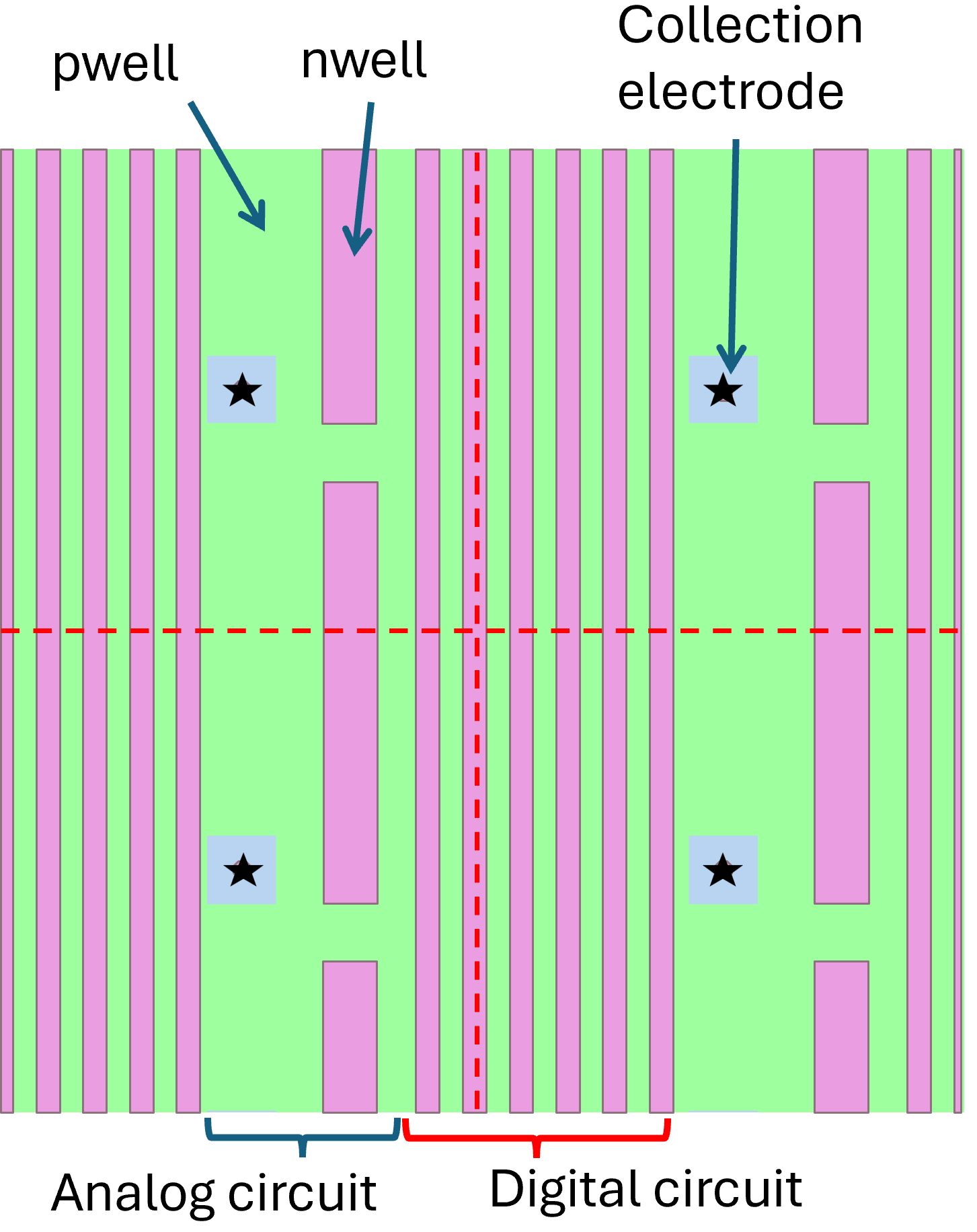}
    \caption{Simplified layout of the n-well and p-well positions within the deep p-well in four pixels. The locations of the analog front-end and digital logic are shown. The collection electrode, placed outside the deep p-well, is also indicated. The pixel cell boundary is marked with a dashed red line.}
    \label{fig:h2m_nwells}
\end{figure}

Each pixel has a small n-type collection electrode and integrates full CMOS circuitry using the quadruple well technology~\cite{quadrupoletech}. This enables the implementation of PMOS transistors in n-wells and NMOS transistors in p-wells, shielded by a deep p-well to prevent charge collection by these n-wells. The positions of these n-wells and p-wells within the deep p-well are sketched in \cref{fig:h2m_nwells}.
The n-type collection electrode is held at a positive potential of \SI{0.8}{V}.

The sensor is reverse-biased from the front side via a contact located outside the pixel matrix, connecting the p-well and p-substrate to the same potential. The limitations of this reverse bias are discussed in \cref{lab:iv}. 
In the following, this reverse bias applied to the p-well and p-substrate is referred to as the \textit{sensor bias voltage}.
Since there is no junction isolation at the location of the gap in the deep n-type implant between the p-well and the p-substrate, current flows through the deep implant gap when the p-well is biased differently from the p-substrate. This prevents the sensor from operating with different p-well and p-substrate bias voltages. 

In this work, the impact of backside thinning (reduction of the substrate thickness) on chip performance will be discussed through laboratory and test beam measurements. In particular, results on samples thinned on single-die level by backside grinding to between \SI{21}{\micro\meter} and \SI{30}{\micro\meter} will be presented.

\subsection{Analog Front-End}
\label{chip:analog}
% how does it work, which parameters are adjustable
% - needed for the equalization part
%   - daq_vthr
%   - tuning_daq

The in-pixel front-end consists of a collection electrode with a charge-sensitive amplifier (CSA) with Krummenacher feedback~\cite{krummenacher}, followed by a continuous-time threshold comparator. 

The global threshold is defined by \texttt{dac\_vthr}, an 8-bit digital-to-analog converter (DAC) located in the chip periphery. Local fine-tuning of the threshold is achieved through an additional 4-bit DAC (\texttt{tuning\_dac}) implemented within each pixel. 
The step size, and thus the dynamic range, of this fine-tuning is determined by another 8-bit DAC in the periphery (\texttt{dac\_itrim}), which defines the current used in the tuning DACs.

\begin{figure}[tbp]
    \centering
    \includegraphics[width=1\linewidth, trim={1cm 0cm 1cm 0.5cm},clip]{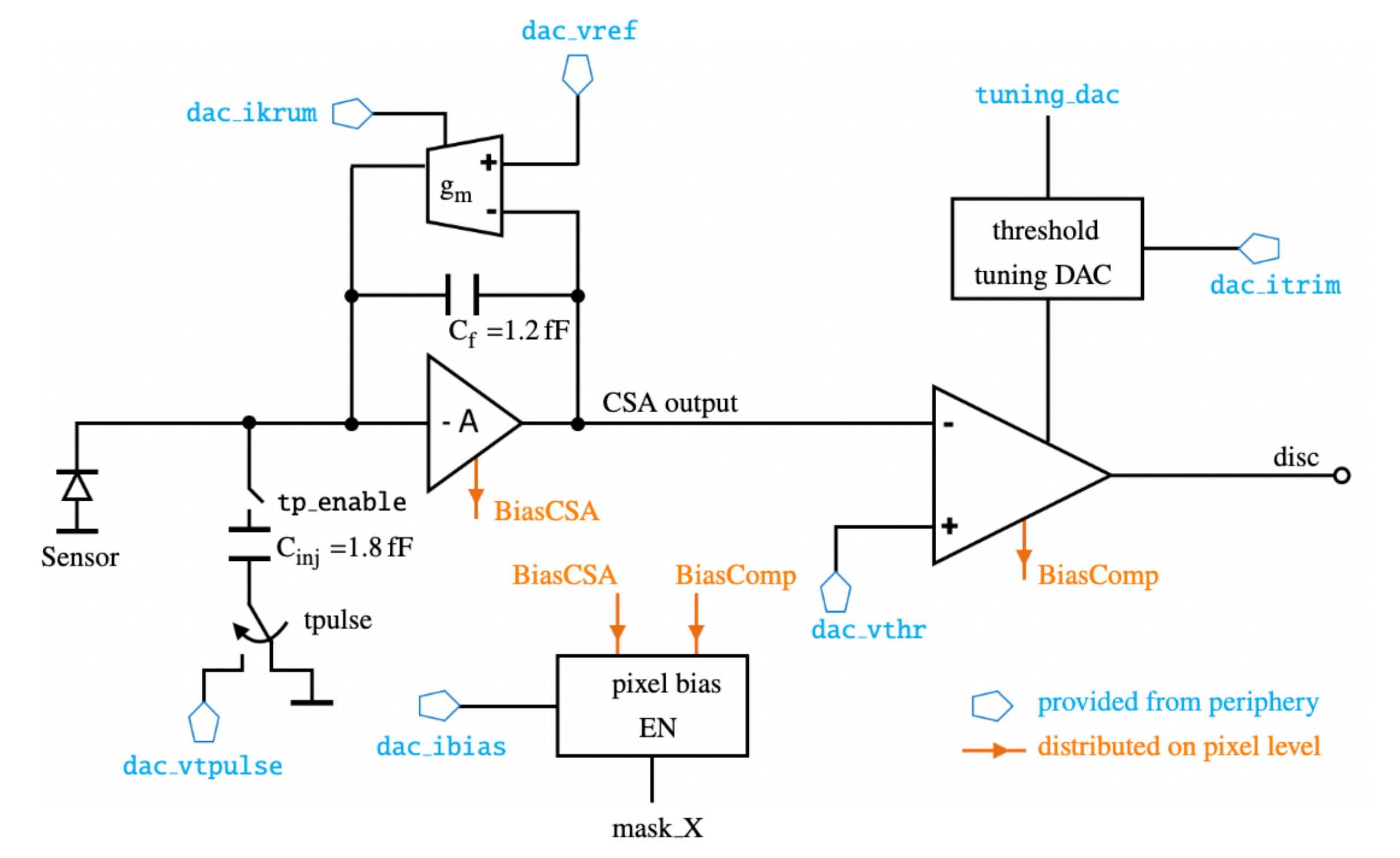}
    \caption[Schematic of the pixel analog front-end]{Schematic of the pixel analog front-end.}
    \label{fig:analog}
\end{figure}

The front-end also offers test pulse injection through a dedicated capacitor connected to the CSA input. Test pulses are enabled via the \texttt{tp\_enable} DAC, and their amplitude is controlled by a global 8-bit DAC (\texttt{dac\_vtpulse}). This amplitude determines the voltage step applied to the injection capacitor, enabling the injection of a controlled amount of charge into the pixel.
Individual pixels can be masked by disabling their digital logic and setting the analog front-end into a low-power mode. 
A schematic of the pixel analog front-end is shown in \cref{fig:analog}.

The inverting amplifier amplifies and converts the negative output of the charge collected by electrons into a positive output. The gain simulations are presented in~\cref{sec:christian}.
The linear discharge rate of the CSA is controlled by a feedback current (\textit{ikrum}), which is set by an 8-bit DAC in the chip periphery. This feedback current defines how fast the amplifier resets after a hit. It compensates for leakage current,  and enables energy measurements, as the time the signal remains above the threshold is proportional to the input charge. \cref{fig:acq_modes} (top) illustrates a triangular signal shape for two different \textit{ikrum} settings. A higher \textit{ikrum} results in a steeper return-to-baseline slope (faster signal decay). 
%The transconductance, \textit{gm}, operating point is approximately \SI{23}{nS}.

Two additional 8-bit DACs are included in the analog periphery, \texttt{dac\_ibias} and \texttt{dac\_vref}, which set the CSA current bias and the Krummenacher reference voltage, respectively.

\subsection{Digital Logic \& Acquisition Modes}
\label{chip:digital}
\begin{figure}[tbp]
    \centering
    \includegraphics[width=1\linewidth]{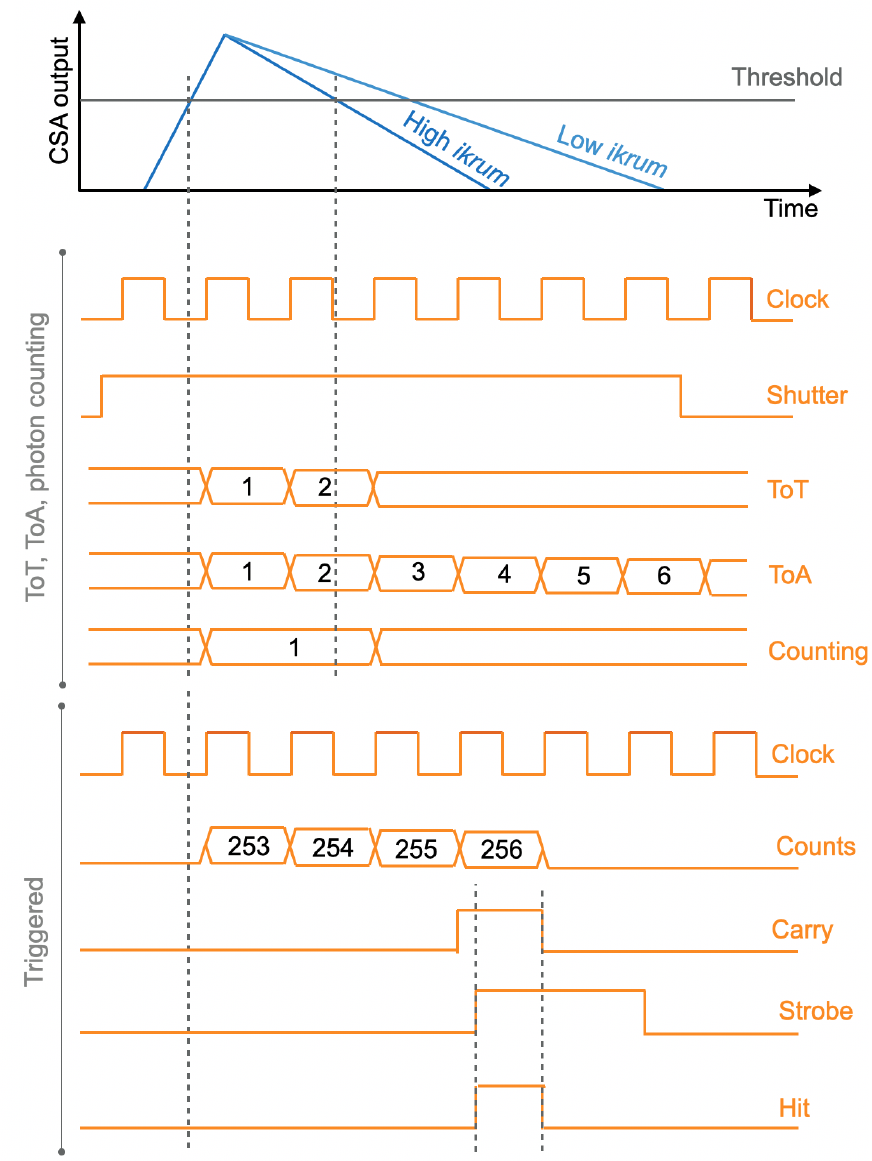}
    \caption{Schematic layout of the four acquisition modes. On top, the CSA output signal is represented with a triangular shape, whose return-to-baseline slope depends on the \textit{ikrum} setting. In the middle, for the high \textit{ikrum}, the signal is compared to a reference voltage threshold, and the corresponding measured counts in ToT, ToA, and photon counting modes are indicated. At the bottom, a readout in triggered mode is depicted, using as an example, a configurable preset of 253 counts and a strobe duration of 2 clock cycles.}
    \label{fig:acq_modes}
\end{figure}

%which operation modes are available
% - needed for the equalization part
%   - counting mode
The H2M test chip can operate in four non-simultaneous acquisition modes: Time-over-Threshold (ToT), Time-of-Arrival (ToA), photon counting, and triggered.
For this purpose, each pixel includes a configurable 8-bit counter, which is used differently in each acquisition mode.
A schematic layout of the four acquisition modes is shown in \cref{fig:acq_modes}.
In ToT, ToA, and photon counting modes, a shutter signal controls the data acquisition, while in triggered mode, it is controlled by a strobe signal.
%Each pixel includes a configurable 8-bit counter, enabling the H2M test chip to operate in four non-simultaneous acquisition modes. These acquisition modes can be categorised into two groups: Timepix-like and LHC-like acquisition modes. The Timepix-like modes are Time-over-Threshold (ToT), Time-of-Arrival (ToA), and photon counting, while the LHC-like is the triggered acquisition mode. 

ToA records the time at which the signal exceeds the threshold during charge collection, while ToT measures how long the signal stays above that threshold.
More specifically, the ToA is defined as the number of clock cycles from the moment the signal crosses the threshold until the shutter closes, and the ToT as the number of clock cycles between the rising and falling edges of the signal crossing the threshold within a single shutter. 
If multiple signals cross the threshold within the same shutter window, their respective ToT values are accumulated for each pixel. 
The acquisition runs at an externally generated \SI{100}{MHz}, corresponding to \SI{10}{\nano\second} per clock cycle, and setting a lower limit on the time/energy resolution in ToA/ToT modes of \SI{10}{\nano\second}$\mathrm{/\sqrt{12} \approx} $\SI{2.9}{\nano\second}.
A time walk correction cannot be applied since a simultaneous ToT and ToA measurement is not possible with the chip.

In the photon counting mode, each pixel registers the number of times that the signal crosses the threshold within a shutter window.

In triggered mode, an external trigger signal is used to validate hits before readout. The in-pixel 8-bit counter is preset with a value corresponding to the expected fixed delay of the trigger latency. 
When a pixel fires, the counter starts incrementing from this preset value and generates a carry signal once it reaches its maximum. This carry signal lasts for one clock cycle.
A strobe signal, provided externally after the trigger signal arrives, is distributed to the pixels using the same line as the shutter in other acquisition modes. This signal defines an adjustable coincidence window, accommodating for the H2M time walk. If the carry and strobe signals coincide, a binary readout is issued. 
In this way, shorter shutter durations can be utilized in test beam measurements compared to ToT and ToA acquisition modes (see \cref{sec:tb_method}), reducing the probability of noise hits and allowing the operation of the chip at lower thresholds.

\subsection{Analog Front-End Simulations}\label{sec:christian}
\begin{figure}[tbp]
    \centering
\includegraphics[width=1\linewidth]{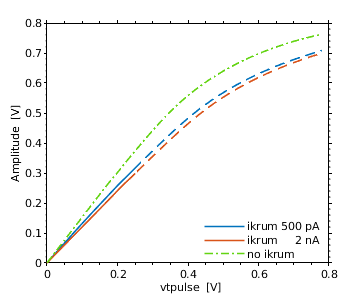}
    \caption{Simulated amplitude response of the CSA as a function of the injected signal \texttt{vtpulse}, for three different \textit{ikrum} settings.}
    \label{fig:Ampl_vtpulse}
\end{figure}
\begin{figure}[tbp]
    \centering
\includegraphics[width=1\linewidth]{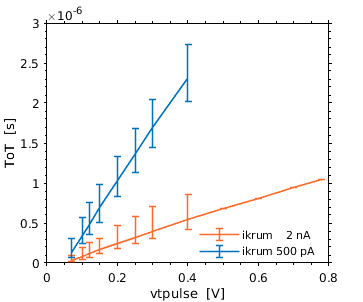}
    \caption{Simulated ToT response to injected signal \texttt{vtpulse}. The deviations (error bars) are primarily caused by variances in \textit{ikrum}.}
    \label{fig:ToT_vtpulse}
\end{figure}

Analog front-end simulations have been performed to validate the circuit performance of the final layout, including all the parasitic elements of the circuit. The sensor is emulated with a \SI{2}{fF} capacitance.

\cref{fig:Ampl_vtpulse} shows the signal amplitude at the output of the CSA as a function of the injected test pulse \texttt{vtpulse} for two different \textit{ikrum} configurations.
Because of the faster discharging of the feedback capacitance C$_\mathrm{{f}}$ for larger Krummenacher bias currents, the gain is reduced from 1.269 V/V to 1.198 V/V~\cite{Ratti}. 
The values are extracted from the linear range, indicated in solid lines. 
Recalculating both gains using the injection capacitance yields a charge sensitivity of \SI{113}{mV/ke^-} and \SI{107}{mV/ke^-}, respectively. 
The maximum gain is theoretically achievable for $\textit{ikrum} = 0$ (no discharging of C$_\mathrm{{f}}$), included in \cref{fig:Ampl_vtpulse}.  It is given by the ratio $\mathrm{C_{inj}/C_{f}  = 1.5}$ when the CSA is operated as a voltage amplifier.

The simulated ToT behavior is shown in \cref{fig:ToT_vtpulse} as a function of the injected test pulse \texttt{vtpulse} for the same \textit{ikrum} settings as in \cref{fig:Ampl_vtpulse}. 
For small \textit{ikrum} values, the CSA pulse width is longer, and with it, the ToT is also increased, resulting in a larger ToT gain. The non-linear region for small input signals is suppressed by setting the comparator threshold to \SI{70}{\milli\volt} (corresponding to \SI{620}{electrons} using the above-mentioned gain) above the baseline. Below this threshold, C$_\mathrm{{f}}$ discharges exponentially with a non-constant \texttt{vtpulse}-dependent discharging current. Whereas for \texttt{vtpulse} values above \SI{70}{\milli\volt}, a constant current of $\mathrm{0.5 \cdot\textit{ikrum}}$~\cite{Ratti} leads to a linear decrease of the CSA output signal. In this region, the CSA output can be approximated as a triangular waveform over time, with a steep rising edge and a slow falling edge, giving:

\begin{equation}\label{eq:tot}
    \mathrm{
    ToT \simeq  2 \ C_f \cdot\frac{Amplitude-vthr}{\textit{ikrum}}
    } .
\end{equation}

The error bars in~\cref{fig:ToT_vtpulse} indicate the ToT variance arising from process statistics in the simulation, with the minimum and maximum deviations shown. The main source of these deviations is transistor mismatch affecting the \textit{ikrum} current distribution from the periphery across the pixel matrix. The error bars are asymmetric because \textit{ikrum} appears in the denominator of \cref{eq:tot}.  

\begin{figure}[tbp]
    \centering
\includegraphics[width=1\linewidth]{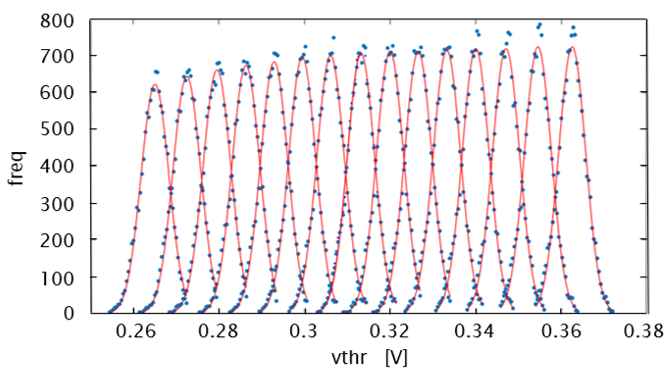}
    \caption{Simulated occupancy as a function of threshold, \texttt{vthr}, for the different \texttt{tuning\_dac} settings. The curves are the fitted Gaussian functions.}
    \label{fig:noise}
\end{figure}

\cref{fig:noise} shows the result of a transient noise~\cite{transient-noise} analysis while scanning the threshold. The simulation is repeated for each of the 15 DAC settings of the threshold \texttt{tuning\_dac}. The RMS of the Gaussian-like distributions represents the CSA noise. It is the same for all of the 15 distributions $\mathrm{\upsigma \approx 3.2 \ mV_{rms}}$, which corresponds to approximately  \SI{28}{e^-_{rms}}. The mean of the distributions is shifted by about \SI{6}{\milli\volt} per DAC step. 
This value can be adjusted by varying the periphery-generated DAC bias current \texttt{itrim}.

\subsection{Configuration \& Readout}
\label{chip:readout}
% DAQ system: how is everything configured, powered, read out
The chip configuration is performed via a slow control interface based on the Medipix4/Timepix4 protocol~\cite{medipix4,timepix4}, featuring a serial communication bus with data and clock inputs and outputs. 
Configuration signals are distributed either globally (to control settings such as acquisition start/stop, shutter, acquisition mode selection, and trigger latency), or locally to individual pixels (enabling features like test pulse injection, pixel masking, and tuning DAC). 

While the clock frequency of the slow control is \SI{40}{MHz}, the readout is performed using a \SI{25}{MHz} clock. 
During readout, the 8-bit pixel counters are connected as a shift register and their contents are sequentially shifted from the top to the bottom of each column across the entire matrix, without zero suppression. 
%This is a synchronous readout system, where data acquisition is gated by the global shutter signal.

%Configuration and data readout are both performed using the Caribou DAQ system.

\refstepcounter{subsubsection}
\subsubsection*{\thesubsubsection\quad Caribou DAQ System}\label{chip:caribou}

The characterization of the H2M chip is performed using the Caribou data acquisition (DAQ) system~\cite{caribou}. This is a modular and flexible platform developed for the prototyping and testing of detectors. It combines a System-on-Chip (SoC) running an embedded Linux Operating System (OS), a user-configurable FPGA firmware, and the \textit{Peary} software to provide a streamlined solution for detector control, configuration, and data readout. Caribou’s hardware, centered around the Control and Readout (CaR) board, offers interfaces, programmable power supplies, and versatile connections to easily integrate a wide range of detector prototypes. 

\begin{figure}[tbp]
  \centering
  \begin{subfigure}{0.45\textwidth}
    \centering
    \includegraphics[width=0.75\linewidth]{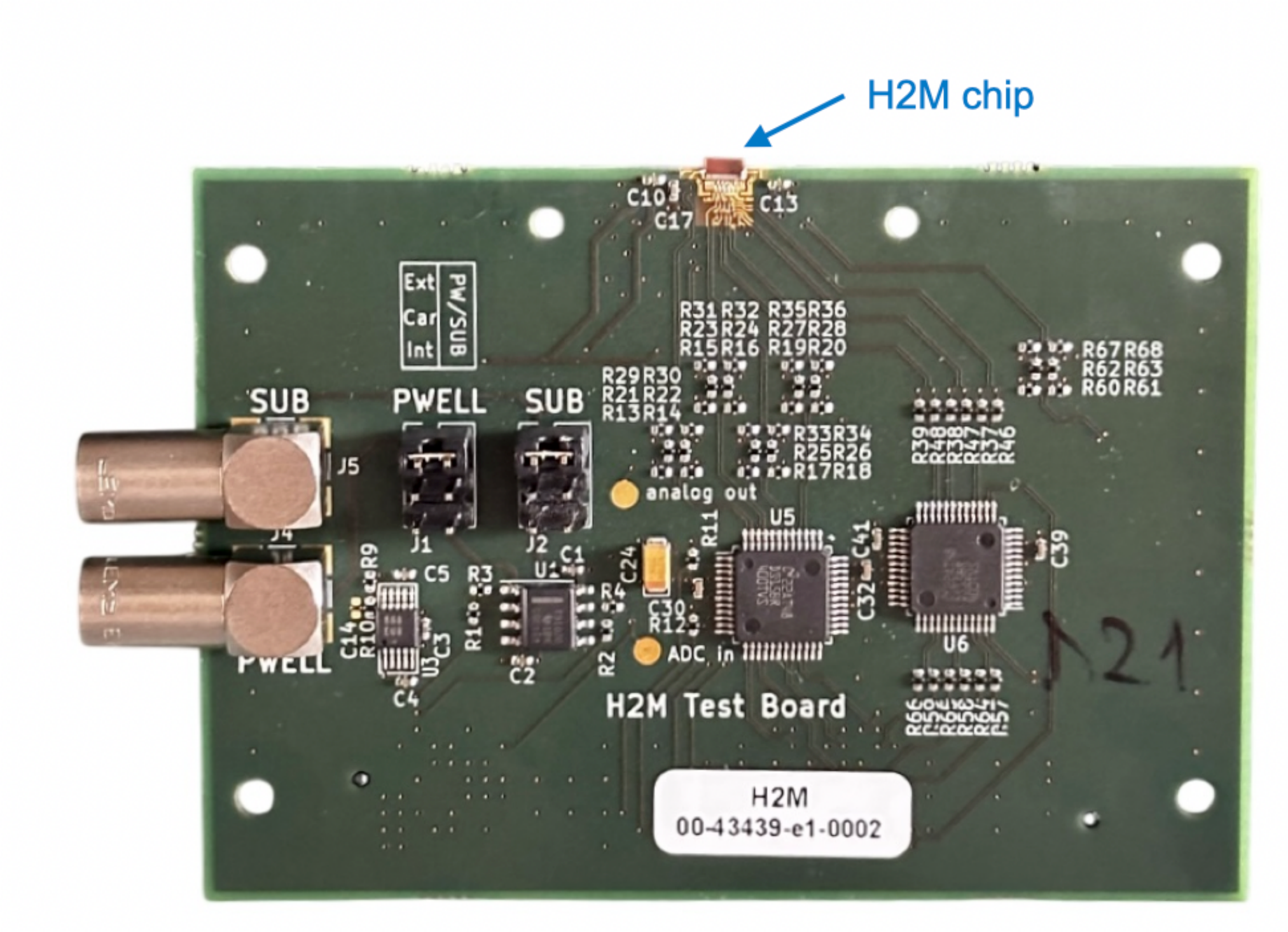}
    \caption{Picture of an H2M chip mounted on a chip board.}
    \label{fig:h2m_chipboard_unzoomed}
  \end{subfigure}
  \hfill
  \vspace{2mm}
  \begin{subfigure}{0.45\textwidth}
    \centering
    \includegraphics[width=0.65\linewidth]{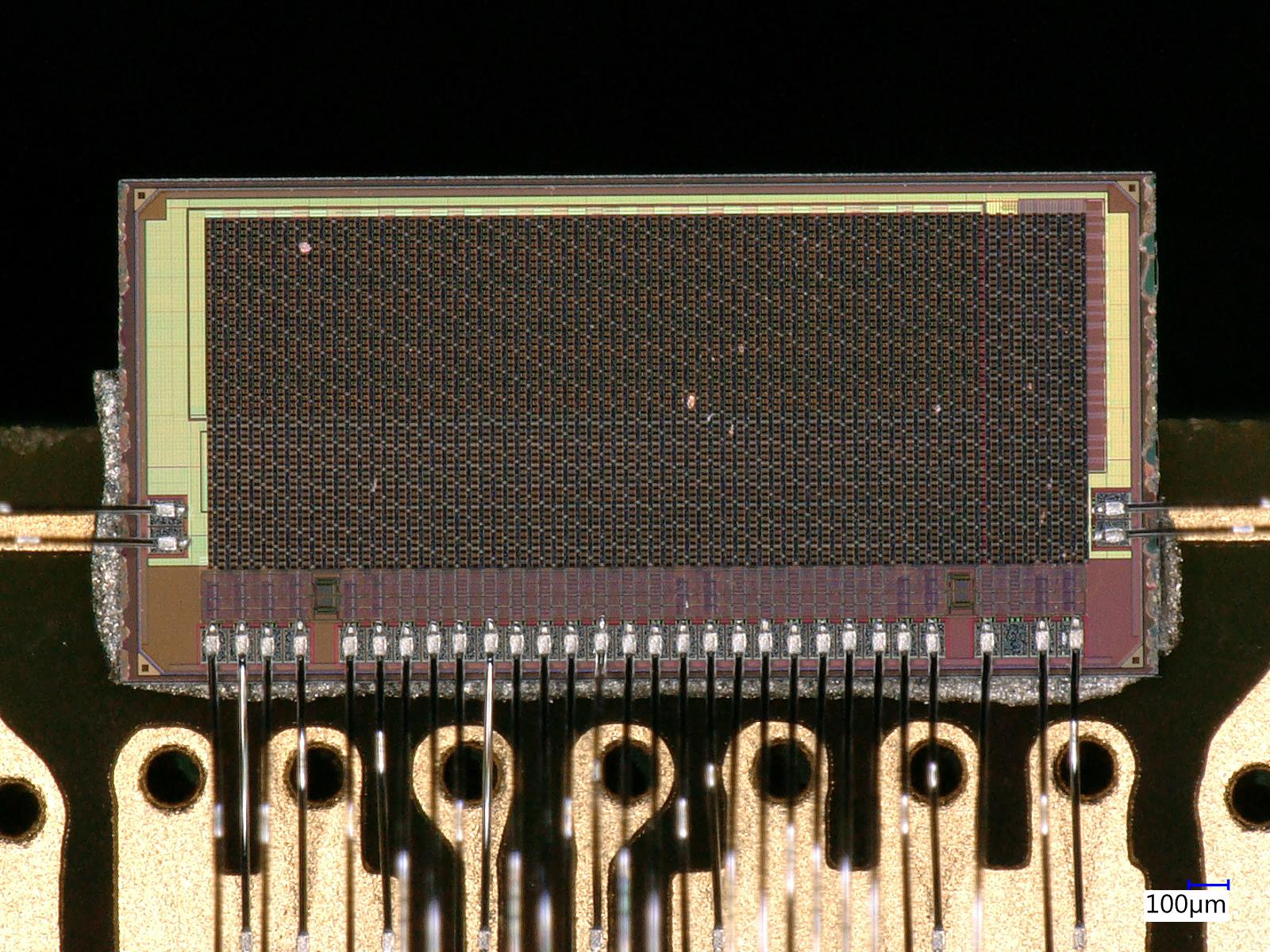}
    \caption{Zoomed-in picture of an H2M chip mounted on a chip board.}
    \label{fig:h2m_chipboard_zoomed}
  \end{subfigure}
  \caption{Picture of an H2M chip glued and wire-bonded at the edge of a chip board.}
  \label{fig:h2m_chipboard}
\end{figure}

For this work, the H2M device was integrated into the Caribou system through a three-step approach. First, a custom printed circuit board (PCB), referred to as the chip board, was developed to host the H2M chip and incorporate the necessary passive components for signal conditioning of all control and readout lines. To allow for backside illumination and reduction of the material budget, the chip is mounted at the edge of the board, with approximately half of its area extending beyond the edge without mechanical support. A picture of an H2M chip mounted on a chip board is shown in \cref{fig:h2m_chipboard}. Second, dedicated FPGA firmware was implemented to provide the device’s slow control interface and handle its custom readout algorithms. Finally, a Peary software interface was developed to manage power-up, configuration, control, and data acquisition in a unified and automated manner. A comprehensive set of test procedures was then established to validate the system's correct operation and to assess the performance of the H2M device under both laboratory and test-beam conditions, as detailed below.

\section{Laboratory Characterization}\label{lab}
This section presents the electrical and functional characterization of the H2M chip, including threshold and ToT calibration using radioactive sources and test pulses.
Unless explicitly stated otherwise, the discussed results correspond to the H2M-3 sample, which has a total thickness of \SI{50}{\micro\meter}. However, all the samples presented in the paper underwent the same measurements and yielded similar results, as will be demonstrated.

% (possibly generic measurements, like IV, DAC scans)

\subsection{Current-Voltage Characteristic}\label{lab:iv}
A current-voltage (IV) measurement has been performed, where the full chip
%sensor leakage current 
p-well and substrate bias currents are recorded as a function of their reverse bias voltages (the sensor bias voltage). The configured chip is kept in the dark, at room temperature, and the p-well and substrate bias voltages are increased simultaneously in steps of \SI{0.02}{\volt}. After each step, a delay of \SI{1.5}{\second} is applied to allow for stabilization, followed by \SI{20}{} measurements from which the mean and RMS are calculated. 

\begin{figure}[tbp]
    \centering
\includegraphics[width=1\linewidth]{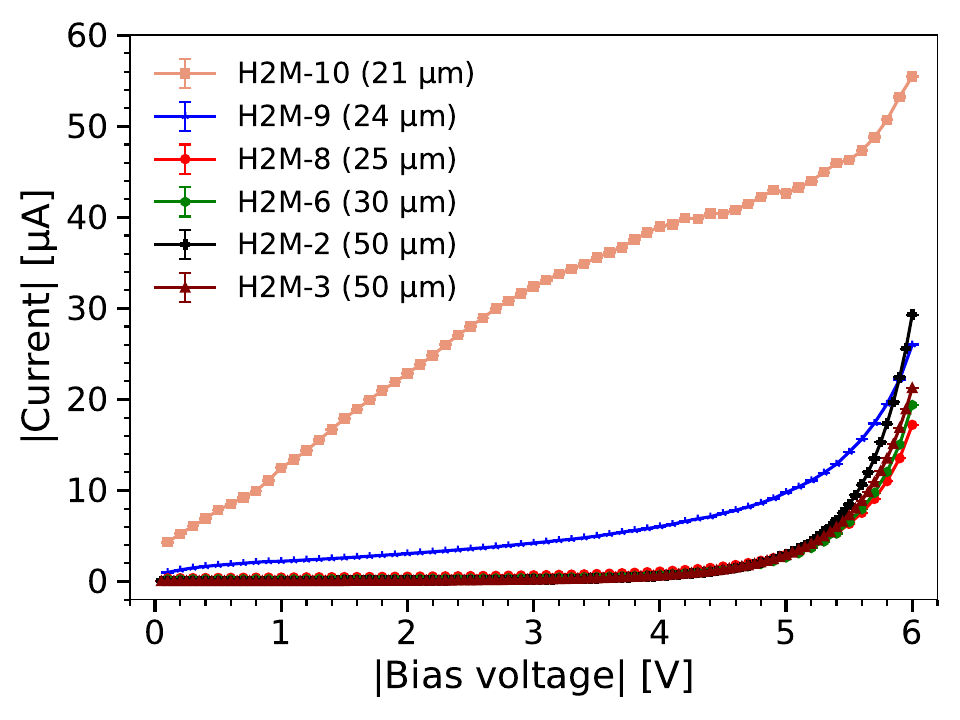}
    \caption{IV measurements for samples with different total chip thicknesses.}
    \label{fig:iv}
\end{figure}

\cref{fig:iv} shows the IV measurement for samples with different total chip thicknesses.
Significantly higher currents are observed in samples with thicknesses below \SI{25}{\micro\meter}, and further systematic studies of thin samples are currently ongoing to have a better understanding of this behavior. 
All samples show a punch-through current at absolute bias voltages above \SI{4.8}{\volt}. 
At this point, the change in current between two consecutive points exceeds \SI{0.2}{\micro\ampere}, and the current begins to increase exponentially. 
An upper limit of \SI{-4.2}{\volt} is therefore defined as a safe operational sensor bias voltage for laboratory and test beam measurements.

\subsection{Threshold Equalisation} \label{lab:equalization}
Variations in the characteristics of circuit elements, and current and voltage supplies over the pixel matrix cause pixel-to-pixel variations of the threshold~\cite{heim2017,rossi2006}. If unaccounted for, these variations limit the minimum achievable threshold for noise-free operation. Hence, the analog front-end allows adjustment of the threshold for each individual pixel (\texttt{tuning\_dac}), as detailed in \cref{chip:analog}. To find the optimal \texttt{tuning\_dac} setting for each pixel, a noise-based equalisation procedure, similar to the one described in~\cite{kremastiotis2020}, is applied. The advantage of this method, compared to those based on calibration-pulse injection, is its insensitivity with respect to gain variations. 

\begin{figure}[tbp]
    \centering
    \includegraphics[width=1\linewidth]{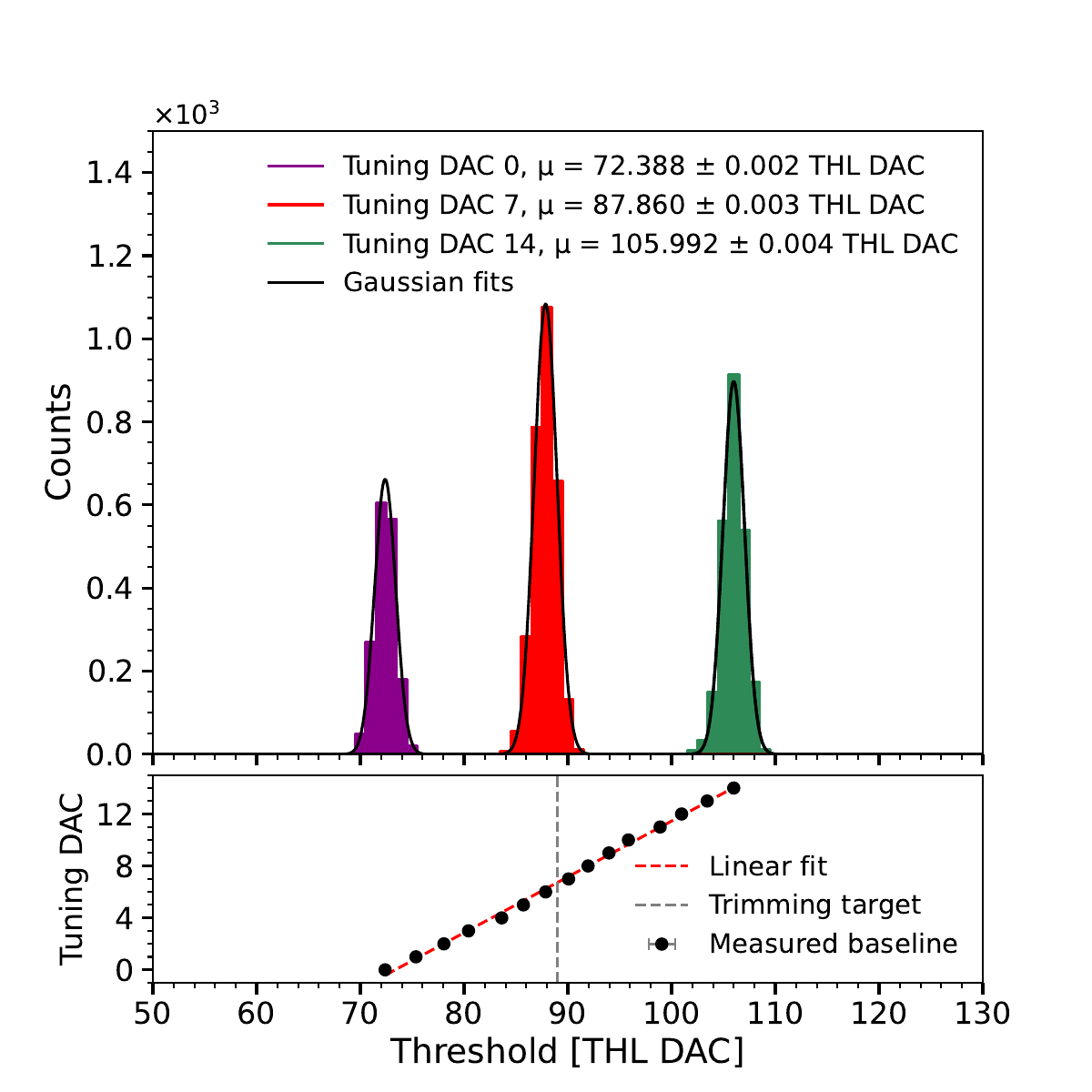}
    \caption{On the top: Occupancy as a function of the threshold for a pixel operated at three different tuning DACs. The mean of the Gaussian fit defines the measured baseline position. On the bottom: the tuning DAC setting as a function of its baseline position. The sensor is biased at \SI{-1.2}{V}.}
    \label{fig:pixel_thl}
\end{figure}

The first step of the equalisation procedure is a scan of the parameter \texttt{dac\_vthr} in units of the DAC (THL DAC), referred to as a threshold scan. 
This is performed while operating the chip in the dark, configured in counting mode, and acquiring data for a fixed duration at each threshold, without test-pulse injection. 
Since applying a masking pattern to avoid instabilities has not been found to improve the results, the procedure is applied to the entire matrix at once. 
Plotting the number of acquired counts as a function of the threshold for each pixel reveals the baseline position of the pixels with respect to the threshold (see top of \cref{fig:pixel_thl}). When the baseline and the threshold are close, the baseline noise is sufficient to toggle the discriminator. Hence, an excess of counts indicates the relative position of the baseline.

This threshold scan is repeated for all possible settings of \texttt{tuning\_dac}. The baseline of each pixel is determined from a Gaussian fit to the above-mentioned plot, as shown at the bottom of \cref{fig:pixel_thl}. 
A linear fit is used to describe the relation between the relative baseline position of each pixel and the tuning DAC setting, in order to find the setting closest to a trimming target. This trimming target is chosen such that it is reachable by the largest possible number of pixels. In particular, a pixel is masked when the Gaussian fit does not converge or if the trimming target cannot be reached with any tuning DAC setting. 

\begin{figure}[tbp]
    \centering
    \includegraphics[width=1\linewidth, trim={0.5cm 4cm 0 4cm},clip]{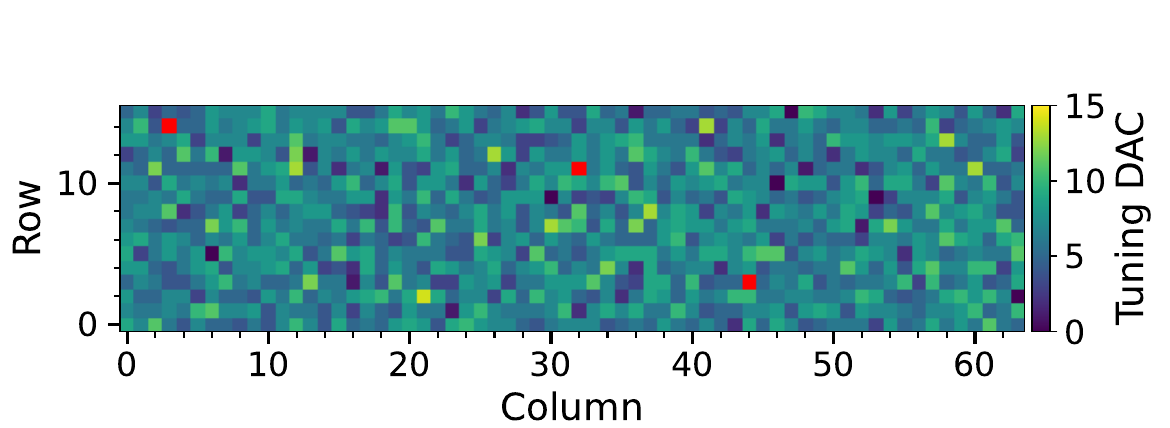}
    \caption{Tuning DAC setting map after equalisation. The three masked pixels are indicated in red, and the sensor is biased at \SI{-1.2}{V}.}
    \label{fig:trim_dacs}
\end{figure}

The distribution of the tuning DAC settings across the matrix, as well as the masked pixels, is shown in \cref{fig:trim_dacs}.  
The uniform distribution, which makes full use of the available tuning range and results in fewer than \SI{0.3}{\%} of pixels being masked, 
%indicates that no regions in the matrix are affected by voltage drops.
indicates that no systematic effects are observed that would lead to localised variations of the pixel baselines or discriminator thresholds. 

\begin{figure}[tbp]
    \centering
    \includegraphics[width=1\linewidth]{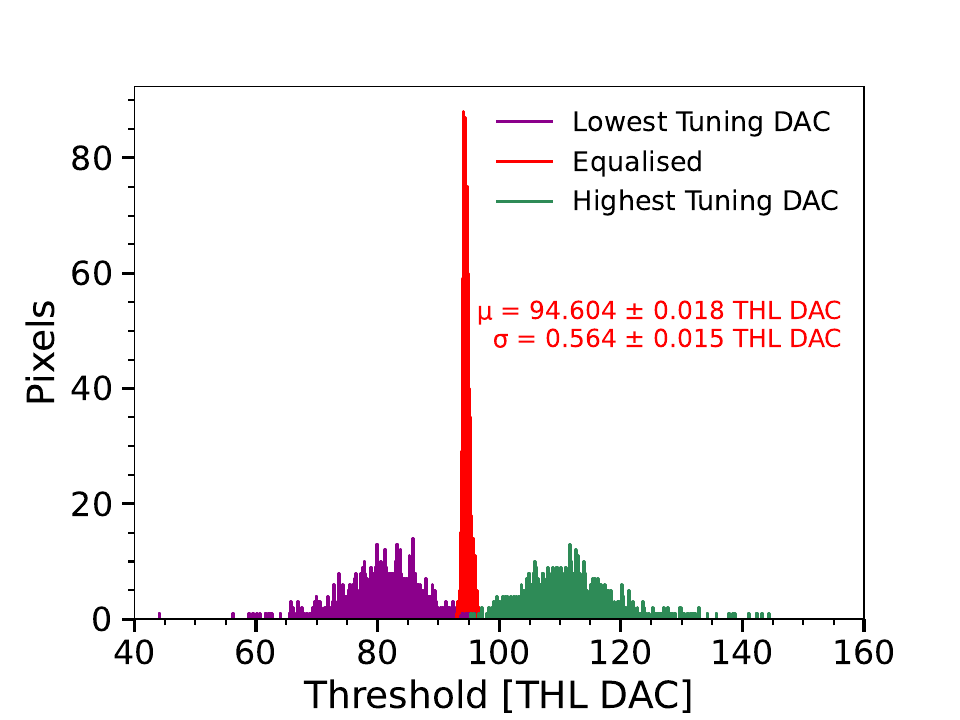}
    \caption{Baseline distribution for the equalised matrix, and the lowest and highest tuning DACs. The sensor is biased at \SI{-1.2}{V}.}
    \label{fig:thl_equalisation}
\end{figure}

Using the equalised matrix, with the pixels configured as in \cref{fig:trim_dacs}, a new threshold scan in counting mode is performed. As before, the count distribution for each pixel is fitted with a Gaussian, where the mean represents its baseline position and the width indicates its single-pixel noise. 
\cref{fig:thl_equalisation} shows the baseline distributions of the entire matrix when all pixels are set to the lowest and highest tuning DAC value, as well as after equalisation.
The mean $\mathrm{\upmu}~=~$\SI{94.604 \pm 0.018}{}~THL~DAC of the equalised distribution corresponds to the chip baseline position, and the width $\upsigma~=~$\SI{0.564 \pm 0.015}{}~THL~DAC represents the threshold dispersion between pixels.  %\cref{fig:singlepixelnoise} shows the single-pixel noise distribution after equalization.
Additionally, the single-pixel noise, shown in~\cref{fig:singlepixelnoise}, is measured as the RMS of the curve obtained in the threshold scan (see \cref{fig:pixel_thl} top), and has a mean of \SI{1.386 \pm 0.004}{}~THL~DAC after equalisation.
This threshold dispersion and single-pixel noise can be converted into electrons after the threshold calibration described in \cref{lab:calibration}. It yields to a threshold dispersion of approximately \SI{17}{} electrons and a mean single-pixel noise of \SI{45}{} electrons RMS when the sensor is biased at \SI{-1.2}{V}.
The measured single-pixel noise is slightly higher than the simulated one (see~\cref{sec:christian}). Although the presented simulations rely on several assumptions, the differences are still under investigation. 
%with Random Telegraph Noise~\cite{randomtelegraphnoise} suspected as a contributing factor. 
In particular, the outliers in the single-pixel noise distribution (\cref{fig:singlepixelnoise}), together with ongoing measurements, suggest that Random Telegraph Noise~\cite{randomtelegraphnoise} is a likely contributing factor.

\begin{figure}[tbp]
  \begin{subfigure}{0.25\textwidth}
    \includegraphics[width=\linewidth]{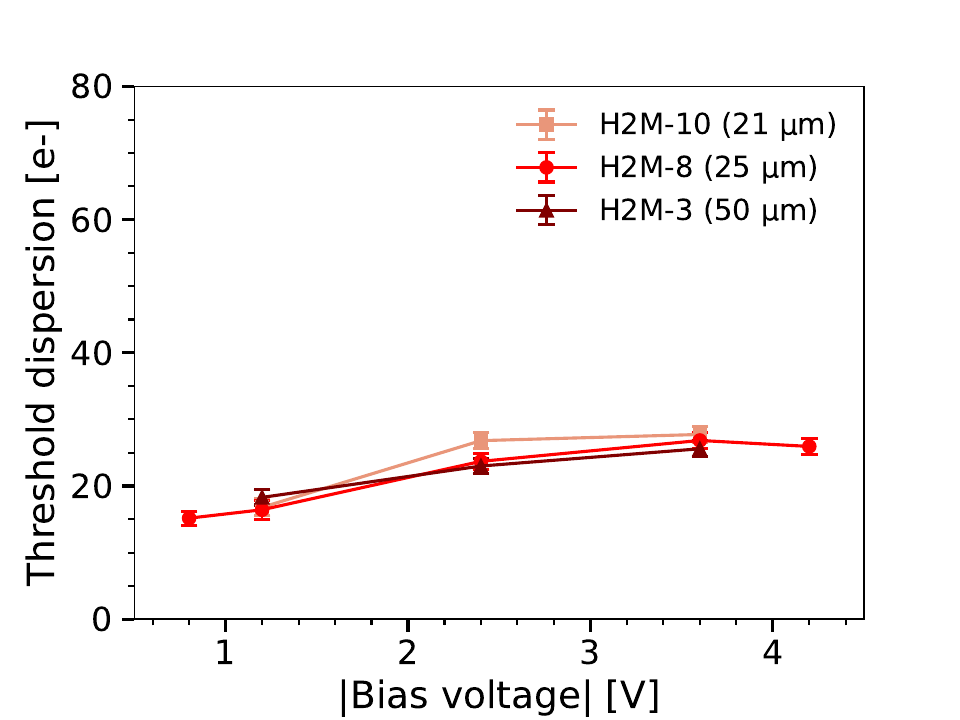}
    \caption{} \label{fig:thl_dis_bias}
  \end{subfigure}%
  \hspace*{\fill}   % maximize separation between the subfigures
  \begin{subfigure}{0.25\textwidth}
    \includegraphics[width=\linewidth]{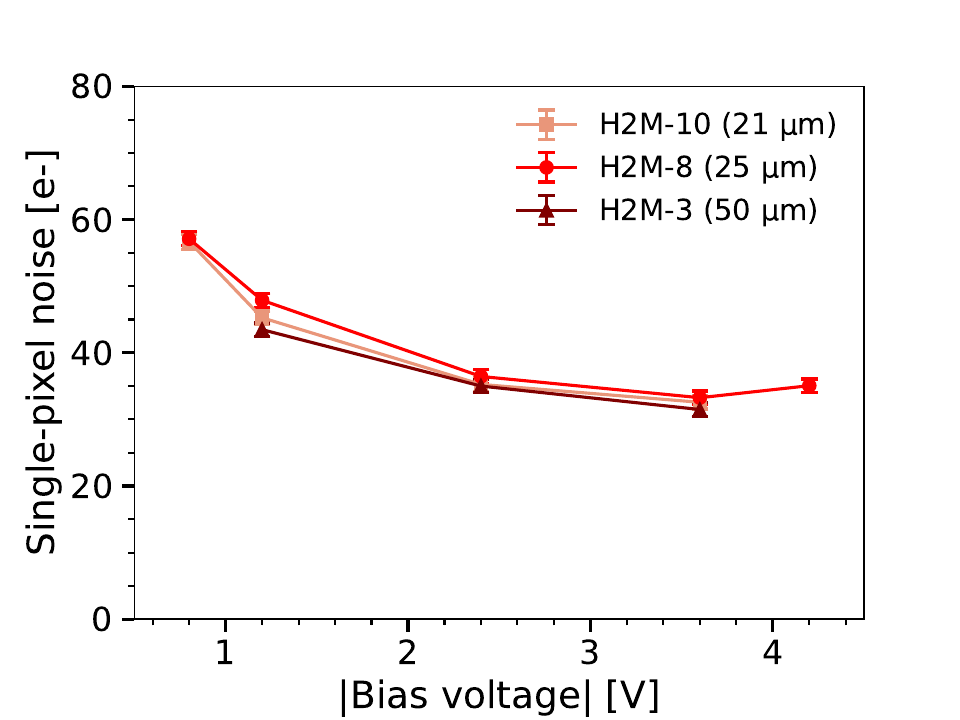}
    \caption{} \label{fig:spn_bias}
  \end{subfigure}%
%\begin{subfigure}{0.33\textwidth}
%\includegraphics[width=\linewidth]{equalisation/totalnoise_vs_bias.pdf}
%\caption{} \label{fig:stotalnoise_bias}
%\end{subfigure}%
\caption{Threshold dispersion (a) and mean single-pixel noise (b) as a function of the bias voltage for samples with different total chip
thicknesses.}\label{fig:total_noise}
\end{figure}

\cref{fig:total_noise} shows the impact of the sensor bias voltages on the threshold dispersion and single-pixel noise for samples with different total chip thicknesses. 
No significant differences are observed between samples.
The combination of noise and threshold dispersion defines the rate of fake hits at a certain threshold level.
Towards lower bias voltages ($< \SI{2}{V}$), the detector capacitance increases and noise dominates the fake hit rate. At higher bias voltages, threshold dispersion among pixels~\footnote{A higher reverse bias voltage reduces the drive current of the NMOS, slowing down digital gates, and reduces the active channel volume, making the transistor more sensitive to dopant fluctuations and thus increasing the threshold dispersion.}, but the single-pixel noise still dominates the fake hit rate.
To overcome the increased threshold dispersion at \SI{-4.2}{V}, a lower value for the \texttt{dac\_itrim} is used at the cost of masking additional pixels. This results in a smaller step size in the tuning process, covering a smaller threshold dynamic range. Consequently, \SI{5}{} extra pixels (8 out of 1024) are masked compared to the other bias voltages. 
Operating the chip at different sensor bias voltages changes the tuning DAC setting by one or two DACs for approximately 20\% of the pixels in the matrix.

\begin{figure}[tbp]
    \centering
    \includegraphics[width=1\linewidth]{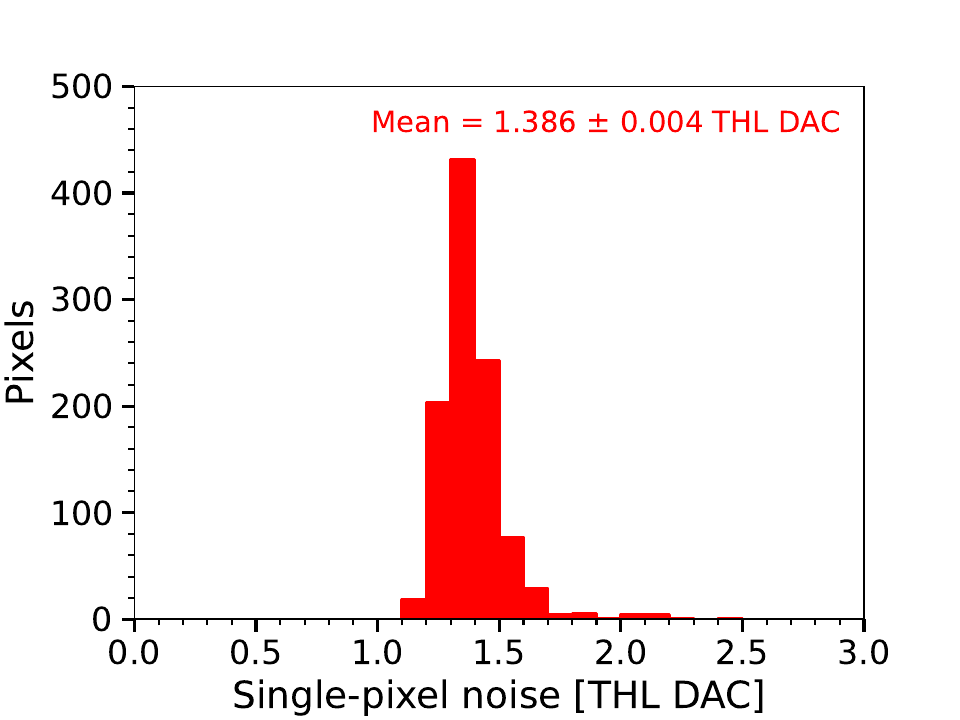}
    \caption{Single-pixel noise distribution of the pixel matrix.}
    \label{fig:singlepixelnoise}
\end{figure}

\iffalse % removing figure
\begin{figure}[tbp]
    \centering
    \includegraphics[width=1\linewidth]{singlepixelnoise_h2m3_1V2.pdf}
    \caption{Single-pixel noise distribution of the equalised matrix. The sensor is biased at \SI{-1.2}{V}.}
    \label{fig:singlepixelnoise}
\end{figure}
\fi

\subsection{Threshold Calibration}
\label{lab:calibration}
%source measurements with iron-55

\begin{figure}[tbp]
    \centering
    \includegraphics[width=1\linewidth]{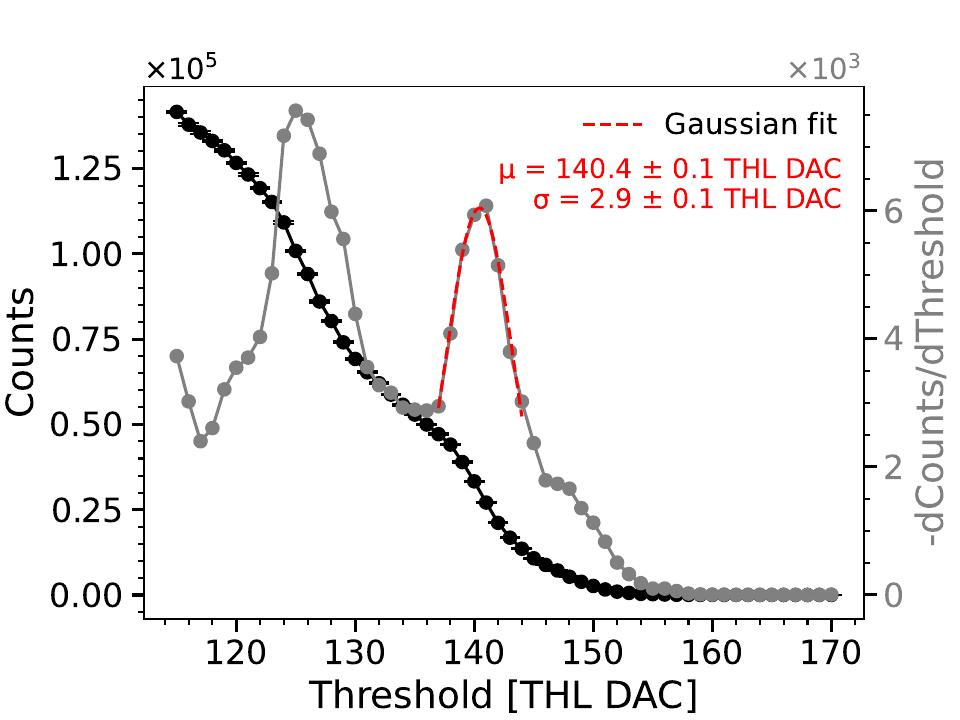}
    \caption{Occupancy (left axis) and its derivative (right axis) for the full matrix as a function of the threshold for the \ce{^{55}Fe} source.}
    \label{fig:thl_calibration}
\end{figure}

\begin{figure}[tbp]
    \centering
    \includegraphics[width=1\linewidth]{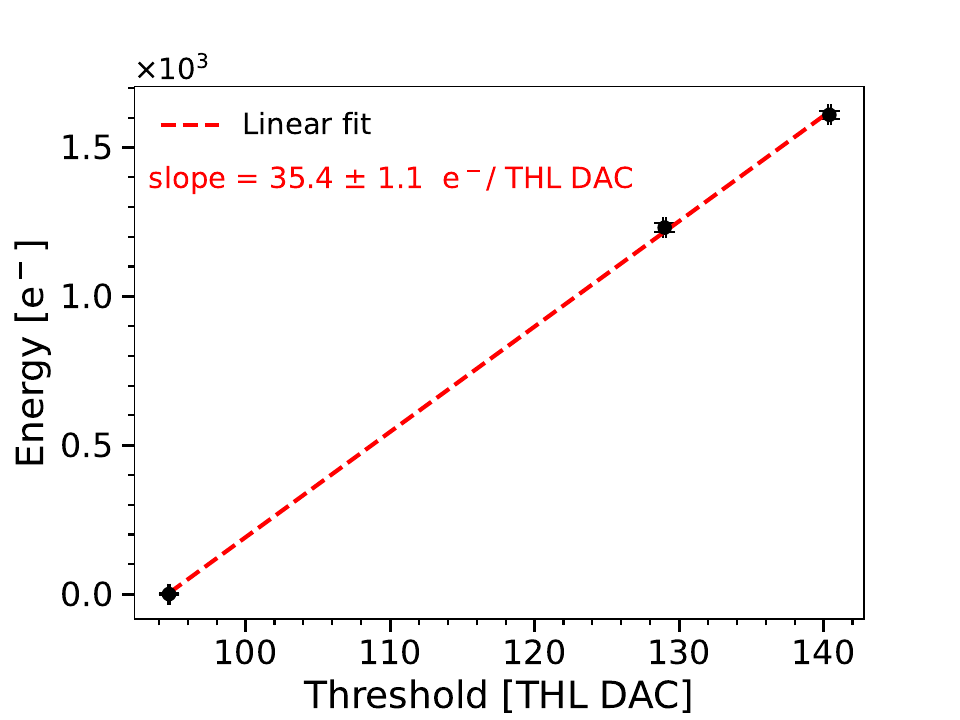}
    \put(-200,30){\footnotesize Baseline}
    \put(-72,120){\footnotesize \ce{^{48}Ti}}
    \put(-45,140){\footnotesize \ce{^{55}Fe}}
    \caption{Energy calibration of the threshold using the determined baseline, and the \ce{^{48}Ti} K$_{\upalpha_1}$ and \ce{^{55}Fe} K$_{\upalpha_1}$ peaks.}
    \label{fig:calibration_factor}
\end{figure}

\begin{figure}[tbp]
    \centering
    \includegraphics[width=1\linewidth]{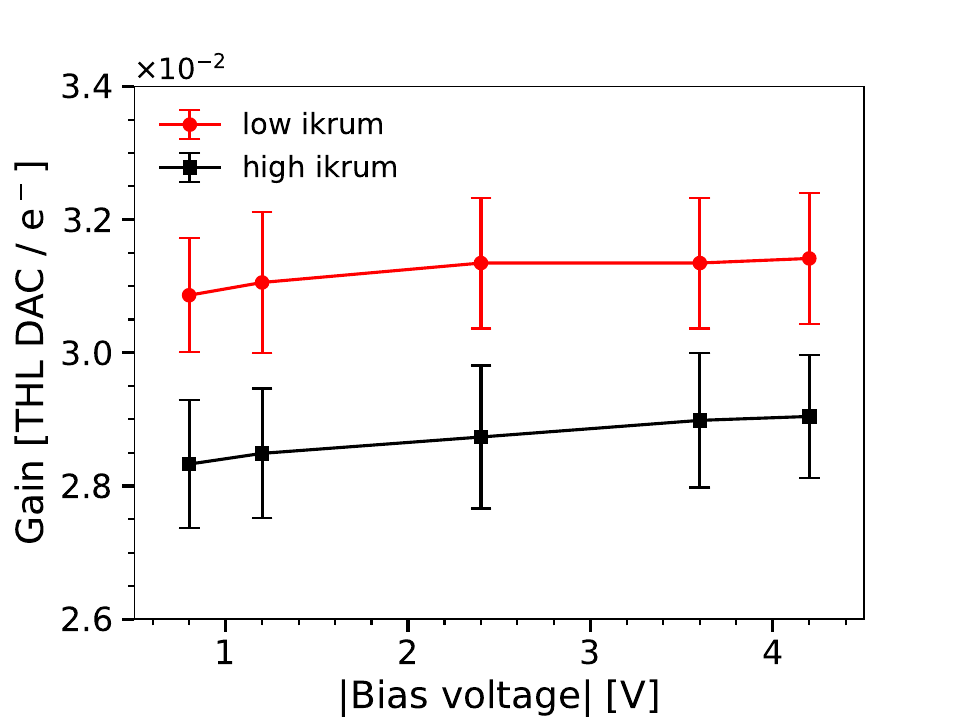}
    \caption{Gain as a function of the bias voltage for low (\SI{\sim 1.65}{\nano\ampere}) and high (\SI{\sim 3.75}{\nano\ampere}) feedback currents.}
    \label{fig:gain_vs_bias}
\end{figure}

\begin{figure}[tbp]
    \centering
    \includegraphics[width=1\linewidth]{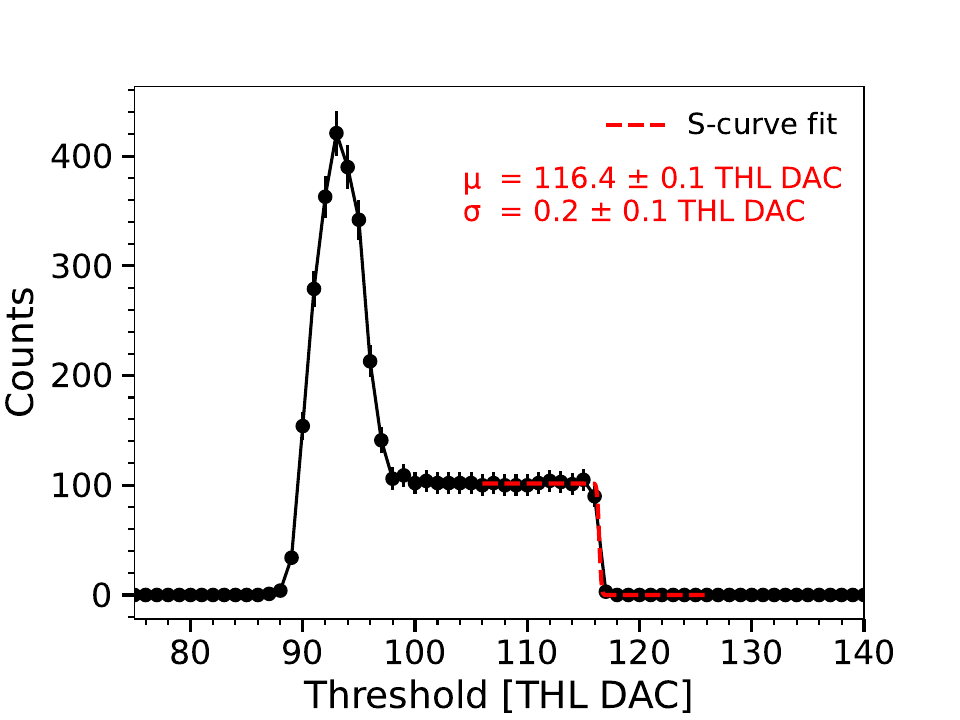}
    \caption{Occupancy for a single-pixel as a function of the threshold with 100 injected test pulses. The injection charge is approximately equivalent to a MIP.}
    \label{fig:onepixel_amplitude}
\end{figure}

To determine the applied threshold in electrons, the chip is calibrated using an iron (\ce{^{55}Fe}) radioactive source. As a cross-check of the measurements, a \SI{75}{\micro\meter} thick titanium (\ce{^{48}Ti}) layer is placed beneath the \ce{^{55}Fe} to produce X-ray fluorescence from \ce{^{48}Ti}. 
Assuming a mean energy of \SI{3.66 \pm 0.03}{eV} is required to produce an electron-hole pair in silicon~\cite{rossi2006}, the characteristic peaks of \ce{^{55}Fe} and \ce{^{48}Ti} K$_{\upalpha_1}$ are expected to correspond to energies of \SI{1609 \pm 14}{} and \SI{1232 \pm 10}{} electrons~\cite{nist}, respectively.
To identify these peaks, the threshold is scanned in counting mode. 
%For each \texttt{dac\_vthr}, \SI{800}{} frames with a \SI{10}{\milli\second} shutter window are captured.

\cref{fig:thl_calibration} shows the obtained occupancy and its derivative as a function of the threshold using the \ce{^{55}Fe} source.  
A significant drop in counts occurs at the threshold corresponding to the K$_{\upalpha_1}$ energy. This threshold is determined by fitting a Gaussian function to the right visible peak of the occupancy derivative. The mean of the Gaussian fit $\upmu~=~$~\SI{140.4 \pm 0.1}{}~THL~DAC represents the energy of the K$_{\upalpha_1}$ peak.
%and the width $\upsigma =$~\SI{2.9 \pm 0.1}{}~THL~DAC represents the noise and threshold dispersion among pixels. 
At a threshold of approximately 148~THL~DAC, a small peak is visible in the derivative, corresponding to the K$_{\upbeta}$ energy. 

\cref{fig:calibration_factor} shows the mean values of the Gaussian fit for the \ce{^{55}Fe} and \ce{^{48}Ti} K$_{\upalpha_1}$ peaks as a function of the energy. It also includes the baseline of the chip as determined in \cref{fig:thl_equalisation}, which corresponds to a charge deposit of zero electrons. A linear function is fitted, obtaining a calibration factor of \SI{35.4 \pm 1.1}{} electrons per THL~DAC. 
This value corresponds to \SI{111}{mV/ke^-}~\footnote{The conversion from voltage to DAC units is obtained by measuring the voltage using the ADCs on the CaR board while scanning the DACs. A conversion factor of \SI{3.928\pm0.001}{mV/THL\ DAC} is determined for the \texttt{dac\_vthr} setting.}, and agrees with the analog front-end simulations presented in~\ref{sec:christian}.
The gain is the inverse of the calibration factor, and it is shown in \cref{fig:gain_vs_bias} for different bias voltages of the sensor and feedback currents. 
The \textit{low ikrum} setting corresponds to \SI{1.65}{\nano\ampere}, while the \textit{high ikrum} setting corresponds to \SI{3.75}{\nano\ampere}. Additionally, no significant differences are observed between samples with different thicknesses. 

%Only a slight increase in the gain is observed when the bias voltage is increased. 

A second peak is visible at lower amplitudes of approximately \SI{126}{} DACs in \cref{fig:thl_calibration}. This effect, which is related to loss of signal height due to ballistic deficit~\cite{ballistic_deficit}, as explained in \cref{sec:eff}, results in two signal amplitudes for the K$_{\upalpha_1}$ peak, depending on the position of the X-ray energy deposition within the pixel cell. 
%Since the ballistic deficit does not strongly affect the ToT measurement, the right peak is utilized for calibration. 
Since the ballistic deficit does not strongly affect the position of the right peak, this is utilized for calibration. 
If spatial information is available, the calibration procedure could be improved by applying different calibration factors to different parts of the pixel cell.

\subsection{ToT Calibration}
\label{lab:tot_calcalibration}
\begin{figure}[tbp]
    \centering
    \includegraphics[width=1\linewidth]{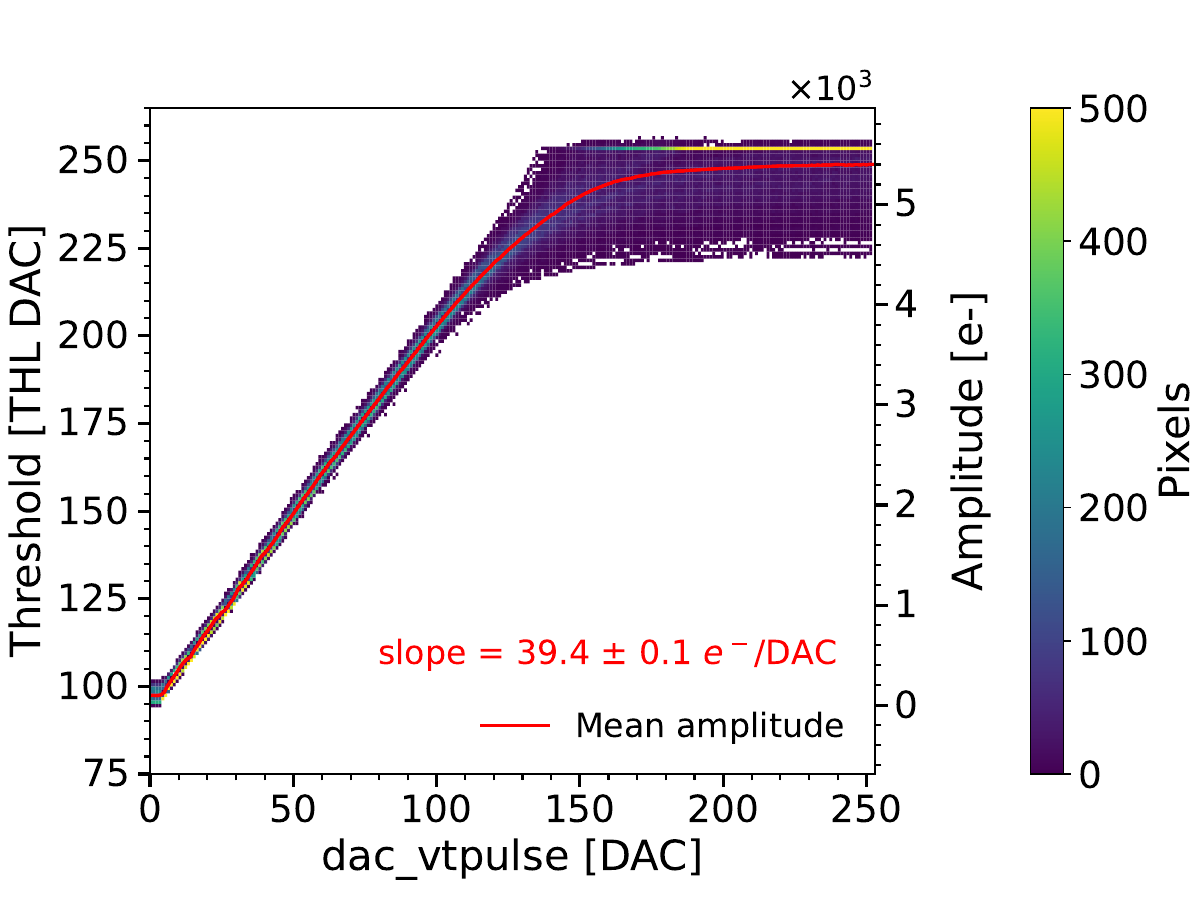}
    \caption{Uncalibrated (left axis) and calibrated (right axis) test pulse amplitudes of the full matrix. The mean amplitude is shown in red.}
    \label{fig:allpixels_amplitude}
\end{figure}
The results of the threshold calibration can now be used to calibrate the ToT response of each pixel individually. To achieve this, test pulses are injected into the analog front-end and the corresponding response is characterised in terms of the amplitude (in threshold units) and ToT.

%Each pixel on the chip is calibrated using analog test pulse injection to determine the ToT value measured in electrons. For an injected test pulse on the analog front-end, both the amplitude and ToT can be measured independently.

Similar to the procedure described in \cref{lab:calibration}, the threshold is scanned in counting mode while injecting test pulses of a given positive amplitude above baseline. A pixel registers one count (i.e., a threshold crossing) per injected test pulse if the threshold lies between the baseline level and the pulse amplitude. If the threshold is lower than the baseline or higher than the pulse amplitude, the pulse is not detected (no threshold crossing), and no count is added. When the threshold is close to the baseline level, noise (or other perturbations such as undershoot in the return to baseline) can cause a higher number of threshold crossings, similar to what is shown in \cref{fig:pixel_thl}.
%After a pulse is injected, a threshold scan in counting mode determines its amplitude. 

\cref{fig:onepixel_amplitude} shows the occupancy as a function of the threshold for 100 test pulses per threshold, using \texttt{dac\_vtpulse} equals 20 DACs (approximate most probable value for a MIP, about 700 electrons), injected into a single pixel. 
Above the Gaussian part (baseline), an occupancy of 100 is expected until the maximum pulse height is reached, after which the occupancy drops to zero. In this region, an S-curve function is fitted, with the mean $\upmu~=~$\SI{118.5 \pm 0.1}{}~THL~DAC representing the test pulse amplitude and the width $\upsigma~=~$~\SI{0.2 \pm 0.1}{}~THL~DAC reflecting the pixel noise and the amplitude variations across the 100 injected pulses. 
This procedure is repeated for each pixel and all possible \texttt{dac\_vtpulse} values, as shown in \cref{fig:allpixels_amplitude}. 
%Since the threshold has already been calibrated in electrons in \cref{lab:calibration}, the amplitude can also be expressed in electrons. 
A linear regression in the region from \texttt{dac\_vtpulse} between \SI{25}{} and \SI{100}{}~DAC is used to extract the conversion factor between electrons and DACs. All pixels have a similar linear gain, with maximum measured amplitude reaching the 8-bit \texttt{dac\_vthr} limit.
%All pixels have a similar linear gain, and saturation at 256 counts is expected from the 8-bit counter.  

\begin{figure}[tbp]
    \centering
    \includegraphics[width=1\linewidth]{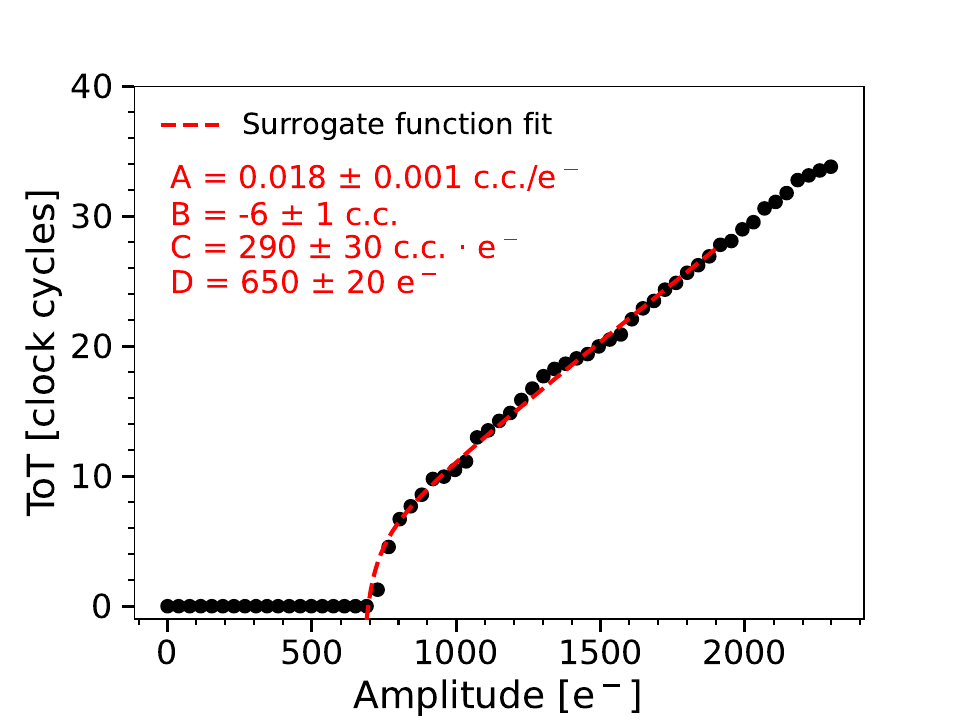}
    \caption{Calibrated ToT response for a single pixel.}
    \label{fig:onepixel_tot_amplitude}
\end{figure}

\begin{figure}[tbp]
    \centering
    \includegraphics[width=1\linewidth]{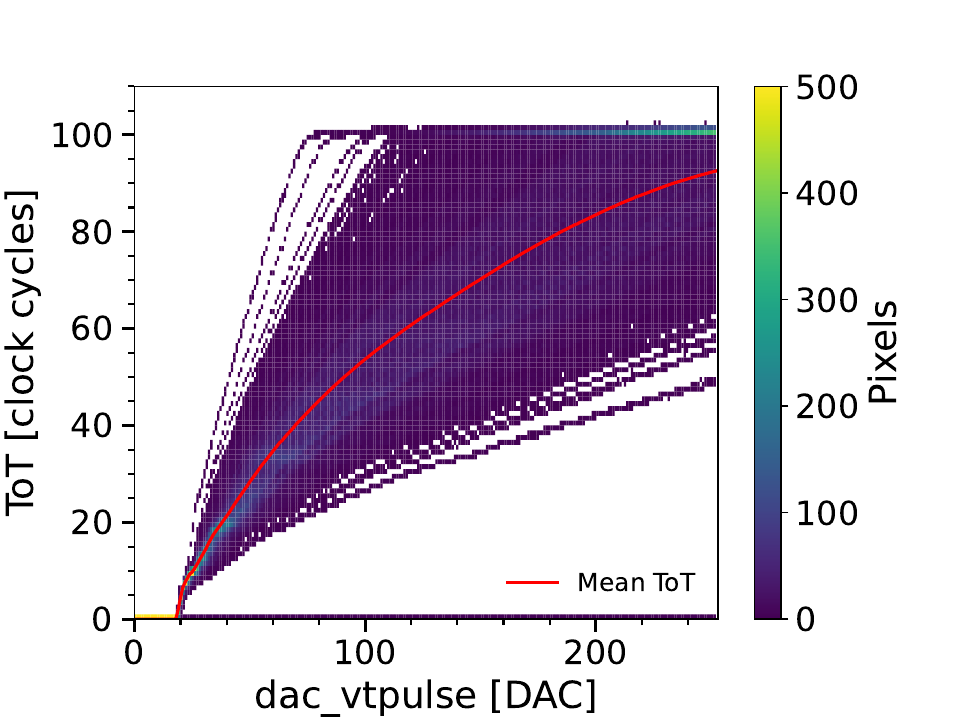}
    \caption{ToT response for the full matrix as a function of the amplitude of the injected test pulses. The threshold is approximately \SI{600}{} electrons. The mean ToT is shown in red.}
    \label{fig:allpixels_tot}
\end{figure}

The ToT value for each test pulse amplitude is calculated from the average of 100 test pulse measurements at a fixed threshold. This procedure is repeated for each pixel and all possible \texttt{dac\_vtpulse} values, as shown in \cref{fig:allpixels_tot} for a threshold of approximately \SI{600}{} electrons. At higher test pulse amplitudes, far from the expected MIP value, the measured ToT values vary between pixels, and the maximum mean value is reached at \SI{1}{\micro\second}, coinciding with the acquisition frame duration. 
This large threshold of \SI{600}{} electrons is chosen in this measurement for consistency with the ToT spectrum of the \ce{^{55}Fe} source measurements presented below.

Next, the ToT and amplitude measurements can be combined. \cref{fig:onepixel_tot_amplitude} shows the ToT as a function of the test pulse amplitude in electrons for a single pixel. A surrogate function
\begin{equation}\label{eq:tot_cal}
    \mathrm{
    ToT \ = \ Ax \ + \ B \ + \ \frac{C}{x-D}
    }
\end{equation}
is used to fit each pixels data, where x is the test pulse amplitude, A modulates the linear component at higher amplitudes, B adjusts the ToT offset, C represents the curvature part at low amplitudes, and D indicates the offset in amplitude~\cite{JAKUBEK2011S262}. The inverse of the surrogate function allows the conversion of measured ToT in clock cycles to signal amplitude in electrons.
\begin{figure}[tbp]
    \centering
    \includegraphics[width=1\linewidth]{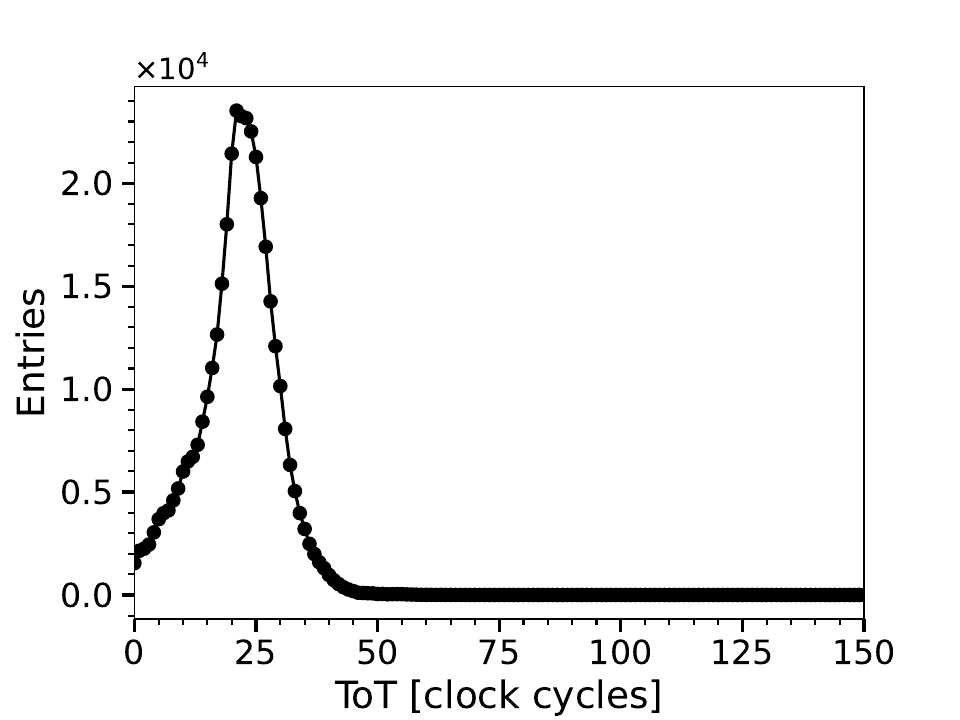}
    \caption{ToT spectrum of the \ce{^{55}Fe} source before calibration.}
    \label{fig:tot_fe55}
\end{figure}
\begin{figure}[tbp]
    \centering
    \includegraphics[width=1\linewidth]{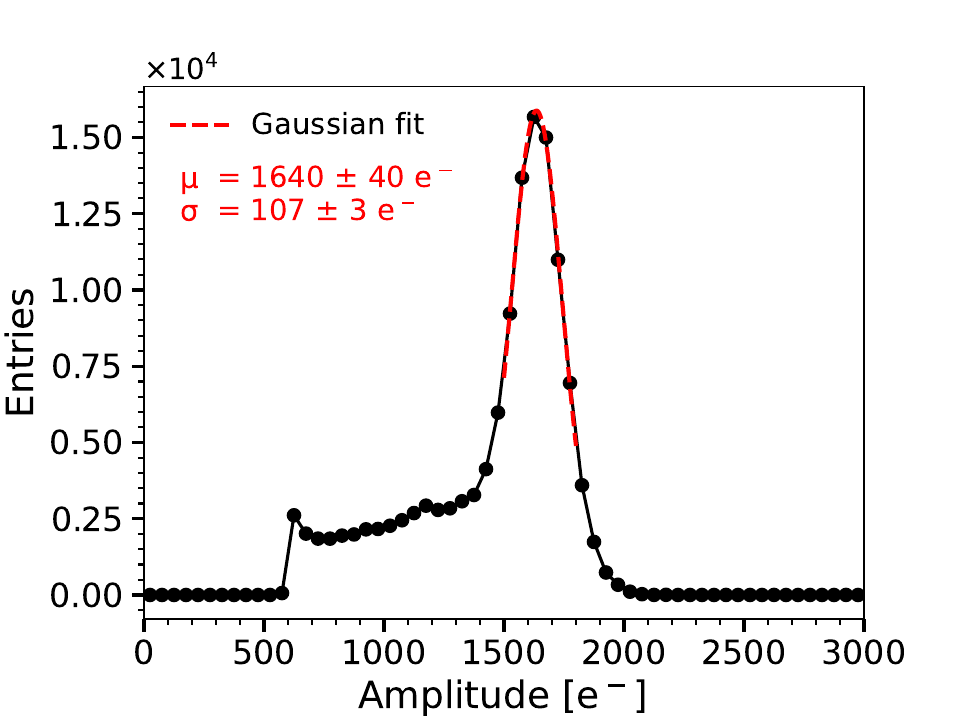}
    \caption{Amplitude spectrum of the \ce{^{55}Fe} source after calibration.}
\label{fig:amplitude_fe55}
\end{figure}

To verify the calibration procedure, a ToT spectrum of \ce{^{55}Fe} is recorded in ToT mode, as shown in \cref{fig:tot_fe55}. The threshold of approximately \SI{600}{} electrons is the same as used during the calibration. This spectrum is then converted into electrons, as shown in 
\cref{fig:amplitude_fe55}, after applying the threshold and ToT calibration.
The peak is described by a Gaussian with $\upmu = 1640 \pm 40$~electrons corresponding to the resolved K$_{\upalpha_1}$ amplitude with an accuracy smaller than \SI{5}{\%}. The dominant systematic uncertainty arises from the threshold calibration uncertainty. 

The ToT calibration is applied to the test beam reconstructed data acquired in ToT mode down to a threshold of \SI{300}{} electrons. For lower thresholds, the surrogate function fits (\cref{eq:tot_cal}) do not converge \textemdash particularly the parameter C, which models the curvature at low thresholds.

\section{Methodology for Test Beam Measurements}\label{sec:tb_method}
This section describes the experimental setup and analysis procedure for the test beam measurements, where the H2M performance in MIP detection is studied. 
% brief intro to SPS and DESY
\subsection{Experimental Setup}\label{sec:tb_setup}

% components and working principle of the setup, integration of H2M into the DAQ system

The data have been recorded at the DESY~II~\cite{desyii} and CERN SPS~\cite{sps} Test Beam Facilities employing the ADENIUM~\cite{adenium} and CLICdp Timepix3~\cite{clicdetector} beam telescopes, respectively. 
Each beam telescope comprises three upstream and three downstream reference planes, with the device under test (DUT) in between.
These planes are arranged and oriented to optimise the pointing spatial resolution of the beam telescope at the DUT position.
With the used configurations and particle beams (\SI{\sim 4.8}{GeV} electrons at DESY II, and \SI{120}{GeV} charged pions at CERN SPS), the track resolution at the DUT position for the ADENIUM telescope is found to be \SI{\sim 3.8}{\micro\meter}, and \SI{\sim 1.5}{\micro\meter} for the CLICdp Timepix3 telescope, determined using the \textit{GBL Track Resolution Calculator}~\cite{gbltrackresolutioncalculator}. 

At DESY II, the Telepix2 HV-CMOS chip~\cite{telepix2} is placed downstream the ADENIUM telescope as a region of interest (ROI) trigger and timing layer, providing a reference time with a resolution below \SI{4}{\nano\second}. When H2M is operated in ToT and ToA modes, the shutter closes \SI{100}{\nano\second} after receiving the trigger signal. It opens again once the readout is completed. This results in a readout frequency of approximately \SI{100}{Hz}, given the selected energy and small trigger area.
The use of Telepix2 as a trigger also enables measurements in triggered mode. 
%In this acquisition mode, the H2M shutter is opened for \SI{500}{\nano\second}, starting approximately \SI{2}{\micro\second} after a hit is received. The binary readout is issued after validation with the trigger signal. The \SI{2}{\micro\second} delay accounts for the trigger latency, while the \SI{500}{\nano\second} shutter window accommodates the H2M time walk. 
In this acquisition mode, the H2M shutter is opened for \SI{500}{\nano\second}, once the external trigger arrives, accounting for the timewalk from H2M.
The preset of the pixel counter compensates for the trigger latency of about \SI{2}{\micro\second}. The binary readout is issued only for hits that are validated by the external trigger. 

The AIDA Trigger Logic Unit~\cite{tlu} ensures synchronization among all devices mentioned, and its signal processing dominates the trigger latency. The EUDAQ2~\cite{eudaq2} data acquisition framework is used for configuring and reading out all these devices. 

At the CERN SPS, the Timepix3 telescope planes provide a timestamp for each track with a resolution of approximately \SI{1}{\nano\second}. The H2M is read out continuously with a fixed shutter duration of \SI{300}{\micro\second} in ToT mode and \SI{2.56}{\micro\second} in ToA mode. With a readout time of \SI{500}{\micro\second} between shutters, the resulting readout frequency can go up to approximately \SI{2}{kHz}. Outside the SPS spill, the shutter remains closed, and no data are acquired.
The SPIDR DAQ system~\cite{spidr} is used for the readout of the Timepix3 reference planes, and EUDAQ2 for configuration and readout of the H2M chip.

\subsection{Reconstruction and Analysis}

% reconstruction procedure and definition of observables

The Corryvreckan framework~\cite{corry} is used for the reconstruction and analysis of the test-beam data. 

The duration of each H2M frame defines the event time window, and the ADENIUM beam telescope provides matching data if its trigger signal has a timestamp within this window. For the Timepix3 beam telescope, all hits with a timestamp within this window are considered. 
%To build these events, the Corryvreckan module \texttt{[EventLoaderEUDAQ2]} loads the data from H2M, ADENIUM, and the TLU, while the module \texttt{[EventLoaderTimepix3]} handles the data from Timepix3.

%Using the center-of-gravity algorithm, the module \texttt{[Clustering4D]} reconstructs the cluster position of recorded adjacent pixel hits.
The cluster position of recorded adjacent pixel hits is reconstructed using the center-of-gravity algorithm.
The number of hits in a cluster determines the cluster size. 
For events with a cluster size of two in data acquired in ToT mode, $\upeta$-correction~\cite{etacorection} is applied to account for nonlinear charge sharing within the pixel cell.

The trajectory of the particle is reconstructed using the General Broken Lines (GBL) track model~\cite{gbl}. This is performed
%with the \texttt{[Tracking4D]} module
requesting a hit in each of the six telescope planes and a track $\upchi^2$ per degree of freedom smaller than \SI{5}{}. Then, the reconstructed tracks are associated with a cluster on H2M if the projected track position on H2M is within a distance of \SI{52.5}{\micro\meter} (\SI{1.5}{} pixel pitches) of the cluster center. 
%This association is performed using the \texttt{[DUTAssociation]} module. 

%To determine the hit detection efficiency of H2M, the \texttt{[AnalysisEfficiency]} module is utilised.
The hit detection efficiency is defined as the ratio of tracks with an associated cluster on H2M to the total number of tracks going through H2M. To avoid sensor-edge effects, tracks that pass through the outermost columns and rows of the pixel matrix are excluded from the calculation. Tracks crossing a masked pixel or any of its neighboring pixels are also excluded. These masked pixels are identified in the equalization procedure described in \cref{lab:equalization} and are disabled during the data taking. The total percentage of masked pixels remains below \SI{0.5}{\%} for all datasets.

The fake-hit rate is also measured in the test beam environment in the absence of beam. For this, the shutter is opened for \SI{100}{\micro\second}, and the number of hits above threshold is counted.

\section{Results} 
\label{sec:tb_results}
The performance of H2M in test beam measurements is discussed in the following. Unless explicitly stated, the results correspond to the H2M-3 (\SI{50}{\micro\meter}) sample, with data recorded at DESY II using the ADENIUM telescope.

\subsection{Efficiency}\label{sec:eff}

\begin{figure}[tbp]
    \centering
    \includegraphics[width=1\linewidth]{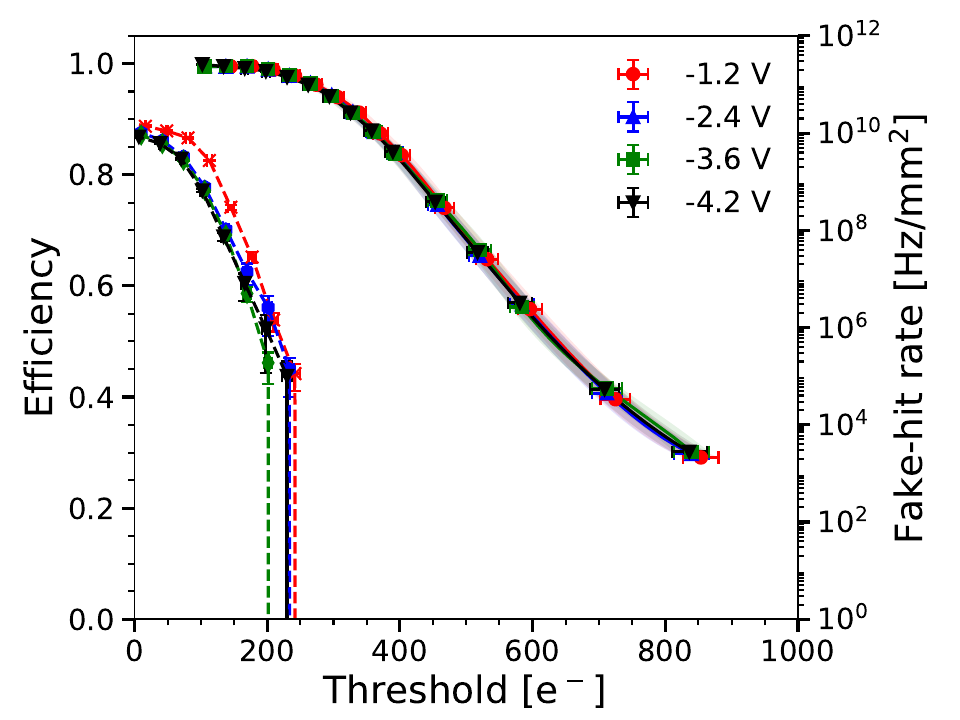}
    \caption{Efficiency (left axis) in solid lines and fake-hit rate (right axis) in dashed lines as a function of the threshold for different sensor bias voltages, obtained in triggered mode.}
    \label{fig:eff_comparebias}
\end{figure}

\cref{fig:eff_comparebias} shows the efficiency and fake-hit rate as a function of the threshold for different sensor bias voltages. 
While the efficiency does not depend significantly on the sensor bias voltage, the fake-hit rate rises at low bias voltages due to the increase in sensor capacitance (see \cref{fig:total_noise}).
An efficiency of \SI{99.6}{\%} is achieved at a threshold of \SI{144}{}~electrons when the sensor is biased at \SI{-3.6}{V}. 
The fake-hit rate at this point is around \SI{1.5e7}{Hz\per\milli\meter^2} (i.e., \SI{18.4}{kHz} per pixel), corresponding to a matrix occupancy of approximately nine pixels hit (less than 1\% of the active pixels) per \SI{500}{\nano\second} frame.
Longer frame durations employed in ToT and ToA acquisition modes result in higher matrix occupancies, limiting the lowest achievable thresholds and, thus, the chip efficiency. 
\begin{figure}[tbp]
    \centering
    \includegraphics[width=1\linewidth]{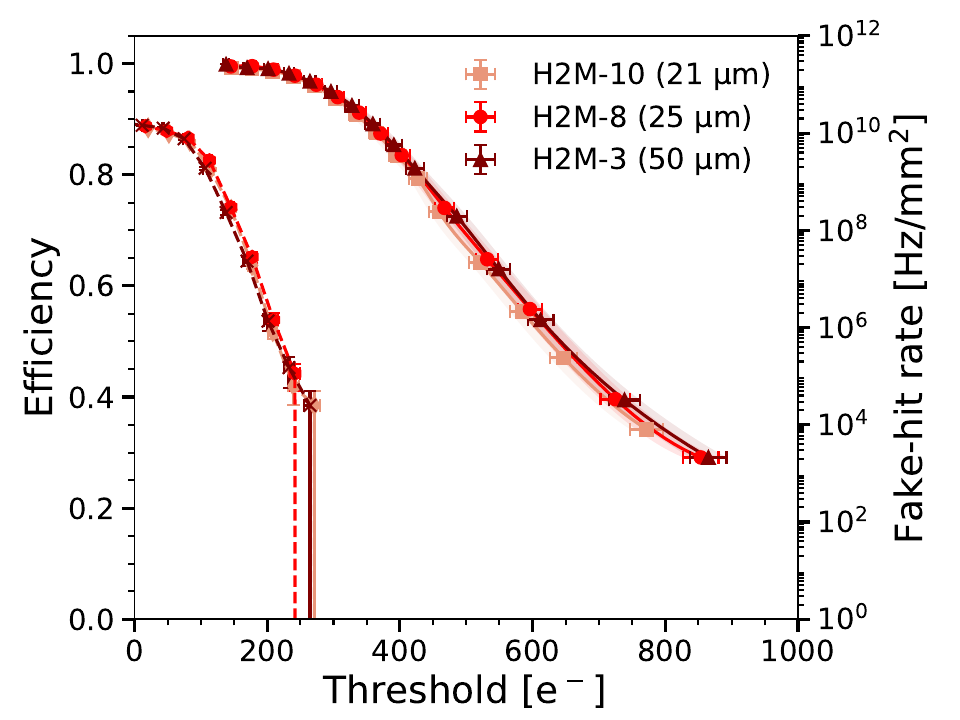}
    \caption{Efficiency (left axis) in solid lines and fake-hit rate (right axis) in dashed lines as a function of the threshold for samples with different total chip thicknesses, obtained in triggered mode. The sensors are biased at \SI{-1.2}{V}.}
    \label{fig:eff_comparesamples}
\end{figure}

%\cref{tab:99eff} summarizes the highest threshold at which efficiencies above \SI{99}{\%} and \SI{99.5}{\%} are achieved for different bias voltages. The uncertainty in the threshold calibration factor dominates the uncertainty. The maximum threshold allowing for \SI{99}{\percent} efficiency is slightly larger compared to those reported by other chips manufactured using the same technology, such as APTS~\cite{apts}, while the fake-hit rate is comparably high, due to the threshold dispersion and single pixel noise discussed in section~\ref{lab:equalization}. At higher thresholds, the measured efficiency is comparably low. The reason for this is investigated in the following sections. 

% REPLACING THE PARAGRAPH ABOVE:
\cref{tab:99eff} summarizes the highest threshold at which efficiencies above \SI{99}{\%} and \SI{99.5}{\%} are achieved for different bias voltages. The uncertainty in the threshold calibration factor dominates the uncertainty. The maximum threshold allowing for \SI{99}{\percent} efficiency is larger than the result from most sensors manufactured in the same process, as shown by comparison with, e.g., the studies on APTS~\cite{apts}, where many pitches, designs and bias voltages are investigated. A reason for that might be the larger pitch, which reduces the relative size of the region where charge sharing occurs, as the studies on APTS suggest. In the meantime, the fake-hit rate of H2M is comparably high, due to the threshold dispersion and single-pixel noise discussed in Section~\ref{lab:equalization}. At higher thresholds, the measured efficiency is comparably low. The reason for this is investigated in the following sections.

\cref{fig:eff_comparesamples} shows the efficiency as a function of the thresholds for three chips thinned to different total thicknesses obtained in triggered mode. No deterioration in efficiency or fake-hit rate is observed between samples for thresholds below \SI{450}{} electrons. In particular, all samples achieve an efficiency above \SI{99}{\%} at a threshold of approximately 200 electrons. At this point, the measured fake-hit rate is around \SI{4e5}{Hz\per\milli\meter^2}, corresponding to a matrix occupancy of fewer than one pixel hit per \SI{500}{\nano\second} frame.

The short shutter durations achieved in triggered mode allowed operation of the chip at low thresholds. However, measurements in ToT and ToA mode have been performed with longer shutters (see \cref{sec:tb_setup}), which increases the probability of fake hits and thus requires a higher operational threshold of around \SI{200}{} electrons. The results achieved at these higher thresholds are compatible with~\cref{tab:99eff}, where an efficiency of \SI{99}{\%} is measured. 

\begin{table}[h!]
\centering
\caption{Highest threshold at which efficiencies above \SI{99}{\%} and \SI{99.5}{\%} are achieved for different bias voltages in triggered mode.}
\begin{tabular}[t]{lcccc}
\toprule
&\SI{-1.2}{V}&\SI{-2.4}{V}&\SI{-3.6}{V}&\SI{-4.2}{V}  \\
\midrule
\SI{99}{\%}&\SI{205 \pm 6}{}&\SI{208 \pm 6}{}&\SI{208 \pm 6}{}&\SI{208 \pm 6}{}\\
\SI{99.5}{\%}&\SI{179 \pm 6}{}&\SI{178 \pm 6}{}&\SI{182 \pm 6}{}&\SI{183 \pm 6}{}\\
\bottomrule
\end{tabular}
\label{tab:99eff}
\end{table}%

\subsubsection{Non-uniform In-pixel Response}
\begin{figure}[!t]
  \centering
  \begin{subfigure}{0.48\textwidth}
    \includegraphics[width=\linewidth, trim={0 1.5cm 0 3.2cm},clip]{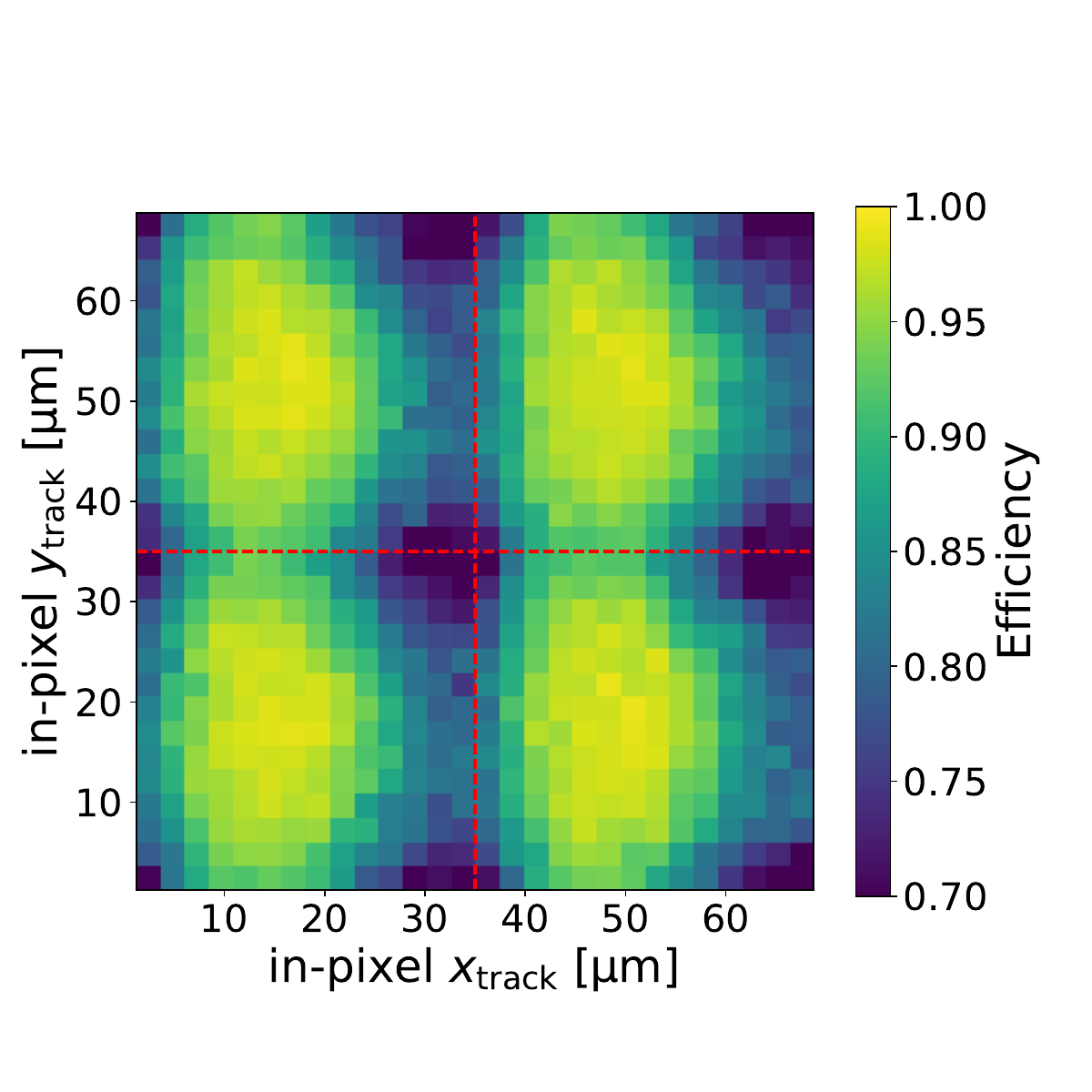}
    \caption{High \textit{ikrum}, \SI{-1.2}{V}, 330 e$^-$.}
    \label{fig:eff_map_worsesetting}
  \end{subfigure}%
  \hfill
  \begin{subfigure}{0.48\textwidth}
    \includegraphics[width=\linewidth, trim={0 1.5cm 0 3.2cm},clip]{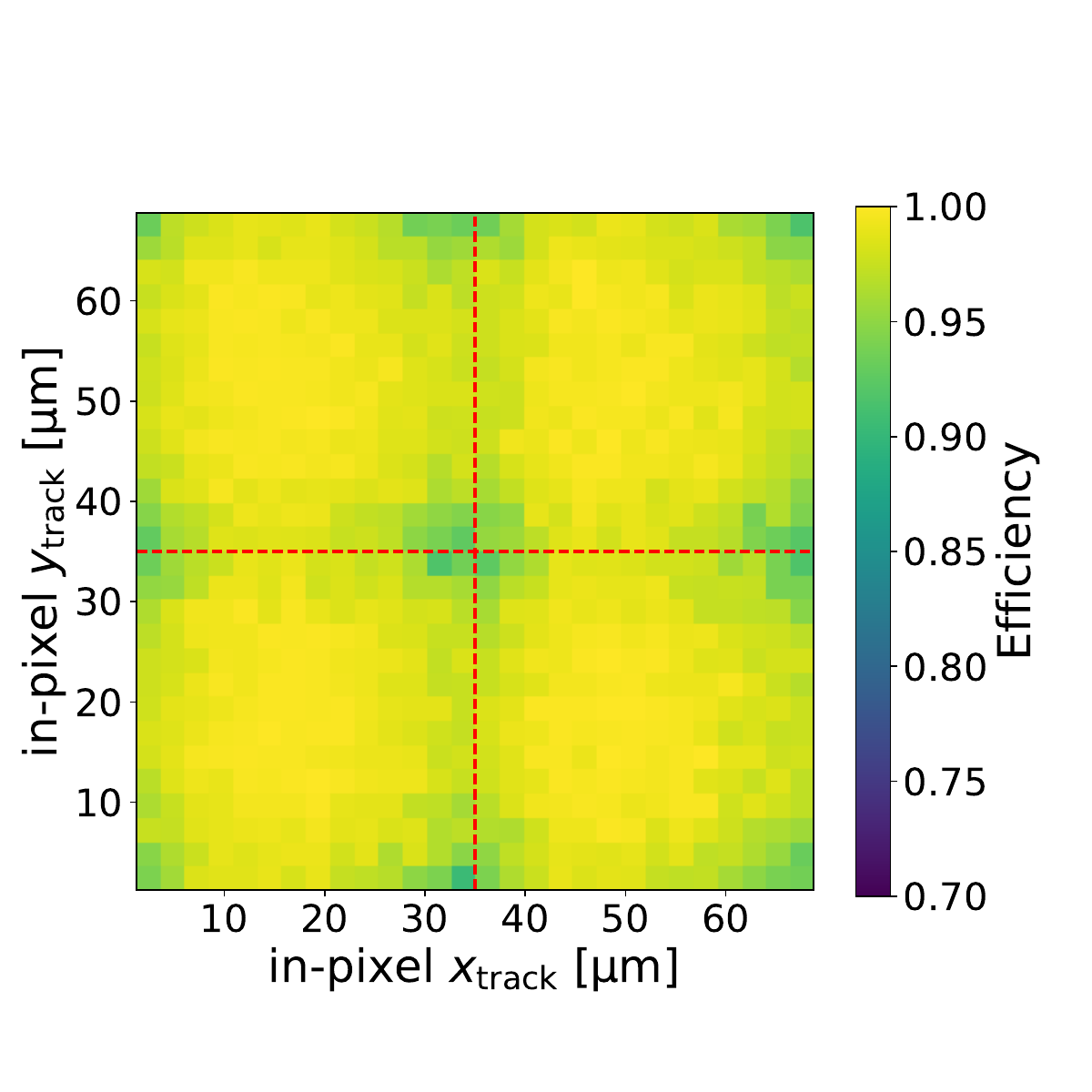}
    \caption{Low \textit{ikrum}, \SI{-3.6}{V}, 224 e$^-$.}
    \label{fig:eff_map_bestsetting}
  \end{subfigure}
  \caption{Efficiency maps projected onto four pixels for different feedback currents, sensor bias voltages, and thresholds. The pixel cell boundary is marked with a dashed red line.}
  \label{fig:inpixel_eff}
\end{figure}
At higher thresholds, a non-uniformity in the in-pixel response is observed. This is shown in \cref{fig:eff_map_worsesetting}, where the right part of the pixel cell shows a lower efficiency than the rest of the pixel.
The efficiency drop location coincides with the position of the \SI{4}{\micro\meter} wide n-well of the analog front-end shown in \cref{fig:h2m_nwells}.
This behavior is attributed to local potential wells of very shallow amplitude at the interface between the low-dose n-implant and the deep p-well~\cite{h2m_measurements, h2m_simulations}. The very low electric field where some of these local potential wells are located, allows them to slow down charge collection, as visible in the in-pixel representation of the mean arrival time discussed in \cref{sec:time-resolution} (\cref{fig:inpixel_toa}). 
This affects not only the charge carriers generated near the interface, but also those originating from the full epitaxial layer due to the drift path of charge carriers (see \cref{fig:h2m_cross_section}).
In contrast to the n-well of the analog front-end, the n-wells of the digital logic are thin, as sketched in Figure \ref{fig:analog}, mitigating this effect. 

This slow charge collection leads to a mismatch between the charge collection time and the fast response of the CSA. As a result, the CSA begins to reset before the full charge is collected, resulting in a reduced signal amplitude (ballistic deficit) and, consequently, a drop in the hit efficiency over a large part of the pixel cell.

In \cref{fig:eff_map_worsesetting}, the sensor is operated at low bias voltage, \SI{-1.2}{V}, with a high threshold of \SI{330}{}~electrons and high \textit{ikrum} to aggravate this effect.
Increasing the bias voltage enhances the electric field within the sensor, and reducing the feedback current effectively increases the integration time of the CSA. Both factors contribute to reducing the ballistic deficit, and improving the hit detection efficiency. \cref{fig:eff_map_bestsetting} shows a homogeneous in-pixel efficiency map at low \textit{ikrum}, a threshold of \SI{224}{electrons}, and a sensor bias of \SI{-3.6}{V}. 

To fully understand this effect, detailed simulations of the sensor and front-end have been performed, which are presented in \cref{sec:simulation}.

\subsection{Cluster Size}
\begin{figure}[btp]
    \centering
\includegraphics[width=1\linewidth]{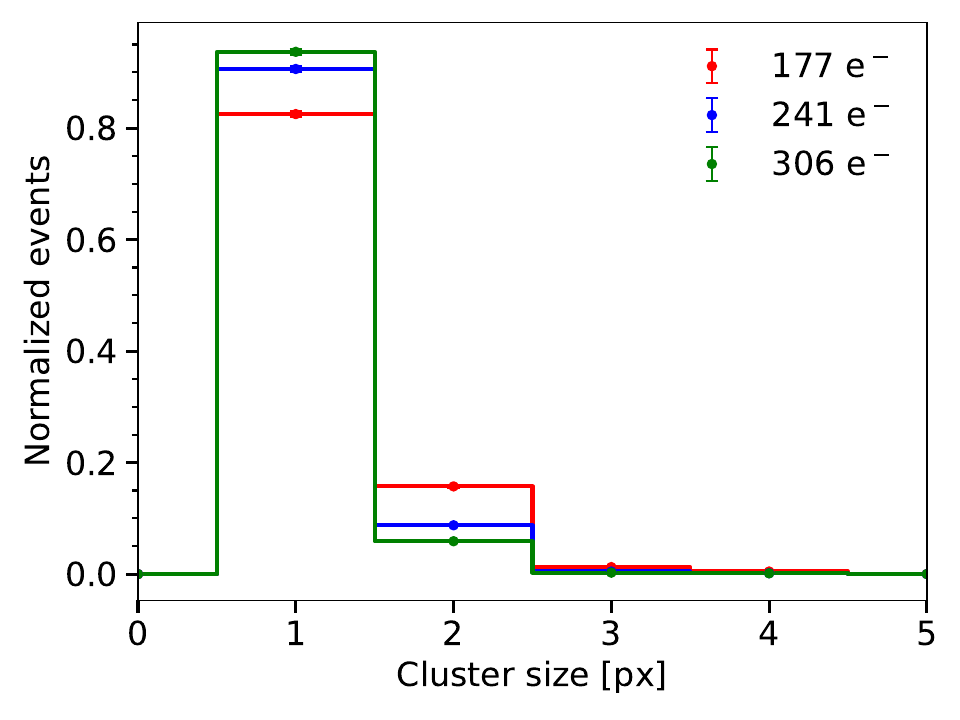}
    \caption{Cluster size distributions for different thresholds. The sensor is biased at \SI{-1.2}{V}.}
    \label{fig:clustersize_thr}
\end{figure}
\begin{figure}[tbp]
    \centering
\includegraphics[width=1\linewidth]{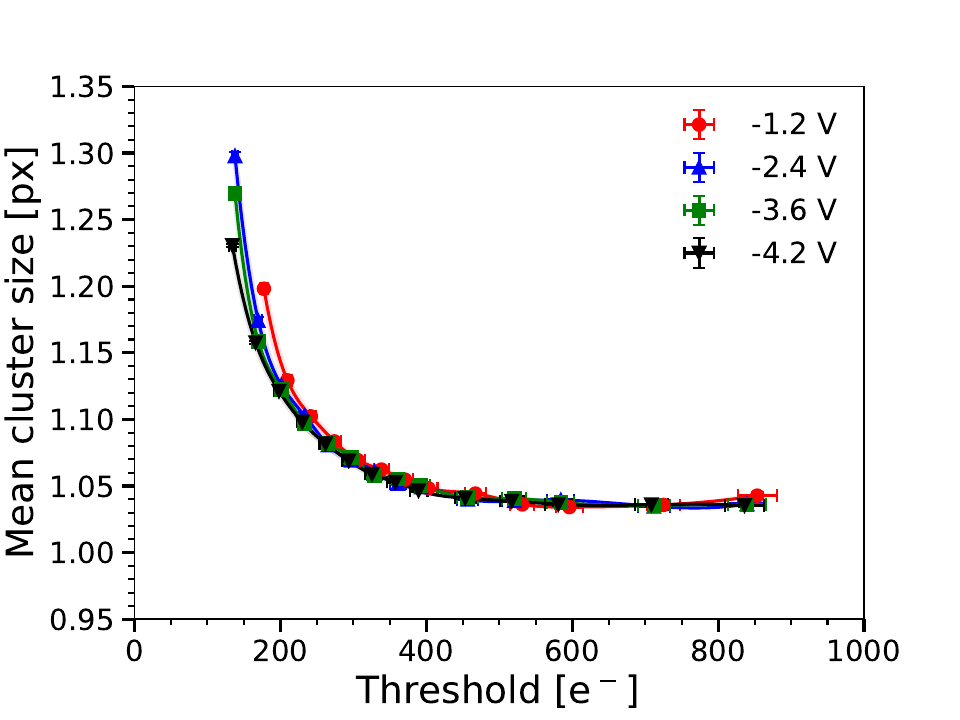}
    \caption{Cluster size as a function of the hit detection
threshold for different sensor bias voltages.}
    \label{fig:clustersize_bias}
\end{figure}

%not sure how much there is to be seen. Might make sense to merge with spatial resolution. Include several operation conditions?
The relatively large pixel pitch, thin epitaxial layer, and the \textit{modified with n-gap} sensor layout minimize charge sharing between neighboring pixels. 
Lowering the threshold increases the probability of inducing a signal in adjacent pixels, making two-pixel clusters more frequent, as shown in \cref {fig:clustersize_thr}. 
For a threshold of \SI{177}{electrons}, the average cluster size is \SI{1.19 \pm 0.01}{}, with 18\% of events with two-pixel clusters. 
\cref{fig:clustersize_bias} shows the mean cluster size as a function of the threshold.
At lower bias voltages, the electric field is reduced, leading to slower charge collection. This leaves more time for charge carrier diffusion, resulting in a slight increase in cluster size.
Towards higher detection thresholds (above 700 electrons), the mean cluster size increases due to the detection of delta-rays, which are more likely to deposit large amounts of energy.

\begin{figure}[tbp]
    \centering
\includegraphics[width=1\linewidth, trim={0cm 1.5cm 0cm 3.4cm},clip]{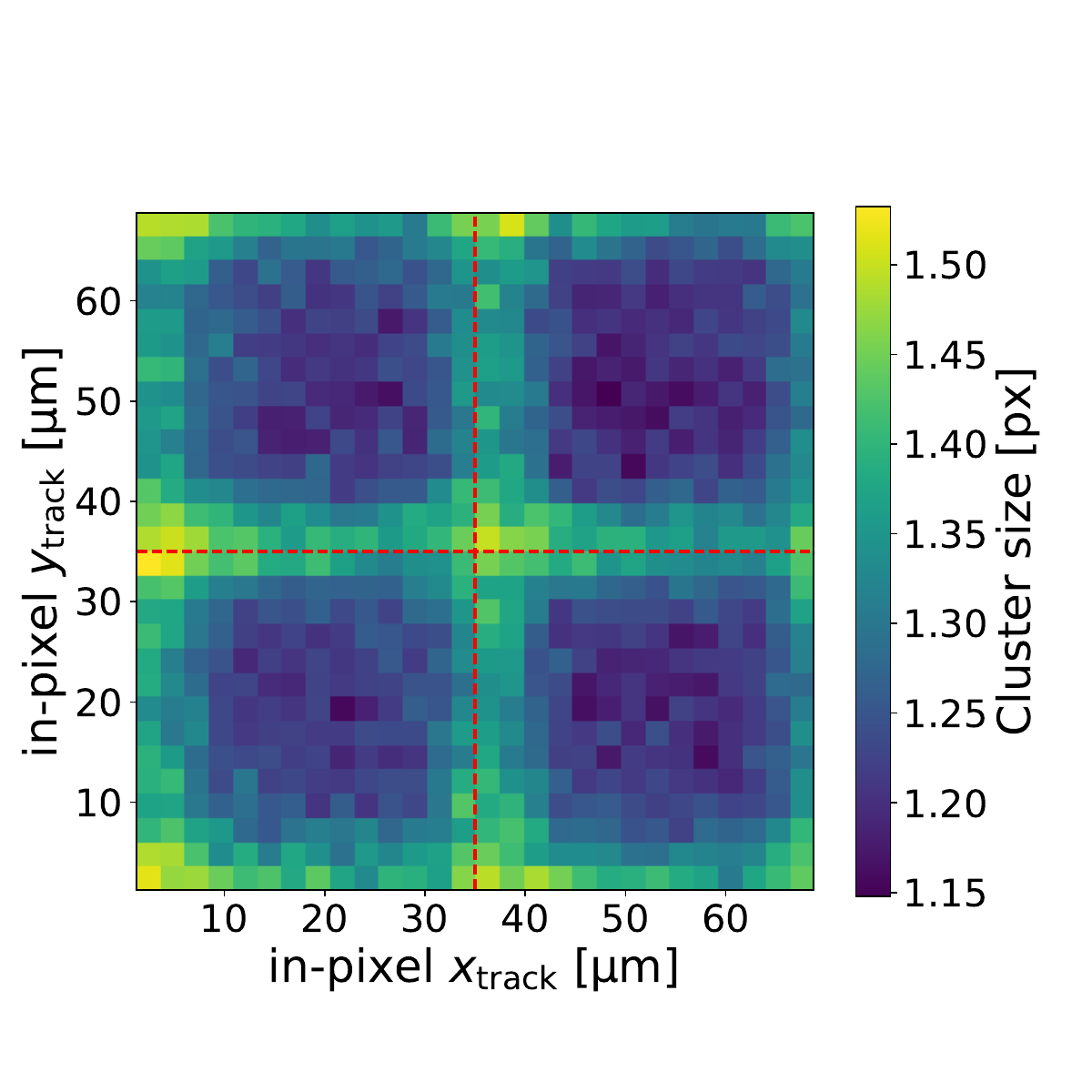}
    \caption{Cluster size map projected onto four pixels. The sensor is biased at \SI{-1.2}{V} and the threshold is \SI{224}{} electrons. The pixel cell boundary is marked with a dashed red line.}
    \label{fig:inpixel_cluster}
\end{figure}

Although the right side of the pixel cell experiences slower charge collection as discussed in \cref{sec:eff}, this does not impact the charge sharing. As a result, the in-pixel cluster-size maps remain symmetric, as shown in \cref{fig:inpixel_cluster}. Near the collection electrode, charge carriers are collected by the nearest pixel. A slight increase in cluster size is observed only at the boundaries of the pixel cell.

\begin{figure}[tbp]
    \centering
    \includegraphics[width=1\linewidth]{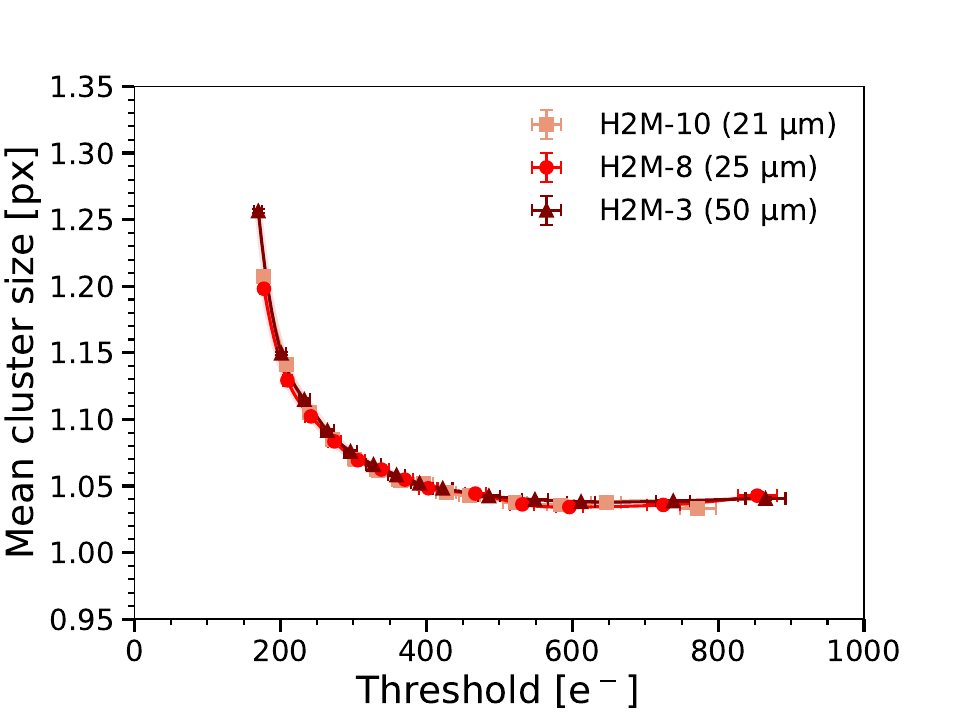}
    \caption{Cluster size as a function of the threshold for samples with different total chip thicknesses. The sensor is biased at \SI{-1.2}{V}.}
    \label{fig:clustersize_comparesamples}
\end{figure}

\cref{fig:clustersize_comparesamples} shows the mean cluster size as a function of the threshold for different total chip thicknesses. Similar to the hit detection efficiency results, there is no effect from thinning on the cluster size.


\subsection{Spatial Resolution}

\begin{figure}[tbp]
    \centering
    \includegraphics[width=1\linewidth]{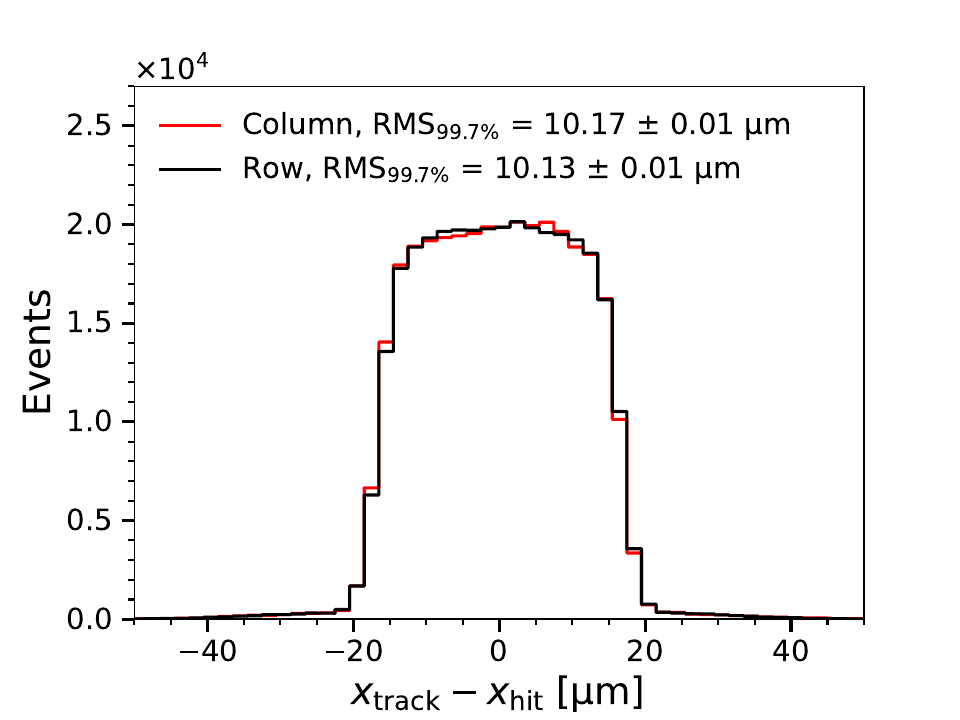}
    \caption{Spatial residuals in column and row direction recorded in ToT mode using the CLICdp Timepix3 beam telescope. The sensor is biased at \SI{-3.6}{V} and the threshold is 224 electrons.}
    \label{fig:residuals}
\end{figure}
% this is bad, probably not worth to make a fuss about telescope resolution. Asymmetry due to bias from efficiency. Include several operation conditions?

\cref{fig:residuals} shows the spatial residuals between the track impact position and the H2M hit position, for a threshold of \SI{224}{electrons}. 
At higher thresholds (\SI{>400}{} electrons), the residual distributions in the column direction become asymmetric due to reduced efficiency in the right part of the pixel cell (see \cref{fig:inpixel_eff}). 
%The reduced RMS, denoted as $\upsigma_{RMS,3\upsigma}$, is computed using the central $3\upsigma$ range of the distribution. 
The RMS is calculated over the central \SI{99.7}{\%} of the shown distribution (RMS$_\mathrm{99.7\%}$).
By subtracting the tracking resolution of the beam telescope from this value, the intrinsic spatial resolution of H2M can be determined. 
%Due to the limited knowledge of the material budget, such as the sensor glue thickness beneath the sensor, or simplifications in the simulation method, such as the assumption of Gaussian scattering widths of the tracks through material, the obtained pointing resolution at the DUT position is likely underestimated. Consequently, the calculated spatial resolution of H2M may be better than the values quoted below. 

\begin{figure}[tbp]
    \centering
    \includegraphics[width=1\linewidth]{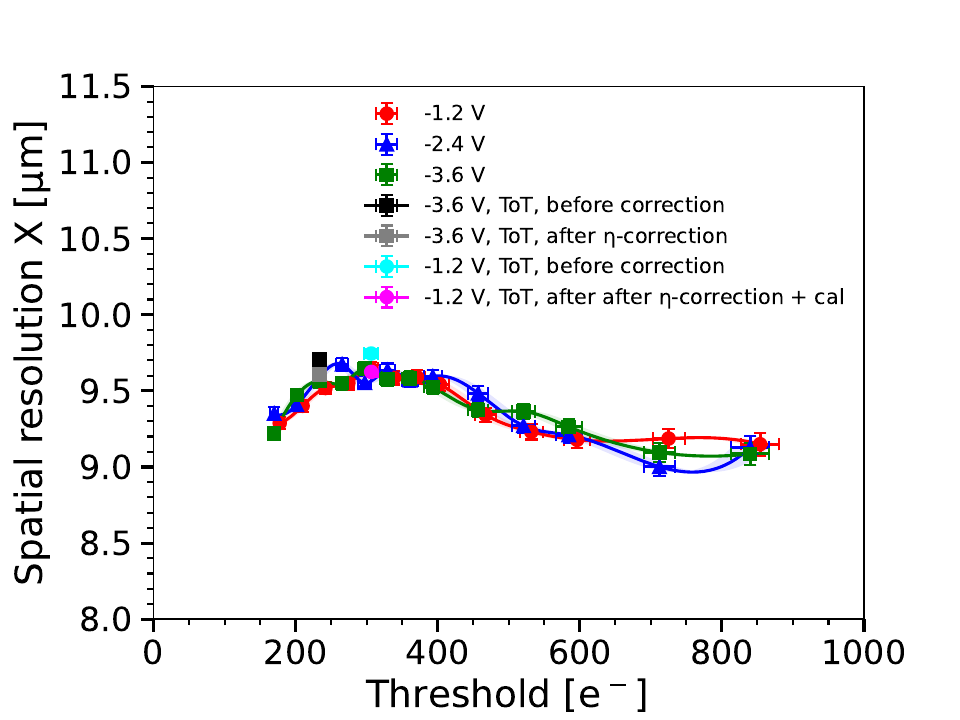}
    \caption{Spatial resolution as a function of the threshold for different sensor bias voltages in triggered and ToT modes.}
    \label{fig:spatialres}
\end{figure}

\cref{fig:spatialres} presents the H2M spatial resolution in the row direction as a function of the threshold for different sensor bias voltages in triggered mode.
A few additional points measured in ToT mode, for which calibration and $\upeta$-correction have been applied, are also included. The pixel-to-pixel variations and nonlinear charge sharing inside the pixel cell lead to a deterioration of the reconstructed cluster position when these corrections are not applied. 
Due to the fact that a majority of the interactions have a cluster size of one,
%Derived from a majority of events with single-cluster,
no significant difference is found between the spatial resolution in ToT mode (with 8 bits) and triggered mode (binary readout), and a spatial resolution close to the pitch~$/ \sqrt{12}~\sim$~\SI{10.1}{\micro\meter} is measured. 
Moreover, no significant differences between sensor bias voltages are observed.

The increase in charge sharing at a threshold below \SI{240}{electrons} leads to an improvement of the spatial resolution. In particular, a spatial resolution of \SI{9.3 \pm 0.1}{\micro\meter} is measured at a threshold of \SI{178 \pm 5}{} electrons.
No significant dependence on the sensor bias voltages is observed.
Between approximately \SI{240}{} and \SI{400}{electrons}, the spatial resolution remains constant at about \SI{9.7}{\micro\meter}, corresponding to a range dominated by single-cluster events. 
Above \SI{400}{} electrons, the above-mentioned asymmetric residuals (caused by the lower efficiency) result in a reduced $\upsigma_{RMS,3\upsigma}$, and therefore in an improvement of the spatial resolution.

\subsection{Signal Distribution}
\begin{figure}[tbp]
    \centering
    \includegraphics[width=1\linewidth]{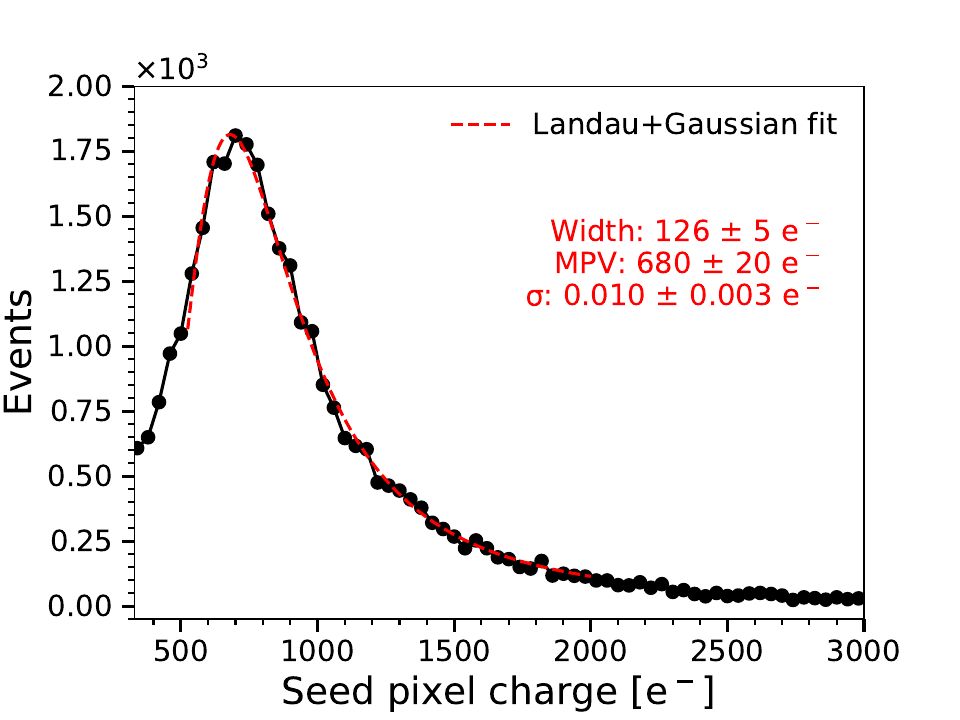}
    \caption{Seed-pixel charge distribution after calibration. The sensor is biased at \SI{-1.2}{V}, and the threshold is approximately 300 electrons. The scale (width) and Most Probable Value (MPV) of the Landau distribution, and the standard deviation of the Gaussian component ($\upsigma$) in the Landau-Gaussian convolution fit, are shown.}
    \label{fig:tb_tot}
\end{figure}

\begin{figure}[tbp]
    \centering
    \includegraphics[width=1\linewidth,trim={0 1.5cm 0 3.2cm},clip]{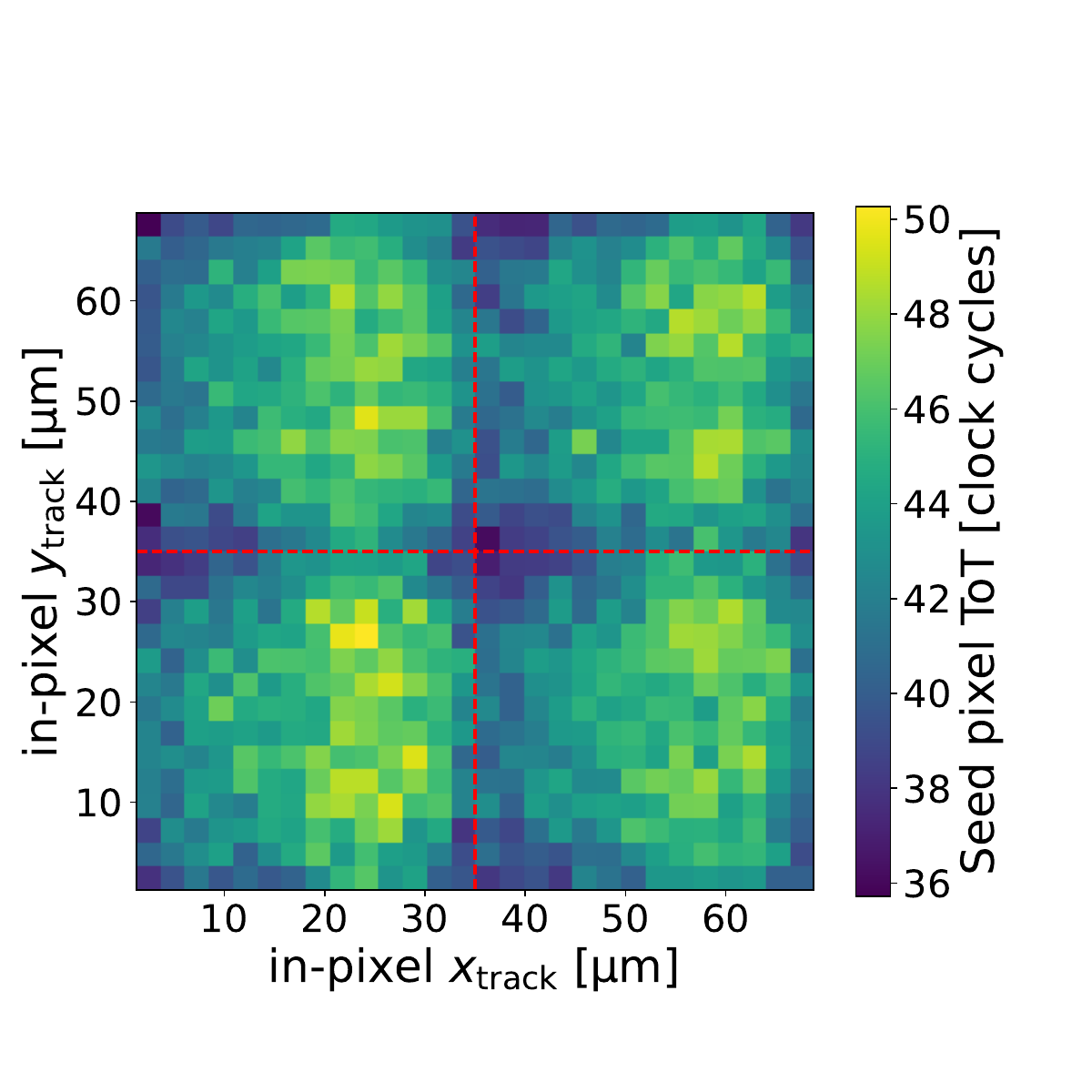}
    \caption{Mean seed pixel ToT map projected onto four pixels. The sensor is biased at \SI{-1.2}{V}, the threshold is 300 electrons, and a low \textit{ikrum} is set. The pixel cell boundary is marked with a dashed red line.}
    \label{fig:inpixel_tot}
\end{figure}

% hints towards slow charge collection opposed to loss of charge carriers (or both? Not sure...). Discuss ballistic deficit and its impact on efficiency. Do we need a ToT calibration for that? Include several operation conditions?

The seed pixel is defined as the pixel with the highest signal within a cluster.
\cref{fig:tb_tot} shows the seed pixel charge distribution fitted 
using a convolution of a Landau and Gaussian function.  
%An MPV of \SI{680 \pm 20}{electrons} has been measured, aligning with the expectations from the \SI{\sim 10}{\micro\meter} epitaxial layer~\cite{apts}.

An MPV of \SI{680 \pm 20}{electrons} has been measured, which is higher than measured in other prototypes from the same submission~\cite{apts}.
The discrepancy could be attributed to the ToT calibration procedure, which assumes a uniform in-pixel response of the sensor. During calibration, test pulses are injected and their amplitudes and corresponding ToT values are measured to obtain a single calibration function per pixel (see Section~\ref{lab:tot_calcalibration}). However, when a charged particle passes through the sensor, the non-uniform in-pixel response, which leads to slower charge collection and smaller amplitudes, also results in larger ToT values in the right part of the pixel. This is shown in the in-pixel seed-pixel ToT map of ~\cref{fig:inpixel_tot}.
This occurs because, for lower signal amplitudes, the feedback current of the CSA is not fully saturated, effectively reducing the \textit{ikrum}, slowing the pulse discharge, and increasing the measured ToT value. As a consequence, the MPV of the seed-pixel charge distribution shifts towards larger values.

\subsection{Time Resolution}
\label{sec:time-resolution}
%this might be again more interesting to look at as a function of operation parameters. Discuss ToA variations, how do they affect the temporal resolution, and how are they related to the efficiency issues?
The time residuals are defined as the difference between the trigger timestamp ($\mathrm{t_{trigger}}$) and the H2M hit timestamp ($\mathrm{t_{hit}}$), measured when the chip operates in ToA mode. For clusters with more than one pixel, $\mathrm{t_{hit}}$ is assigned to the earliest pixel timestamp within the cluster.

\begin{figure}[tbp]
    \centering
    \includegraphics[width=1\linewidth, trim={0 1.5cm 0 3.2cm},clip]{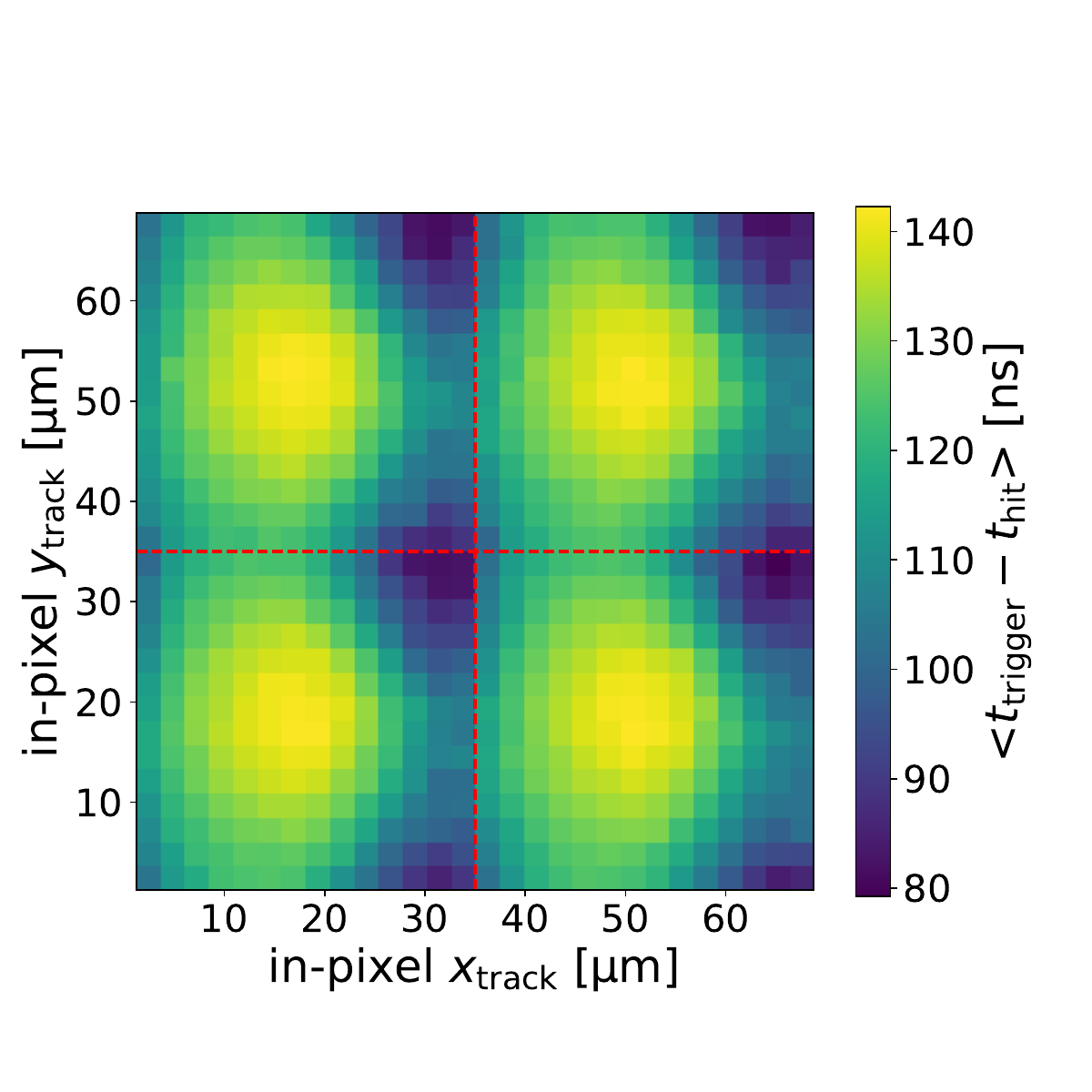}
    \caption{ToA map projected onto four pixels. The sensor is biased at \SI{-3.6}{V}, the threshold is 224 electrons, and a low \textit{ikrum} is set. The pixel cell boundary is marked with a dashed red line.}
    \label{fig:inpixel_toa}
\end{figure}

\cref{fig:inpixel_toa} shows the in-pixel ToA projected onto four pixels. Under the position of the n-wells of the analog front-end, there is a slower charge collection, leading to lower efficiency due to the ballistic deficit, as explained in \cref{sec:eff}. 
In contrast to the efficiency pattern, the ToA asymmetry remains visible even at the lowest thresholds and high sensor bias voltages. 

\begin{figure}[tbp]
    \centering
    \includegraphics[width=1\linewidth, trim={0 0cm 0 0.5cm},clip]{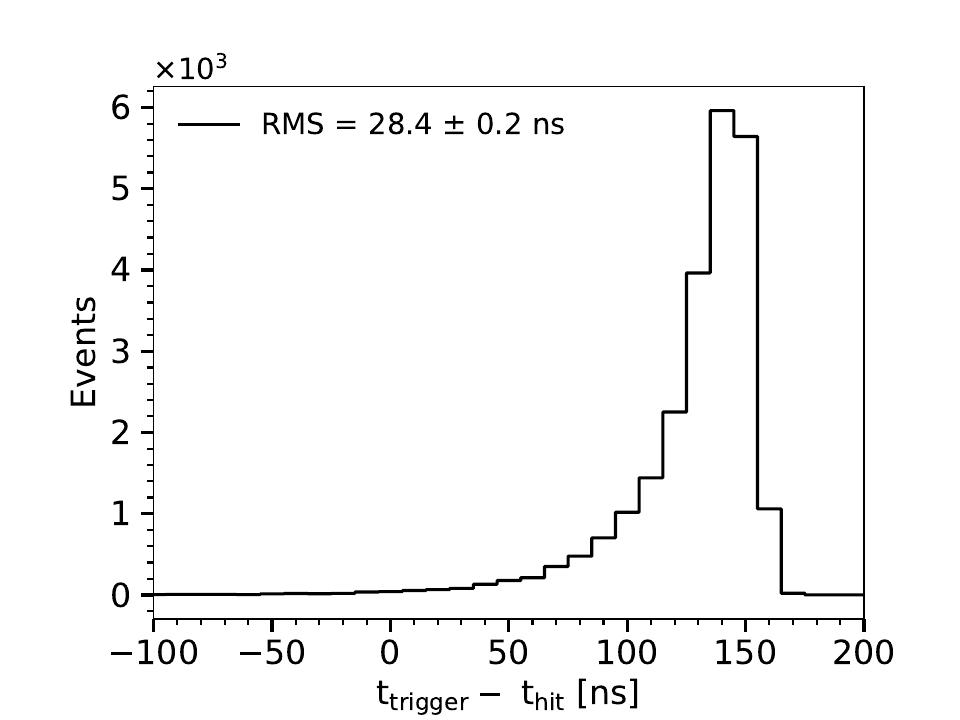}
    \caption{Time residuals distribution. The sensor is biased at \SI{-3.6}{V}, the threshold is 224 electrons, and a low \textit{ikrum} is set.}
    \label{fig:toa_distribution}
\end{figure}

This non-uniform in-pixel response results in non-Gaussian time residuals. An example is shown in \cref{fig:toa_distribution} for a threshold of 224 electrons and a sensor bias voltage of \SI{-3.6}{V}, where a long left-side tail, more pronounced than what would be expected from time walk, is visible. 
The truncated RMS ($\mathrm{\upsigma_{t_{trigger} - t_{hit}}}$), computed within the histogram range shown, is \SI{28.4 \pm 0.2}{\nano\second}. Since the time resolution of the reference detector is negligible compared to that of the DUT resolution, the measured $\mathrm{\upsigma_{t_{trigger} - t_{hit}}}$ is attributed directly to the DUT time resolution.

\begin{figure}[tbp]
    \centering
    \includegraphics[width=1\linewidth, trim={0 1.5cm 0 3.2cm},clip]{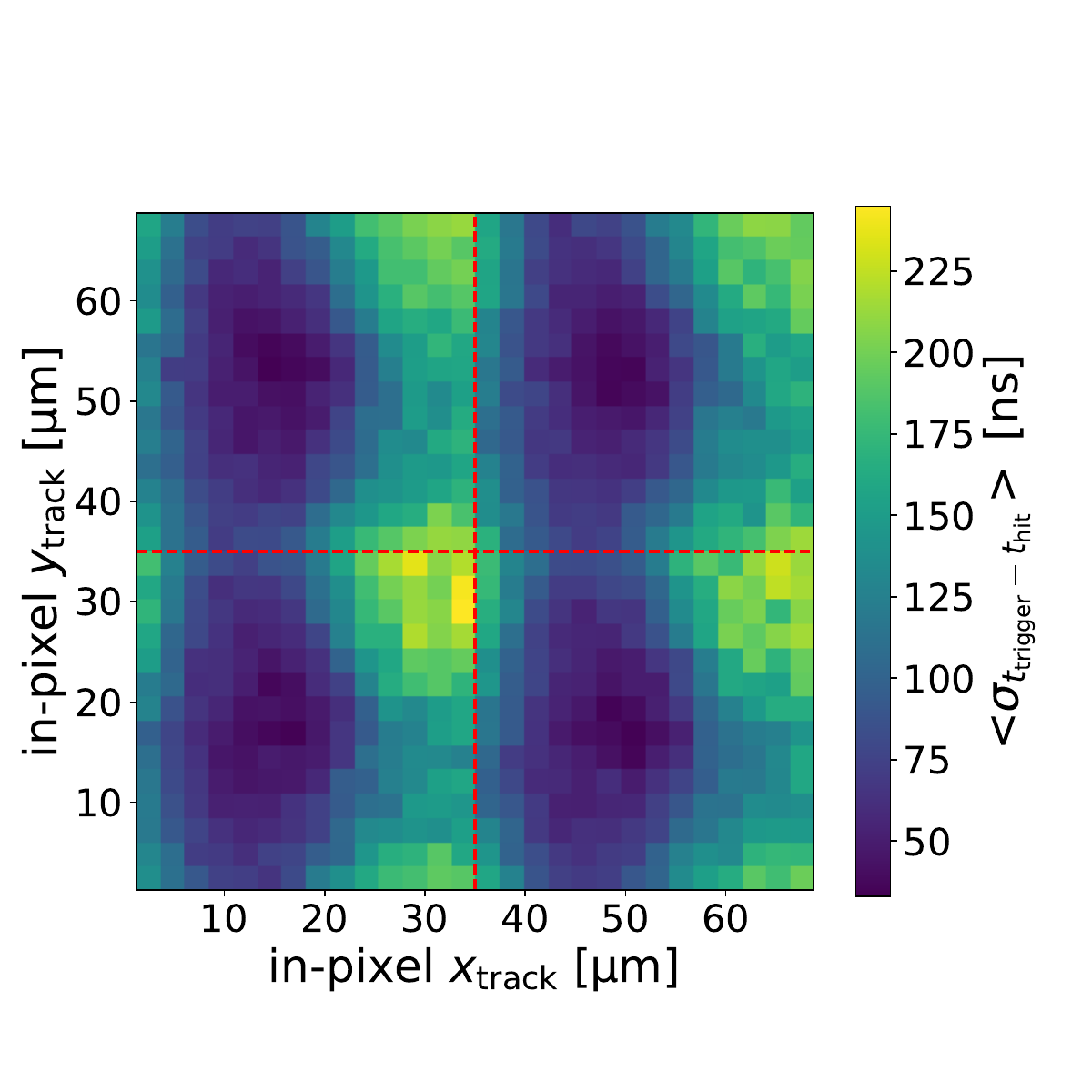}
    \caption{Time residuals RMS map projected onto four pixels. The sensor is biased at \SI{-3.6}{V}, the threshold is 224 electrons and a low \textit{ikrum} is set. The pixel cell boundary is marked with a dashed red line.}
    \label{fig:inpixel_timeres}
\end{figure}

\cref{fig:inpixel_timeres} shows the $\mathrm{\upsigma_{t_{trigger} - t_{hit}}}$ as a function of the in-pixel position. 
Regions with slower charge collection exhibit worse time resolution, whereas a faster and more uniform charge collection is observed near the pixel center. 
The opening in the n-well hosting the analog front-end (see~\cref{fig:h2m_nwells}) affects the time resolution locally, and an improvement in the time resolution is visible in that region.

%The time resolution can be obtained after the quadratic subtraction of the time reference resolution from $\mathrm{\upsigma_{t_{trigger} - t_{hit}}}$. 

\begin{figure}[tbp]
    \centering
    \includegraphics[width=1\linewidth, trim={0 0cm 0 0.5cm},clip]{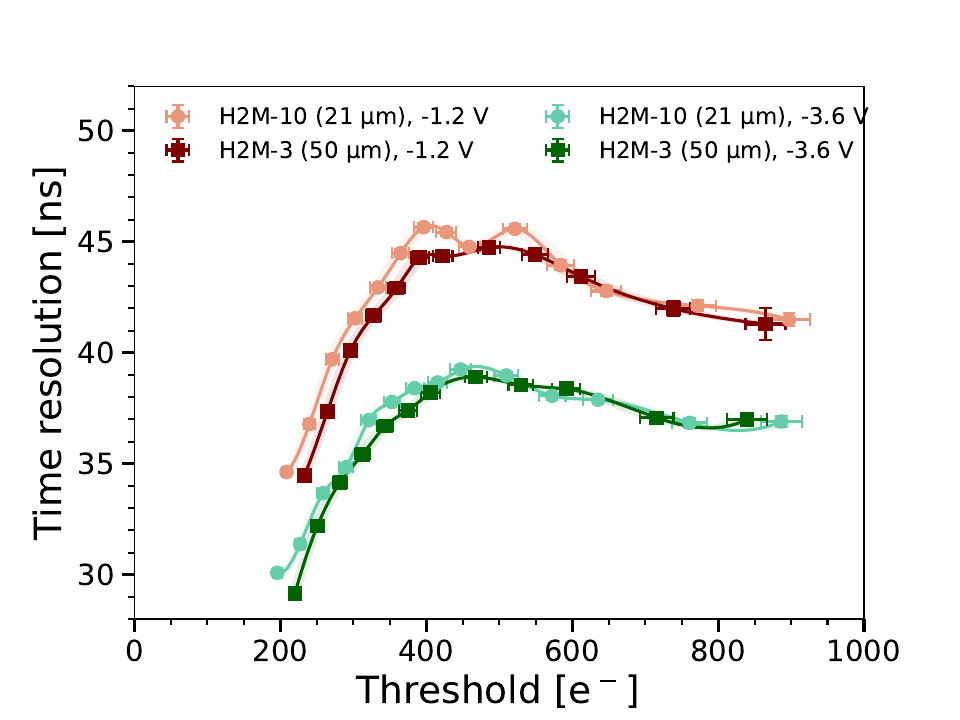}
    \caption{Time resolution as a function of the threshold for two different total chip thicknesses and sensor bias voltages.}
    \label{fig:timeres_vs_thr}
\end{figure}

\cref{fig:timeres_vs_thr} shows the time resolution of H2M as a function of the threshold, for two different total chip thicknesses and sensor bias voltages.
While no significant difference is observed between samples, increasing the sensor bias voltages improves the time resolution by enhancing the charge collection speed. 
%A time resolution of \SI{28.4 \pm 0.2}{\nano\second} is measured at a threshold of \SI{224}{}~electrons and a sensor bias of \SI{-3.6}{V}, limited by the non-uniformity of charge collection across the pixel. -> FOR THE CONCLUSION

\section{Simulation}\label{sec:simulation}
%\textcolor{orange}{State: intro to be redone + clarify the goals/ what should be added/should we compare with measurement (Corentin)}\\
% about the benefits of combining a simulations and measurement program
%To better understand the second peak at low amplitude in the iron spectrum of \cref{fig:thl_calibration} as well as the asymmetric pattern in efficiency and ToA observed in figures~\ref{fig:inpixel_eff} and \ref{fig:inpixel_toa}, simulations of both sensor and front-end were performed. A simulation flow combining technology computer-aided design (TCAD), Monte Carlo and circuit simulations is employed.
To gain a better understanding of the sensor and its behaviour when combined with the circuit of the H2M, simulations were performed. A simulation flow combining technology computer-aided design (TCAD), Monte Carlo and circuit simulations is employed.

This simulation procedure focuses on qualitatively reproducing the behaviour of the chip by using generic configuration such as typical transistors or capacitors without considering variations due to e.g. manufacturing. Though the method could in theory be extended and fitted with measurement results, results presented here are only qualitative and obtained independently from measurement in order to demonstrate root cause and understand general trends. 

\subsection{TCAD Sensor Simulations}
% TCAD, little detail, many reference to e.g. Haakan, +diff
The first step of the simulation consists of computing the electric field inside the sensor with a TCAD simulation. The Sentaurus~\cite{sentaurus} framework from Synopsys is used to first build a 3D finite element model of the sensor and solve Poisson and continuity equations on this mesh.
Considering the goal of understanding the asymmetric response, the TCAD simulation includes not only the collection electrode, epitaxial layer, p-well and deep p-well but also the n-wells of the circuitry inside the deep p-well, as it represents the main asymmetric component of the layout. The layout of these n-wells used in the simulation is shown in \cref{fig:h2m_nwells}; it is slightly simplified compared to the chip layout to eliminate small features of the design. The n-wells hosting the in-pixel circuit are biased at \SI{1.2}{V} while the p-well and substrate are biased at \SI{-1.2}{V} and the collection electrode at \SI{0.8}{V}.

To reproduce the asymmetric features of the sensor design, the boundary conditions of the TCAD simulation need to be carefully set. Ideally, to model the real device included in a large matrix, a pixel with periodic boundary conditions should be simulated, but to achieve numeric convergence, a 2 by 2 pixel matrix composed of a central pixel surrounded by parts of its 8 neighbors was simulated with mirror boundary conditions instead and later cropped to the central pixel. A fine mesh was used in the central pixel, especially close to the collection electrode and in the deep n-type implant, to allow accurate electric field simulation in these sensitive regions.

Transient TCAD simulations in~\cite{h2m_simulations} demonstrated that including the n-wells from the in-pixel circuitry leads to a significant slow down of the charge collection due to low amplitude potential wells in the deep n-type implants underneath the n-wells as can be inferred from the electrostatic potential. This effect appears in combination with the large pixel pitch creating a region of very low lateral electric field and the specific layout of the n-wells. It will be shown later in \cref{sec:optimized_layout} that minor layout modifications can have a significant impact on charge collection slowdown.

The electric field and doping concentration (for computing mobility and lifetime) are extracted from the output of the TCAD simulation to be used in a Monte Carlo simulation.

\subsection{Monte Carlo Sensor Simulations}
\label{sec:monte_carlo}
% AP2, little detail, many reference to e.g. Haakan, +diff
Monte Carlo simulation of the H2M is executed with the Allpix Squared framework~\cite{spannagel2018_apsq}. Different configurations are simulated, either using a uniform charge deposition in order to ignore stochastic effects and emphasize the intrinsic response of the sensor or using Geant4~\cite{geant4_1,geant4_2,geant4_3} to simulate realistic energy deposition corresponding to a perpendicularly incident \SI{5}{GeV} electron beam or X-rays produced by a \ce{^{55}Fe} radioactive source. %The electric field and doping concentration from the TCAD simulation is imported into the framework to allow realistic simulation of charge carrier motion in the sensor\cite{dort2022}. %
The charge carrier mobility is computed using the extended Canali model~\cite{canali} and recombination with the combined Shockley-Read-Hall-Auger model~\cite{shockley1952,hall1959,fossum1982}.
To increase computation speed, the signal induced on the collection electrode is not computed by applying the Shockley-Ramo theorem~\cite{ramo,shockley1938currents} but simply by registering the time of arrival of electrons at the collection electrode which is a reasonable assumption for small collection electrode sensors. 

Several studies~\cite{dort2022,simulationWorkflow2025} demonstrated that Allpix Squared can accurately simulate silicon sensors.  A dedicated comparison of transient TCAD and Allpix Squared simulations of the H2M showed that Allpix Squared reproduces the charge-collection slow down induced by the n-well layout~\cite{h2m_simulations}.

\begin{figure}[tbp]
    \centering
    \includegraphics[width=1\linewidth, trim={0 2cm 0 3cm},clip]{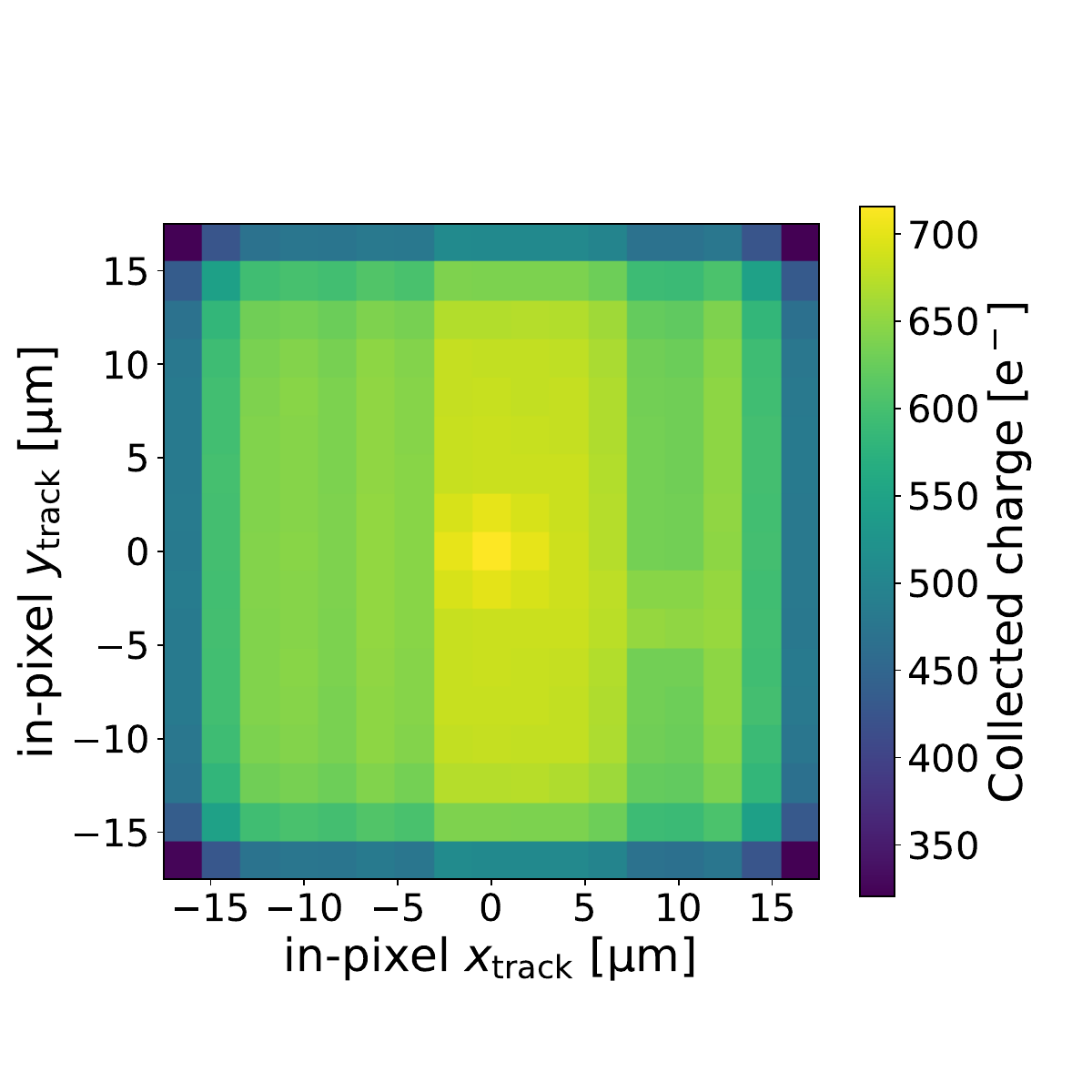}
    \caption{Simulated seed pixel average collected charge map projected onto one pixel for a uniform $\SI{63}{\electron\per\um}$ deposition. The sensor is biased at \SI{-1.2}{V} and an integration time of $\SI{500}{\nano\second}$ is used.}
    \label{fig:simu_MIPscan_charge}
\end{figure}

\begin{figure}[tbp]
    \centering
    \includegraphics[width=1\linewidth, trim={0 2cm 0 3cm},clip]{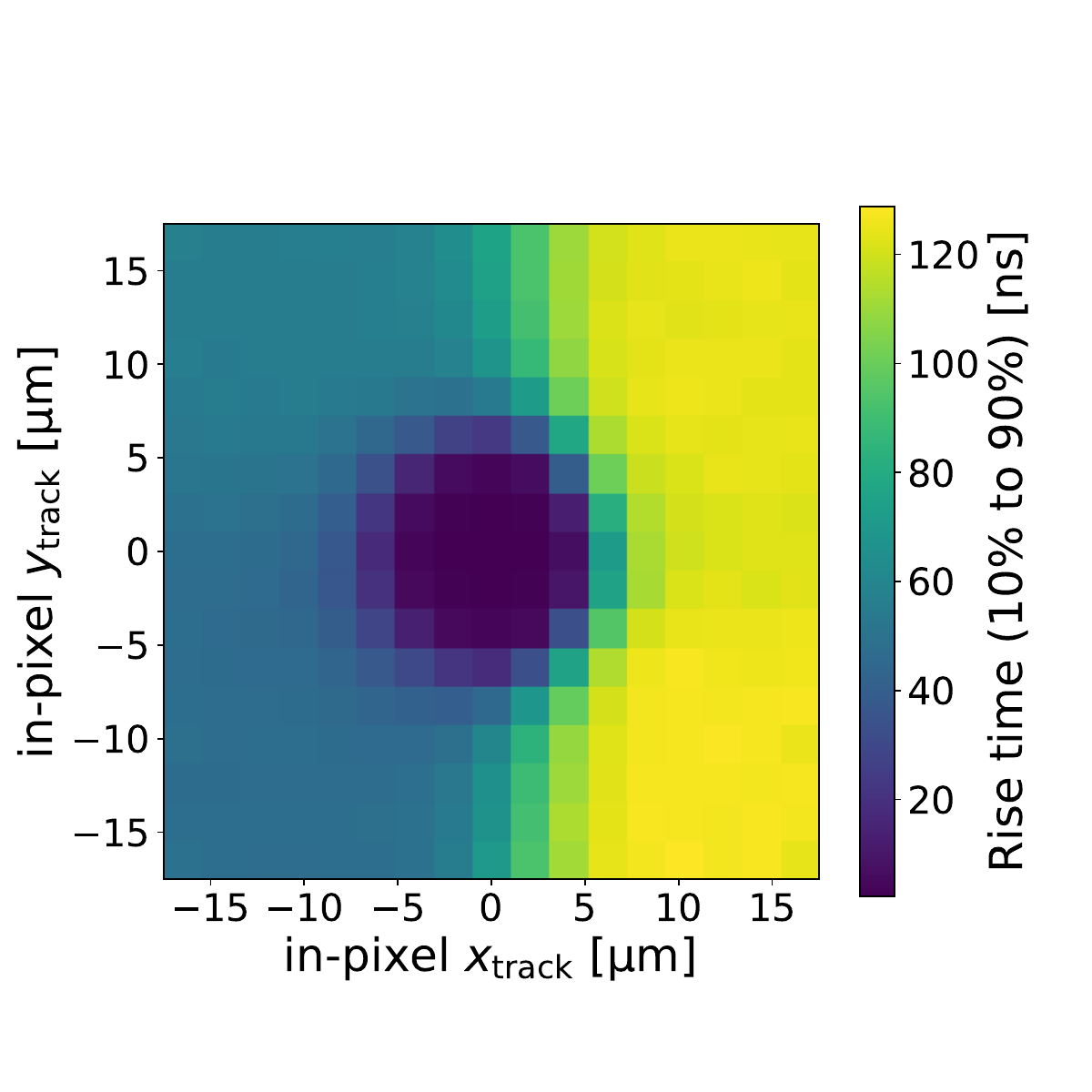}
    \caption{Simulated seed pixel average charge collection time (10\% to 90\%) map projected onto one pixel for a uniform $\SI{63}{\electron\per\um}$ deposition. The sensor is biased at \SI{-1.2}{V}.}
    \label{fig:simu_MIPscan_rise_time}
\end{figure}

\cref{fig:simu_MIPscan_charge} and \ref{fig:simu_MIPscan_rise_time} show respectively the average collected charge and the average collection time (10\% to 90\%) of the seed signal as a function of the particle impinging position in the pixel. For this simulation, the sensor is biased at \SI{-1.2}{V} and $\SI{63}{\electron\per\um}$ are deposited uniformly along lines orthogonal to the sensor surface. Deposition steps are small enough to consider the deposition as perfectly uniform in depth, and such injection is performed with a step of $\SI{0.1}{\micro\meter}$ on both $x$ and $y$ axes before averaging it over bins of $\SI{1}{\micro\meter}$ in \cref{fig:simu_MIPscan_charge,fig:simu_MIPscan_rise_time}. Except in the pixel edge where charge sharing occurs, the collected charge after the $\SI{500}{\nano\second}$ simulation varies between $\SI{640}{\electron}$ and $\SI{700}{\electron}$, with the visible pattern corresponding to the part of the charge deposited within the circuit n-wells (and thus not collected by the collection electrode). On the other hand, the collection time exhibits a clear asymmetric pattern, with the right part of the pixel where the large analog n-well is located having a significantly slower rise time. The combination of both information suggests that the observed asymmetric efficiency pattern might not be caused by charge loss in the sensor but rather by the slower collection leading to partial reset of the amplifier output before full charge collection, i.e.\ the ballistic deficit.

\subsection{Analog Front-End Simulations}
Ballistic deficit is an effect intrinsically related to the front-end circuit and to the shape of the signal induced by charge carrier motion. Considering the non-linearity of the H2M front-end, a simulation of the circuit was executed in a dedicated electronic circuit simulator, Spectre~\cite{spectre}.

In order to simulate a comparable number of events as in measurements in a reasonable time, only the charge sensitive amplifier with Krummenacher feedback is simulated, without including noise. The signal at the input of the front-end is extracted from the Allpix Squared simulation with a script and injected in the form of a current source in parallel to a \SI{3}{\femto\farad} capacitance representing the sensor capacitance plus parasitic capacitance~\cite{apts}.% (this value is not critical due to the use of a CSA).

\begin{figure}[tbp]
    \centering
    \includegraphics[width=1\linewidth, trim={0 0cm 0 0.0cm},clip]{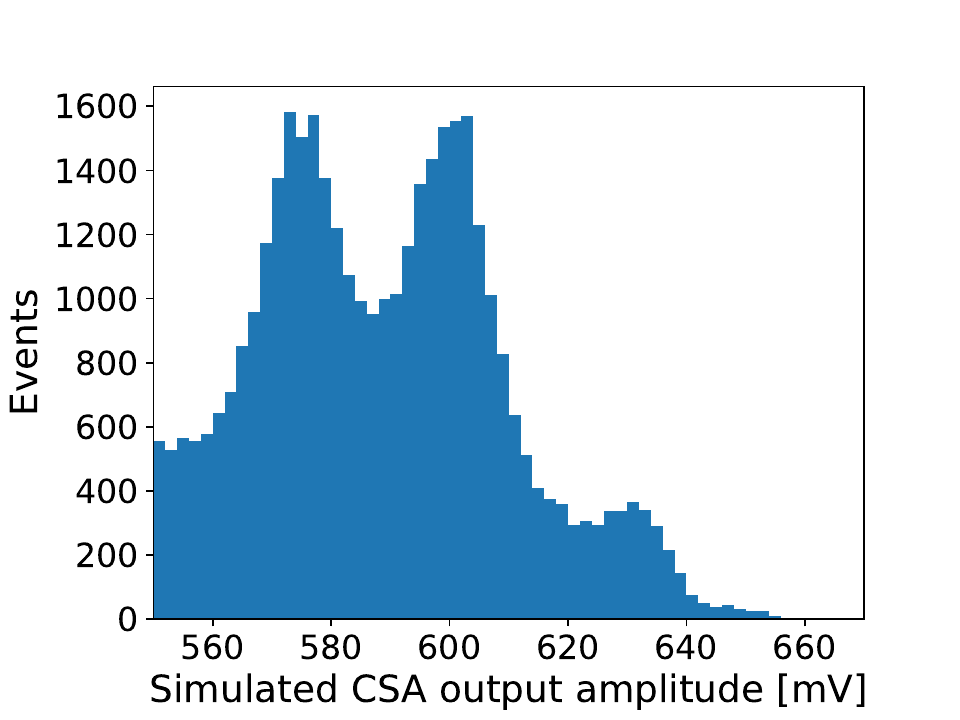}
    \caption{Simulated single pixel amplitude spectrum at the output of the CSA for a \ce{^{55}Fe} source. The sensor is biased at $\SI{-1.2}{\volt}$ and the front-end with high \textit{ikrum}. }
    \label{fig:simu_Fe_spectrum}
\end{figure}

By simulating the analog front-end for two inputs corresponding to an \ce{^{55}Fe} simulation in Allpix Squared, one from the ‘fast’ left part of the pixel and one from the ‘slow’ right part, \cite{h2m_simulations} showed that even though both events collect the same number of charges, the amplitude at the output of the CSA is significantly smaller for the ‘slow’ event than the ‘fast’ one. This simulation confirms that ballistic deficit is a relevant effect in this chip. 
Additionally, the two different collection times identified in \cref{fig:simu_MIPscan_rise_time} will lead to two different magnitudes of ballistic deficit, thus causing a double peak in the iron spectrum. This double peak is indeed present in the measured \ce{^{55}Fe} spectrum of \cref{fig:thl_calibration} but also in the simulated one in \cref{fig:simu_Fe_spectrum}. As expected by the use of a typical simulation without any adjustments, more in-depth comparison of the simulated and measured spectra yields significant differences both in the peak amplitude and overall shape of the spectrum. 

The ballistic deficit impacts mainly the amplitude of the signal, but since the total integrated charge is quite uniform (see \cref{fig:simu_MIPscan_charge}), it has no large effect on the time over threshold (for thresholds low enough compared to the input charge). Therefore only a single peak is observed in the iron spectrum acquired in TOT mode (\cref{fig:amplitude_fe55}).

\subsection{H2M Efficiency Simulation}
\label{sec:simu_efficiency}
\begin{figure}[tbp]
    \centering
    \includegraphics[width=\linewidth,clip]{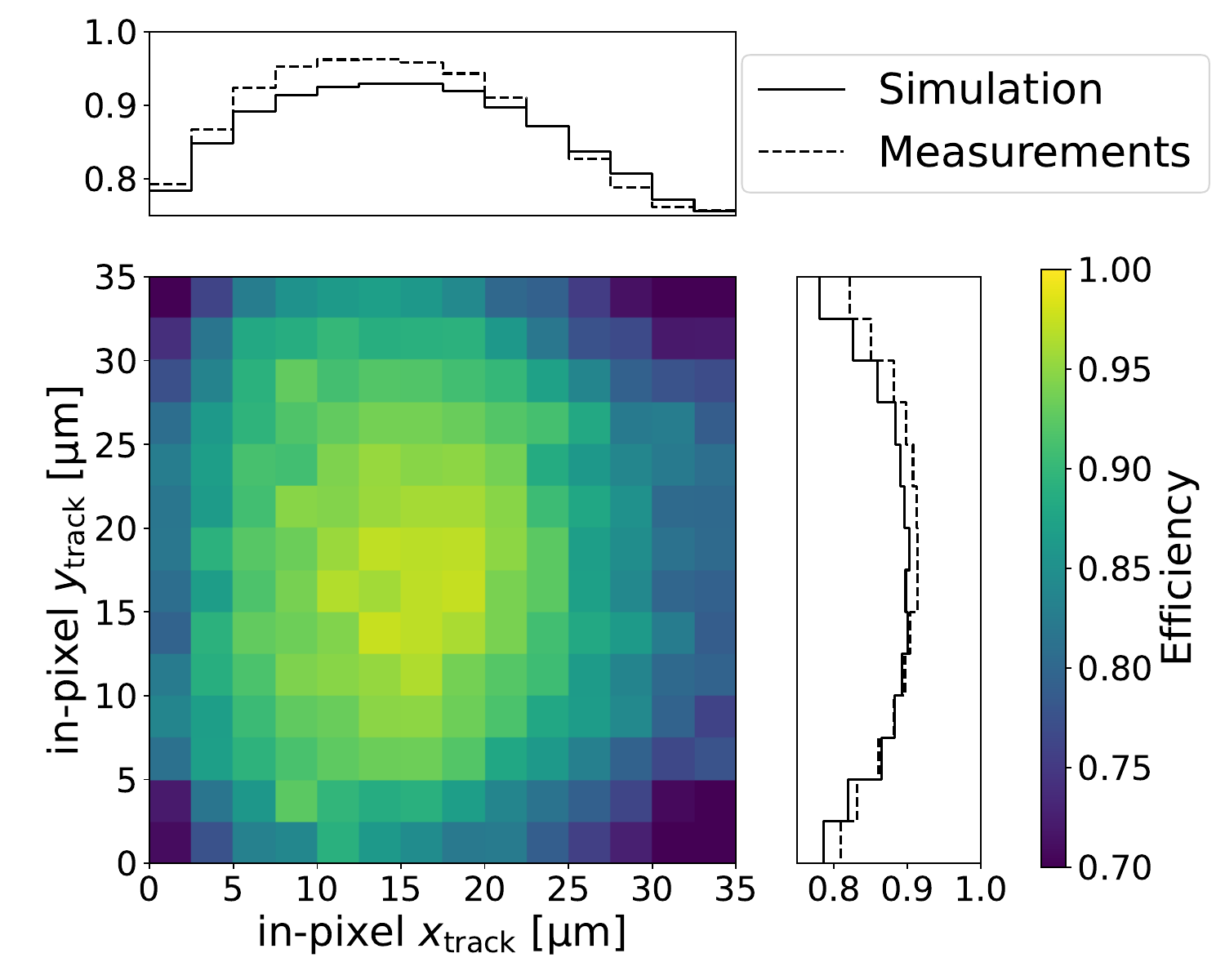}
    \caption{Simulated efficiency maps of the H2M for one pixel and its average projected on each axis. The sensor is biased at \SI{-1.2}{V}, and a high \textit{ikrum} is used. The projected average is compared to measured data from \cref{fig:eff_map_worsesetting}.}
    \label{fig:simu_efficiency_comparison}
\end{figure}

The H2M efficiency in a test beam can be estimated by combining TCAD, Monte-Carlo and analog circuit simulations. The fit of the right $K_\alpha$ peak (at $\approx \SI{600}{\milli\volt}$) of the iron spectrum from \cref{fig:simu_Fe_spectrum} is used for calibration of the simulated amplitude into electrons. Signals are smeared with a noise corresponding to the measured one in \cref{lab:equalization} ($\SI{45}{\electron}$ single-pixel noise and $\SI{17}{\electron}$ for threshold dispersion) and a threshold of $\SI{330}{\electron}$ is used. A smearing of $\SI{3}{\micro\meter}$ on the position of the particle is also applied to account for the telescope resolution. The obtained efficiency map is shown in \cref{fig:simu_efficiency_comparison} and the average efficiency is projected along each axis and compared to measurements. The efficiency pattern reproduces approximately the measured one from \cref{fig:eff_map_worsesetting}. The simulated average efficiency is $86\%$ while the measured one is $88\%$. The simulated efficiency pattern and average efficiency agree with the measurements within 3\%. %Uncertainties on the simulated value are difficult to estimate because of the large number of dependencies (such as doping profiles, models, mesh, timestep...).

%\begin{figure}[tbp]
%    \centering
%    \includegraphics[width=\linewidth, trim={0cm 1cm 0cm 2cm},clip]{simu/H2M_TB_efficiency_330e_1p2V.pdf}
%    \caption{Simulated efficiency maps of the H2M projected onto four pixels. The simulated sensor is biased at \SI{-1.2}{V}, high \textit{ikrum} and a threshold of $\SI{330}{\electron}$ are used. The pixel cell boundary is marked with a dashed red line (all 4 pixel contain the same simulated data).}
%    \label{fig:simu_efficiency}
%\end{figure}

\subsection{Optimization of Circuit Layout}
\label{sec:optimized_layout}

\begin{figure}[tbp]
    \centering
    \includegraphics[width=0.7\linewidth]{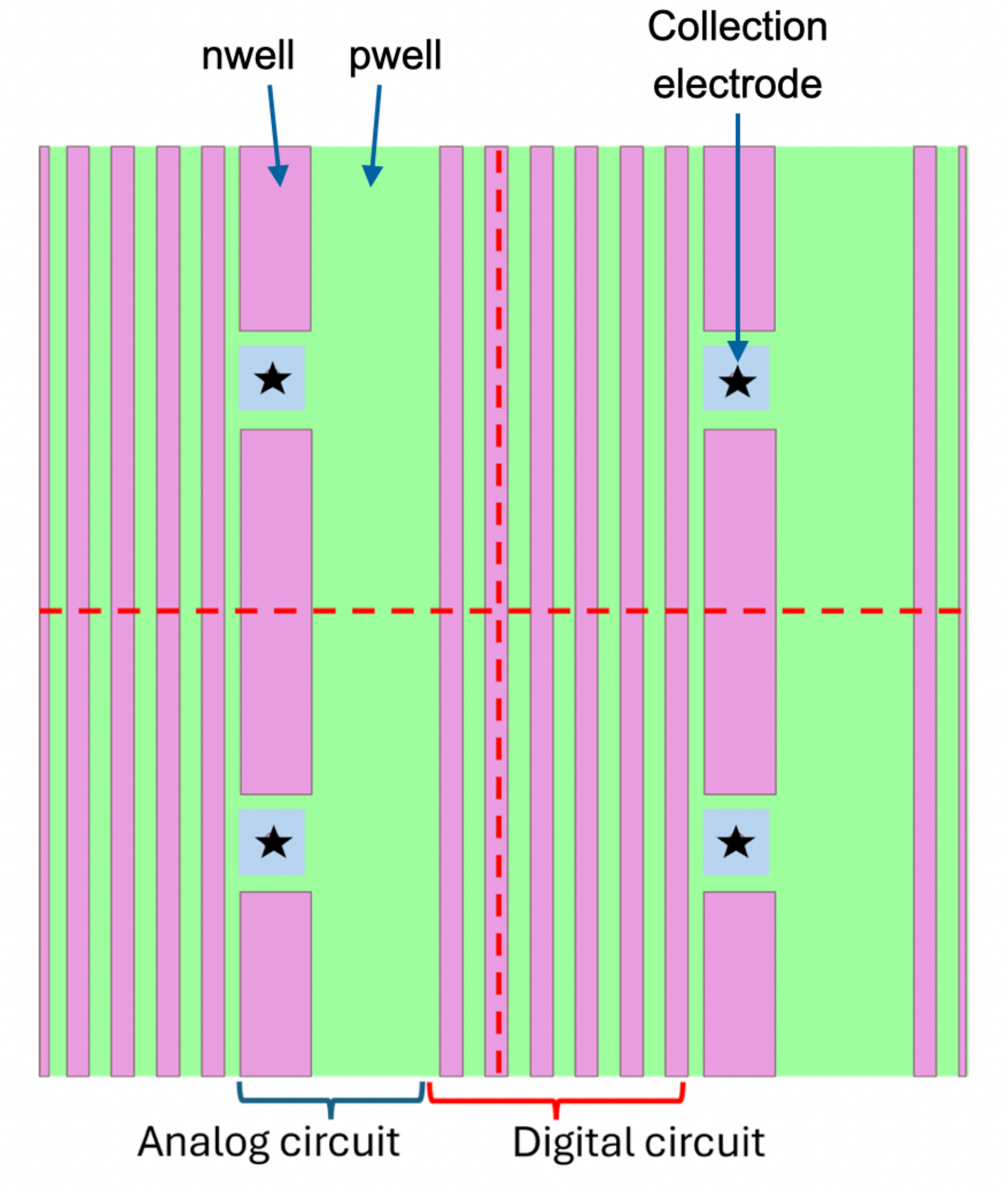}
    \caption{Schematic of the n-wells and p-wells positions within the deep p-well in four pixels for the proposed layout discussed in \cref{sec:optimized_layout}. The position of analog front-end, and digital logic is also indicated.}
    \label{fig:H2Mopt_nwells}
\end{figure}

The process described above, which leads to slower charge collection, is expected to be dependent on the layout of the in-pixel electronics. A design optimization was thus performed in simulation to determine if a more uniform charge collection can be achieved while maintaining the pitch and functionalities. This simulation employed the same simulation workflow, and the only change is the position and shape of the n-wells in the analog circuitry, as shown in Figure~\ref{fig:H2Mopt_nwells}. The area of n-wells is preserved compared to the H2M but no redesign of the analog front-end to fit with this new n-well shape was attempted and simulation use the same circuit as before.

Shape and position are chosen such that the junction between the analog circuit n-well and the p-well in the direction of the collection electrode is located in a region of high electric field (as close as possible to the collection electrode). Other variations of the layout not following this condition were also simulated and produced as expected worse results compared to the one presented here.

\begin{figure}[tbp]
    \centering
    \includegraphics[width=0.9\linewidth, trim={0 2cm 0 0},clip]{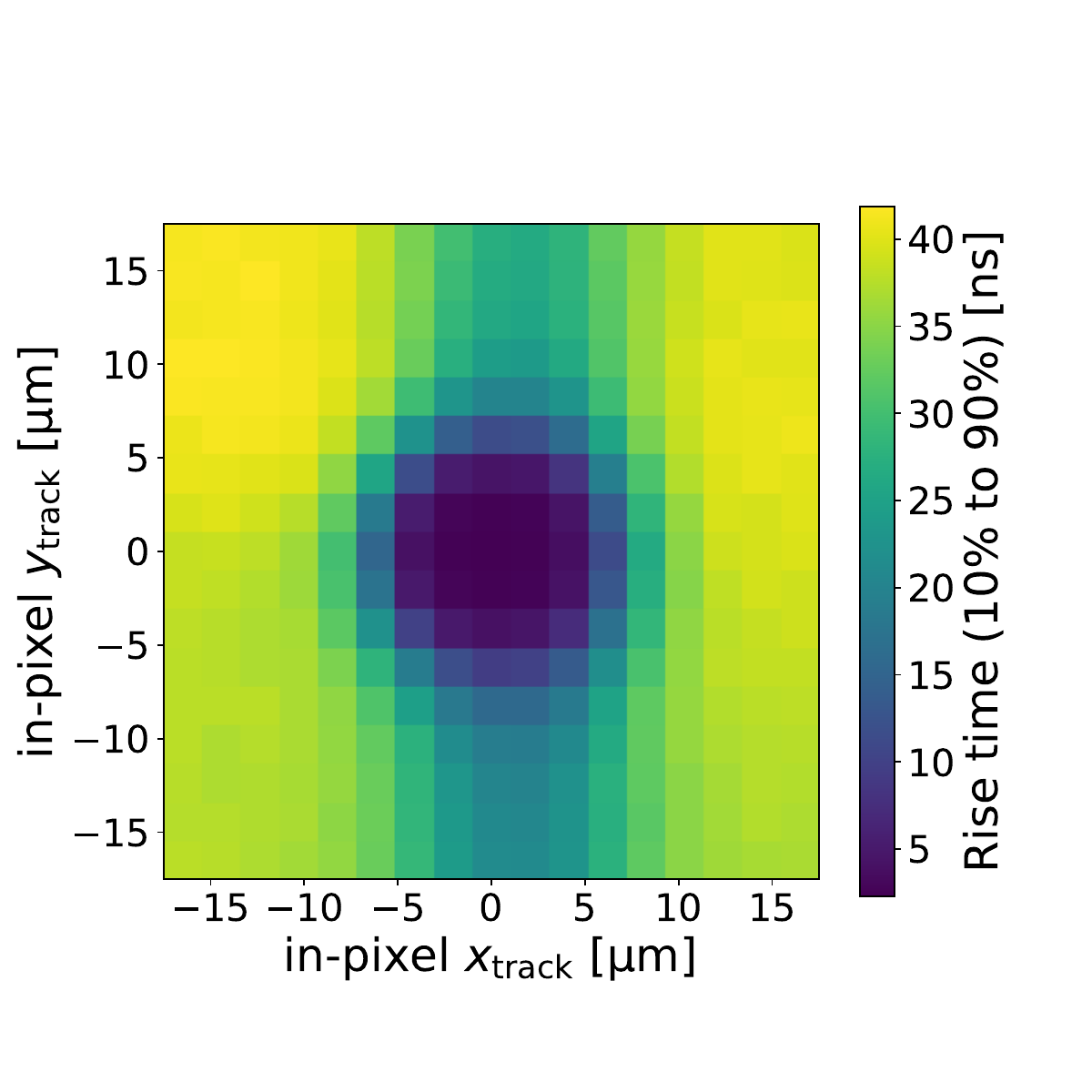}
    \caption{Simulated seed pixel average charge collection time (10\% to 90\%) map projected onto one pixel for a uniform $\SI{63}{\electron\per\um}$ deposition. The n-well layout presented in~\ref{sec:optimized_layout} is used and the sensor is biased at \SI{-1.2}{V}.}
    \label{fig:optimized_layout_MIPscan_rise_time}
\end{figure}

The resulting signal rise time is shown in \cref{fig:optimized_layout_MIPscan_rise_time}. Despite the analog n-well being even wider than in the original H2M, the proposed layout features a three times lower maximal rise time: $\SI{42}{\nano\second}$ to be compared to the $\SI{129}{\nano\second}$  from \cref{fig:simu_MIPscan_rise_time}.

\begin{figure}[tbp]
  \centering
  \includegraphics[width=\linewidth,trim={0 1.5cm 0 3.2cm},clip]{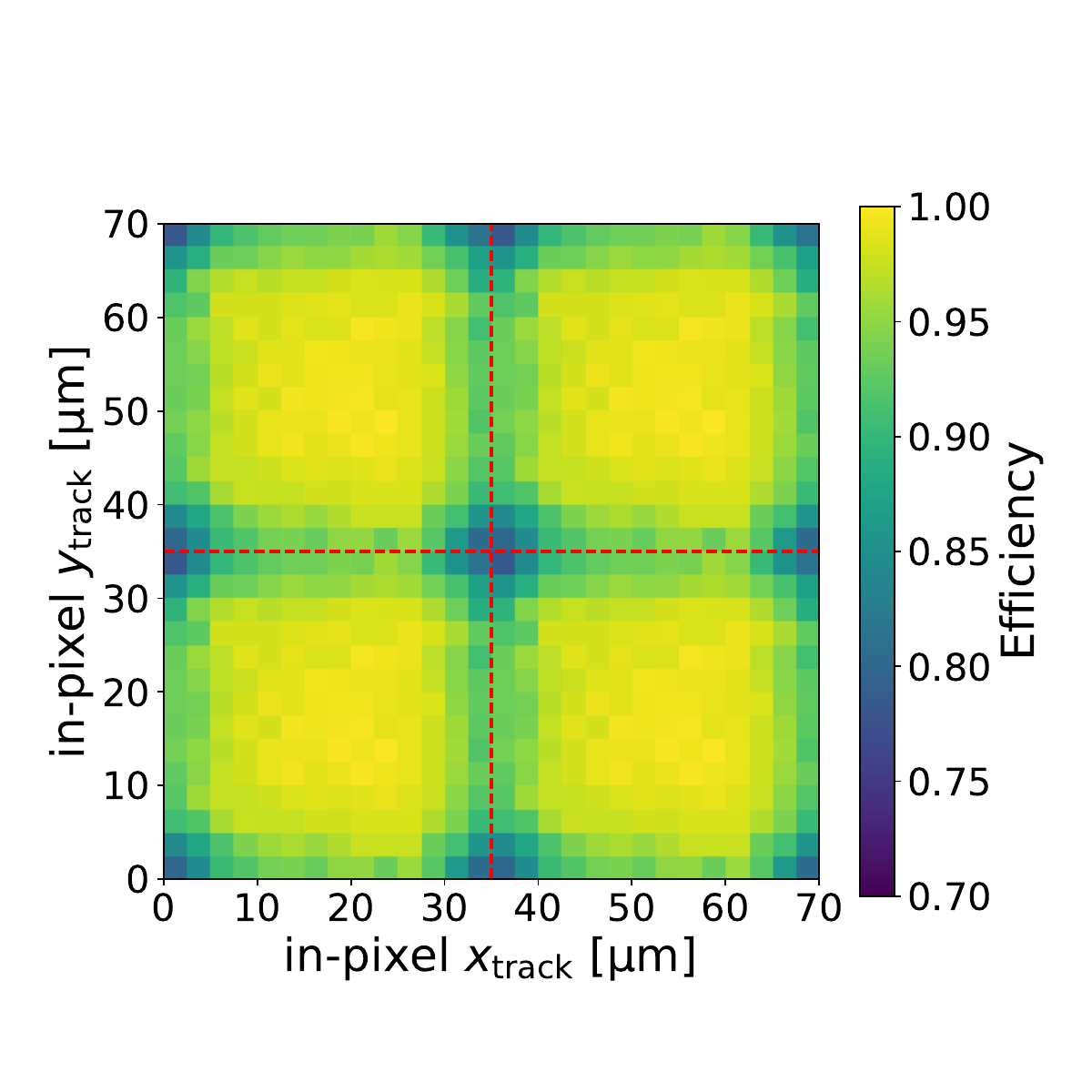}
  \caption{Simulated efficiency maps projected onto four pixels for the proposed modified n-well layout. The simulated sensor is biased at \SI{-1.2}{V}, high \textit{ikrum} and a threshold of $\SI{330}{\electron}$ are used. The pixel cell boundary is marked with a dashed red line (all four pixels contain the same simulated data).}
  \label{fig:simu_efficiency_opt}
\end{figure}

The simulated performance of this new n-well layout can be estimated with the same method as in \cref{sec:simu_efficiency}, the results are presented in \cref{fig:simu_efficiency_opt}. As expected, significant improvement can be observed, the efficiency is more uniform than in \cref{fig:simu_efficiency_comparison} and the average simulated efficiency reaches $95\%$, to be compared to the $86\%$ simulated with the H2M layout.

\section{Conclusion \& Outlook}
\label{sec:summary}

The H2M chip has proven to be fully functional in all four data acquisition modes. All relevant circuit elements work in the foreseen way, and their performance matches specifications and expectations from analog front-end simulations. This shows that the strategy behind the project --- porting the architecture of a hybrid pixel detector readout chip into a MAPS, and applying the digital-on-top workflow ---  works, so that the development of future MAPS may profit from the gained experience. It also validates the performance of the building blocks designed to be a part of the new, compact digital cell library.

The characterization of the H2M chip has shown that the chip can be thinned down to a total thickness of \SI{21}{\micro\meter}, without significant effect on the key performance characteristics. A MIP detection efficiency above \SI{99}{\percent}, is reached for thresholds below 205 electrons, which is among the best results from prototypes produced in the same process, and attributed to the large pixel pitch. However, single-pixel noise and threshold dispersion are found to be on the order of 40 and 25 electrons, respectively, and depend on the bias voltage. This will limit the lowest attainable operation threshold, depending on limits on the fake-hit rate. The spatial hit resolution is \SI{9.3 \pm 0.1}{\micro\meter} at a threshold of 178 electrons, and limited by a combination of the pitch and the cluster size, as expected for sensors employing the modified with a gap layout. The temporal hit resolution goes down to \SI{28.4 \pm 0.2}{\nano\second} at a threshold of 224 electrons, and is limited by the non-uniformity of the in-pixel response.

The cause for this non-uniformity is thought to be
local potential wells slowing down the charge collection in regions of very low electric field, caused by the layout of n-wells within the deep p-well. This leads to an asymmetric temporal response as a function of the in-pixel position. Due to the fast response of the CSA this reduces the pulse amplitude, which results in an accelerated loss of efficiency as a function of threshold. The presented simulation procedure shows to be capable of reproducing this effect qualitatively and is an indispensable tool to understand the underlying mechanism. The effect can be mitigated by reducing the feedback current of the CSA, or by re-positioning the large n-wells in a new design, as the simulations suggest.

%The cause for this non-uniformity are local potential wells slowing down the charge collection in regions of very low electric field. As a first consequence, this leads to a non-uniform temporal response as a function of the in-pixel position. The longer collection time leads to a reduction of the pulse amplitude, due to the ballistic deficit caused by the fast response of the CSA. As a second consequence, this results in an accelerated loss of efficiency as a function of threshold, accompanied by non-uniform in-pixel efficiency. This second effect can be mitigated by reducing the feedback current of the CSA. The presented simulation procedure shows to be capable of reproducing this effect qualitatively and is an indispensable tool to understand the underlying mechanism. The simulation suggests, that the impact of the local potential wells is amplified by the small lateral electric field in the affected area (which is a consequence of the large pitch), and suggests that re-positioning the large n-wells will significantly improve efficiency and charge-collection time. \textcolor{orange}{Paragraph too long?}

Overall, the characterization of the H2M prototype led to a better understanding of MAPS produced in the \SI{65}{\nano\meter} CMOS imaging process of Tower Partner Semiconductor Co. (TPSCo). This knowledge is fundamental to optimize the performance of future sensors --- especially those with a comparably large pitch.

\section*{Acknowledgments}

Some measurements leading to these results have been performed at the Test Beam Facility at DESY Hamburg (Germany), a member of the Helmholtz Association (HGF).

The developments presented in this contribution are performed in collaboration with the CERN EP R\&D program on technologies for future experiments.

This project has received funding from the European Union’s Horizon 2020 Research and Innovation program under GA no 101004761.

 \bibliographystyle{elsarticle-num} 
 \bibliography{bibliography}

@article{snoeys2012,
    title = {Monolithic pixel detectors for high energy physics},
    journal = {Nucl. Instrum. Methods Phys. Res. A},
    volume = {731},
    pages = {125-130},
    year = {2013},
    note = "{Pixel 2012}",
    doi = {10.1016/j.nima.2013.05.073},
    author = {W. Snoeys}
}

@article{heim2017,
    title = {Self-adjusting threshold mechanism for pixel detectors},
    journal = {Nucl. Instrum. Methods Phys. Res. A},
    volume = {867},
    pages = {209-214},
    year = {2017},
    doi = {10.1016/j.nima.2017.06.040},
    author = {Timon Heim and Maurice Garcia-Sciveres}
}

@book{rossi2006,
    author = {Leonardo Rossi and Peter Fischer and Tilman Rohe and Norbert Wermes},
	  title = {Particle Detectors --- From Fundamentals to Applications},
	  publisher = {Springer Berlin, Heidelberg},
	  year = {2006},
    doi = {10.1007/3-540-28333-1}
}

@article{kremastiotis2020,
    author = "Kremastiotis, I.  and  Ballabriga, R.  and  Dort, K.  and  Egidos, N.  and  Munker, M.",
    title = "{CLICTD: A monolithic HR-CMOS sensor chip for the CLIC silicon tracker}",
    doi = "10.22323/1.370.0039",
    journal = "PoS",
    year = "2020",
    volume = "{Topical Workshop on Electronics for Particle Physics (TWEPP2019)}",
    pages = {039}
}

@article{JAKUBEK2011S262,
    title = {Precise energy calibration of pixel detector working in time-over-threshold mode},
    journal = {Nucl. Instrum. Methods Phys. Res. A},
    volume = {633},
    pages = {S262-S266},
    year = {2011},
    note = {11th International Workshop on Radiation Imaging Detectors (IWORID)},
    doi = {10.1016/j.nima.2010.06.183},
    author = {Jan Jakubek}
}

@article{simulationWorkflow2025,
    title = {Simulating monolithic active pixel sensors: A technology-independent approach using generic doping profiles},
    journal = {Nucl. Instrum. Methods Phys. Res. A},
    volume = {1073},
    pages = {170227},
    year = {2025},
    issn = {0168-9002},
    doi = {10.1016/j.nima.2025.170227},
    author = {Håkan Wennlöf and Dominik Dannheim and Manuel {Del Rio Viera} and Katharina Dort and Doris Eckstein}
}

@article{spannagel2018_apsq,
    title = {Allpix2: A modular simulation framework for silicon detectors},
    journal = {Nucl. Instrum. Methods Phys. Res. A},
    volume = {901},
    pages = {164-172},
    year = {2018},
    issn = {0168-9002},
    doi = {10.1016/j.nima.2018.06.020},
    author = {S. Spannagel and K. Wolters and D. Hynds and N. {Alipour Tehrani} and M. Benoit and others}
}

@article{desyii,
    title = {The {DESY II} test beam facility},
    journal = {Nucl. Instrum. Methods Phys. Res. A},
    volume = {922},
    pages = {265-286},
    year = {2019},
    issn = {0168-9002},
    doi = {10.1016/j.nima.2018.11.133},
    author = {R. Diener and J. Dreyling-Eschweiler and H. Ehrlichmann and I.M. Gregor and U. Kötz and others}
}

@article{adenium,
    title="{ADENIUM — A demonstrator for a next-generation beam telescope at DESY}",
    volume={18},
    DOI={10.1088/1748-0221/18/06/p06025},
    number={06},
    journal={J. Instrum.},
    author={Liu, Yi and Feng, Changqing and Gregor, Ingrid-Maria and Herkert, Adrian and Huth, Lennart and others},
    year={2023},
    pages={P06025}
}

@misc{telepix2,
    title = "{TelePix2: Full scale fast region of interest trigger and timing for the EUDET-style telescopes at the DESY II test beam facility}",
    journal = {Nucl. Instrum. Methods Phys. Res. A},
    volume = {1080},
    pages = {170720},
    year = {2025},
    doi = {10.1016/j.nima.2025.170720},
    author = {L. Huth and H. Augustin and L. Dittmann and S. Dittmeier and J. Hammerich and others}
}

@article{tlu,
    title="{The AIDA-2020 TLU: a flexible trigger logic unit for test beam facilities}",
    volume={14},
    DOI={10.1088/1748-0221/14/09/p09019},
    number={09},
    journal={J. Instrum.}, 
    author={Baesso, P. and Cussans, D. and Goldstein, J.},
    year={2019},
    pages={P09019–P09019}
}

@article{eudaq2,
    title="{EUDAQ2—A flexible data acquisition software framework for common test beams}",
    volume={14},
    DOI={10.1088/1748-0221/14/10/p10033},
    number={10},
    journal={J. Instrum.},
    author={Liu, Y. and Amjad, M.S. and Baesso, P. and Cussans, D. and Dreyling-Eschweiler, J. and others}, 
    year={2019},
    pages={P10033–P10033}
}

@article{corry,
    title="{Corryvreckan: a modular 4D track reconstruction and analysis software for test beam data}",
    volume={16},
    DOI={10.1088/1748-0221/16/03/p03008},
    number={03},
    journal={J.Instrum.},
    author={Dannheim, D. and Dort, K. and Huth, L. and Hynds, D. and Kremastiotis, I. and others},
    year={2021},
    pages={P03008}
}

@article{spidr,
    doi = {10.1088/1748-0221/10/12/C12028},
    year = {2015},
    volume = {10},
    number = {12},
    pages = {C12028},
    author = {Visser, J. and Beuzekom, M. van and Boterenbrood, Henk and Heijden, B. van der and Muñoz, J.I. and others},
    title = "{SPIDR: a read-out system for Medipix3 \& Timepix3}",
    journal = {J. Instrum.}
}

@article{gbl,
    title = {A new fast track-fit algorithm based on broken lines},
    journal = {Nucl. Instrum. Methods Phys. Res. A},
    volume = {566},
    number = {1},
    pages = {14-17},
    year = {2006},
    note = {{TIME} 2005: Proceedings of the 1st Workshop on Tracking in High Multiplicity Environments},
    doi = {10.1016/j.nima.2006.05.156},
    author = {V. Blobel}
}

@misc{sps,
    author       = {{CERN}},
    title        = "{Secondary Beam Areas of the PS, SPS machines}",
    year         = {2019},
    note         = {Accessed: 2025-05-13},
    url          = {http://sba.web.cern.ch/sba/}
}

@article{h2m_measurements,
    doi = {10.1088/1748-0221/20/06/C06037},
    year = {2025},
    volume = {20},
    number = {06},
    pages = {C06037},
    author = {Ruiz Daza, S. and Ballabriga, R. and Buschmann, E. and Campbell, M. and Casanova Mohr, R. and others},
    title = "{The H2M Monolithic Active Pixel Sensor — characterizing non-uniform in-pixel response in a 65 nm CMOS imaging technology}",
    journal = {J. Instrum.}
}

@article{h2m_simulations,
doi = {10.1088/1748-0221/20/06/C06052},
url = {https://dx.doi.org/10.1088/1748-0221/20/06/C06052},
year = {2025},
month = {jun},
publisher = {IOP Publishing},
volume = {20},
number = {06},
pages = {C06052},
author = {Lemoine, C. and Ballabriga, R. and Buschmann, E. and Campbell, M. and Casanova Mohr, R. and others},
title = {Impact of the circuit layout on the charge collection in a monolithic pixel sensor},
journal = {Journal of Instrumentation}
}

@article{etacorection,
    title = {Spatial resolution of silicon microstrip detectors},
    journal = {Nucl. Instrum. Methods Phys. Res. A},
    volume = {335},
    number = {1},
    pages = {44-58},
    year = {1993},
    doi = {https://doi.org/10.1016/0168-9002(93)90255-G},
    author = {R. Turchetta}
}

@misc{gbltrackresolutioncalculator,
    title = "{GBL Track Resolution Calculator v2.0}",
    year = {2016},
    doi = {10.5281/zenodo.48795},
    author = {S. Spannagel and H. Jansen}
}

@misc{sentaurus,
    author = {Synopsys},
    title = {Sentaurus},
    url = {https://www.synopsys.com/manufacturing/tcad.html},
    year = {2025},
    note = {Accessed: 2025-07-30}
}

@misc{spectre,
    author = {Cadence},
    title = {Spectre},
    url = {https://www.cadence.com/en\_US/home/tools/custom-ic-analog-rf-design/circuit-simulation.html},
    year = {2025},
    note = {Accessed: 2025-07-30}
}

@article{geant4_1,
    title="{GEANT4--a simulation toolkit}",
    author={Agostinelli, Sea and Allison, John and Amako, K al and Apostolakis, John and Araujo, Henrique and others},
    journal={Nucl. Instrum. Methods Phys. Res. A},
    volume={506},
    number={3},
    pages={250--303},
    year={2003},
    doi={10.1016/S0168-9002(03)01368-8}
}

@article{geant4_2,
    title={Geant4 developments and applications},
    author={Allison, John and Amako, Katsuya and Apostolakis, JEA and Araujo, HAAH and Dubois, P Arce and others},
    journal={IEEE Trans. Nucl. Sci.},
    volume={53},
    number={1},
    pages={270--278},
    year={2006},
    doi={10.1109/TNS.2006.869826}
}

@article{geant4_3,
    title="{Recent developments in GEANT4}",
    author={Allison, John and Amako, Katsuya and Apostolakis, John and Arce, Pedro and Asai, Makoto and others},
    journal={Nucl. Instrum. Methods Phys. Res. A},
    volume={835},
    pages={186--225},
    year={2016},
    doi={10.1016/j.nima.2016.06.125}
}

@article{dort2022,
    title="{Transient Monte Carlo simulations for the optimisation and characterisation of monolithic silicon sensors}",
    author={Ballabriga, Rafael and Braach, Justus and Buschmann, Eric and Campbell, Michael and Dannheim, Dominik and others},
    journal={Nucl. Instrum. Methods Phys. Res. A},
    volume={1031},
    pages={166491},
    year={2022},
    doi={10.1016/j.nima.2022.166491}
}

@misc{nist,
    author    = {R.~D. Deslattes and E.~G. {Kessler Jr.} and P. Indelicato and L. de Billy and E. Lindroth and others},
    title        = {X-ray Transition Energies (version 1.2)},
    year         = {2005},
    note         = {Accessed: 2025-05-13},
    url          = {http://physics.nist.gov/XrayTrans}
}

@article{canali,
    author={Canali, C. and Majni, G. and Minder, R. and Ottaviani, G.},
    journal={IEEE Trans. Electron Devices}, 
    title={Electron and hole drift velocity measurements in silicon and their empirical relation to electric field and temperature},
    year={1975},
    volume={22},
    number={11},
    pages={1045-1047},
    doi={10.1109/T-ED.1975.18267}
}

@article{shockley1952,
    title={Statistics of the recombinations of holes and electrons},
    author = {Shockley, W. and Read, W. T.},
    journal={Phys. Rev.},
    volume={87},
    number={5},
    pages={835},
    year={1952},
    doi={10.1103/PhysRev.87.835}
}

@article{hall1959,
    title={Recombination processes in semiconductors},
    author={Hall, R. N.},
    journal={Proceedings of the IEE - Part B: Electronic and Communication Engineering},
    volume={106},
    number={17S},
    pages={923--931},
    year={1959},
    doi={10.1049/pi-b-2.1959.0171}
}

@article{fossum1982,
    title={A physical model for the dependence of carrier lifetime on doping density in nondegenerate silicon},
    author={Fossum, J. G. and Lee, D. S.},
    journal={Solid-State Electron.},
    volume={25},
    number={8},
    pages={741--747},
    year={1982},
    doi={10.1016/0038-1101(82)90203-9}
}

@article{ramo,
    title={Currents induced by electron motion},
    author={S. Ramo},
    journal={Proc. IRE},
    volume={27},
    number={9},
    pages={584--585},
    year={2006},
    doi={10.1109/JRPROC.1939.228757}
}

@article{shockley1938currents,
    title={Currents to conductors induced by a moving point charge},
    author={W. Shockley},
    journal={J. Appl. Phys.},
    volume={9},
    number={10},
    pages={635--636},
    year={1938},
    doi={10.1063/1.1710367}
}

@article{ballistic_deficit,
    title={Ballistic deficits in pulse shaping amplifiers},
    author={Loo, B.W. and Goulding, F.S. and Gao, D.},
    journal={IEEE Trans. Nucl. Sci.},
    volume={35},
    number={1},
    pages={114--118},
    year={1988},
    doi={10.1109/23.12686}
}

@misc{quadrupoletech,
    AUTHOR = {Ballin, Jamie Alexander and Crooks, Jamie Phillip and Dauncey, Paul Dominic and Magnan, Anne-Marie and Mikami, Yoshiari and others},
    TITLE = "{Monolithic Active Pixel Sensors (MAPS) in a Quadruple Well Technology for Nearly 100\% Fill Factor and Full CMOS Pixels}",
    JOURNAL = {Sensors},
    VOLUME = {8},
    YEAR = {2008},
    NUMBER = {9},
    PAGES = {5336--5351},
    DOI = {10.3390/s8095336}
}

@article{walter,
    author = "Snoeys, Walter  and  Aglieri Rinella, Gianluca  and  Andronic, Anton  and  Antonelli, Matias  and  Baccomi, Roberto  and others",
    title = "{Optimization of a 65 nm CMOS imaging process for monolithic CMOS sensors for high energy physics}",
    doi = "10.22323/1.420.0083",
    journal = "PoS",
    volume = "Pixel2022",
    pages = "083",
    year = "2023"
}

@article{krummenacher,
    title="{Pixel detectors with local intelligence: an IC designer point of view}",
    volume={305},
    journal={Nucl. Instrum. Methods Phys. Res. A},
    author={F. Krummenacher},
    year={1991},
    pages={527-532},
    doi={10.1016/0168-9002(91)90152-G}
}

@article{caribou,
    doi = {10.1088/1748-0221/20/07/C07043},
    year = {2025},
    volume = {20},
    number = {07},
    pages = {C07043},
    author = {Otarid, Y. and Benoit, M. and Buschmann, E. and Chen, H. and Dannheim, D. and others},
    title = {Caribou — A versatile data acquisition system for silicon pixel detector prototyping},
    journal = {J. Instrum.}
}

@misc{medipix4,
    doi = {10.1088/1748-0221/19/02/P02024},
    year = {2024},
    volume = {19},
    number = {02},
    pages = {P02024},
    author = {Sriskaran, V. and Alozy, J. and Ballabriga, R. and Campbell, M. and Christodoulou, P. and others},
    title = "{High-rate, high-resolution single photon X-ray imaging: Medipix4, a large 4-side buttable pixel readout chip with high granularity and spectroscopic capabilities}",
    journal = {J. Instrum.}
}

@article{LLOPART2007485,
title = {Timepix, a 65k programmable pixel readout chip for arrival time, energy and/or photon counting measurements},
journal = {Nucl. Instrum. Methods Phys. Res. A},
volume = {581},
number = {1},
pages = {485-494},
year = {2007},
issn = {0168-9002},
doi = {10.1016/j.nima.2007.08.079},
author = {X. Llopart and R. Ballabriga and M. Campbell and L. Tlustos and W. Wong}
}

@phdthesis{clicpix2,
    author = "Williams, Morag Jean",
    title = "{Evaluation of fine-pitch hybrid silicon pixel detector prototypes for the CLIC vertex detector in laboratory and test-beam measurements}",
    school = "University of Glasgow",
    year = "2021"
}

@article{clicpix,
doi = {10.1088/1748-0221/9/01/C01012},
year = {2014},
month = {jan},
publisher = {},
volume = {9},
number = {01},
pages = {C01012},
author = {P Valerio and J Alozy and S Arfaoui and R Ballabriga and M Benoit and others},
title = "{A prototype hybrid pixel detector ASIC for the CLIC experiment}",
journal = {J. Instrum.},
}

@article{timepix4,
    doi = {10.1088/1748-0221/17/01/C01044},
    year = {2022},
    volume = {17},
    number = {01},
    pages = {C01044},
    author = {Llopart, X. and Alozy, J. and Ballabriga, R. and Campbell, M. and Casanova, R. and others},
    title = {Timepix4, a large area pixel detector readout chip which can be tiled on 4 sides providing sub-200 ps timestamp binning},
    journal = {J. Instrum.}
}

@book{clicdetector,
    title        = "{Detector Technologies for CLIC}",
    editor       = {Dannheim, Dominik and Krüger, Katja and Levy, Aharon and Nürnberg, Andreas and Sicking, Eva},
    year         = {2019},
    publisher    = {CERN},
    series       = {CERN Yellow Reports: Monographs},
    volume       = {1},
    doi          = {10.23731/CYRM-2019-001},
    institution  = {CERN},
    address      = {Geneva}
}

@article{apts,
    title = "{Characterization of analogue Monolithic Active Pixel Sensor test structures implemented in a 65 nm CMOS imaging process}",
    journal = {Nucl. Instrum. Methods Phys. Res. A},
    volume = {1069},
    pages = {169896},
    year = {2024},
    doi = {10.1016/j.nima.2024.169896},
    author = {Gianluca {Aglieri Rinella} and Giacomo Alocco and Matias Antonelli and Roberto Baccomi and Stefania Maria Beole and others}
}

@article{dpts,
    title = "{Digital pixel test structures implemented in a 65 nm CMOS process}",
    journal = {Nucl. Instrum. Methods Phys. Res. A},
    volume = {1056},
    pages = {168589},
    year = {2023},
    doi = {10.1016/j.nima.2023.168589},
    author = {Gianluca {Aglieri Rinella} and Anton Andronic and Matias Antonelli and Mauro Aresti and Roberto Baccomi and others}
}

@article{Ratti,
    title= "{A Front-End Channel in 65 nm CMOS for Pixel Detectors at the HL-LHC Experiment Upgrades}",
    author={L.Ratti and F.DeCanio and M.Manghisoni and V.Re and G.Traversi},
    journal={IEEE Trans. Nucl. Sci.},
    volume={64},
    number={2},
    pages={789--799},
    year={2017},
    doi={10.1109/TNS.2016.2646908}
}

@INPROCEEDINGS{randomtelegraphnoise,
  author={Puglisi, Francesco Maria and Padovani, Andrea and Larcher, Luca and Pavan, Paolo},
  booktitle={2017 IEEE 24th International Symposium on the Physical and Failure Analysis of Integrated Circuits (IPFA)}, 
  title={Random telegraph noise: Measurement, data analysis, and interpretation}, 
  year={2017},
  volume={},
  number={},
  pages={1-9},
  doi={10.1109/IPFA.2017.8060057}}

@Inbook{transient-noise,
author="Vaseghi, Saeed V.",
title="Transient Noise",
bookTitle="Advanced Signal Processing and Digital Noise Reduction",
year="1996",
publisher="Vieweg+Teubner Verlag",
address="Wiesbaden",
pages="314--327",
isbn="978-3-322-92773-6",
doi="10.1007/978-3-322-92773-6_12",
}

@article{fccee,
  author       = {Abada, A. and Abbrescia, M. and AbdusSalam, S. S. and Abdyukhanov, I. and Abelleira Fernandez, J. and et al.},
  title        = {{FCC-ee: The Lepton Collider}},
  journal      = {Eur. Phys. J. Spec. Top.},
  volume       = {228},
  pages        = {261--623},
  year         = {2019},
  doi          = {10.1140/epjst/e2019-900045-4}
}

@article{clic-report, 
title={{CLIC CDR - physics and detectors: CLIC conceptual design report}}, DOI={10.2172/1035023}, abstractNote={This report forms part of the Conceptual Design Report (CDR) of the Compact LInear Collider (CLIC). The CLIC accelerator complex is described in a separate CDR volume. A third document, to appear later, will assess strategic scenarios for building and operating CLIC in successive center-of-mass energy stages. It is anticipated that CLIC will commence with operation at a few hundred GeV, giving access to precision standard-model physics like Higgs and top-quark physics. Then, depending on the physics landscape, CLIC operation would be staged in a few steps ultimately reaching the maximum 3 TeV center-of-mass energy. Such a scenario would maximize the physics potential of CLIC providing new physics discovery potential over a wide range of energies and the ability to make precision measurements of possible new states previously discovered at the Large Hadron Collider (LHC). The main purpose of this document is to address the physics potential of a future multi-TeV e+e- collider based on CLIC technology and to describe the essential features of a detector that are required to deliver the full physics potential of this machine. The experimental conditions at CLIC are significantly more challenging than those at previous electron-positron colliders due to the much higher levels of beam-induced backgrounds and the 0.5 ns bunch-spacing. Consequently, a large part of this report is devoted to understanding the impact of the machine environment on the detector with the aim of demonstrating, with the example of realistic detector concepts, that high precision physics measurements can be made at CLIC. Since the impact of background increases with energy, this document concentrates on the detector requirements and physics measurements at the highest CLIC center-of-mass energy of 3 TeV. One essential output of this report is the clear demonstration that a wide range of high precision physics measurements can be made at CLIC with detectors which are challenging, but considered feasible following a realistic future R and D program.}, author={Berger, E. and Demarteau, M. and Repond, J. and Xia, L. and Weerts, H. and Argonne National Laboratory (United States)}, year={2012}, month={Feb} }

\end{document}